\let\latexKernelLabel\label
\documentclass[aps,pra,10pt,twocolumn,preprintnumbers,amsmath,amssymb,superscriptaddress,showpacs,longbibliography,floatfix]{revtex4-2}

\usepackage{graphicx}
\usepackage{amssymb,amsmath}
\usepackage{upgreek}
\usepackage[usenames, dvipsnames]{color}
\usepackage[svgnames]{xcolor}
\usepackage{booktabs}
\usepackage{tabularx}

\usepackage{makecell}

\usepackage{enumitem}
\setlist[enumerate]{leftmargin=4pt, itemindent=*, labelsep=0.4em, align=left}

\usepackage{placeins}
\usepackage{wasysym}

\newcommand{\ii}{{\rm i}}
\newcommand{\U}{{\rm U}}

\usepackage{tikz}
\usepackage{etoolbox}

\makeatletter
\let\label\latexKernelLabel
\let\ltx@label\label
\patchcmd{\@bibdataout@aps}{author="08"}{author="48"}{}{\PackageError{3450}{Cannot patch APS bibliography control}{Check the REVTeX version.}}
\patchcmd{\@bibdataout@aps}{author="08"}{author="48"}{}{\PackageError{3450}{Cannot patch APS bibliography control}{Check the REVTeX version.}}
\makeatother

\usepackage[colorlinks,bookmarks=true,citecolor=blue, urlcolor=blue, linkcolor=black]{hyperref} 

\newcommand{\tensorSix}{
\begin{tikzpicture}[baseline=-0.1ex, scale=0.2]
    \filldraw[black] (0,0) circle (2pt);
    \draw[semithick] (0,0) -- (0,1);
    \draw[semithick] (0,0) -- (1,1);
    \draw[semithick] (0,0) -- (1,-1);
    \draw[semithick] (0,0) -- (0,-1);
    \draw[semithick] (0,0) -- (-1,-1);
    \draw[semithick] (0,0) -- (-1,1);
\end{tikzpicture}
}

\newcommand{\tensorSixLadder}{
\begin{tikzpicture}[baseline=-0.1ex, scale=0.2]
    \filldraw[black] (0,0) circle (2pt);
    \draw[semithick] (0,0) -- (0,1);
    \draw[semithick] (0,0) -- (1,1);
    \draw[semithick] (0,0) -- (2,1);
    \draw[semithick] (0,0) -- (0,-1);
    \draw[semithick] (0,0) -- (1,-1);
    \draw[semithick] (0,0) -- (2,-1);
\end{tikzpicture}
}

\newcommand{\dd}{\mathrm{d}}

\DeclareMathOperator{\Tr}{Tr}

\begin{document}

\title{Charge Pumping Leptogenesis 
from the Symmetric Mass Generation Phase\\[2mm]
on a Four-Leg Ladder Lattice}

\author{Runze Chi}
\email{rzchi@caltech.edu}
\affiliation{Division of Chemistry and Chemical Engineering,
  California Institute of Technology, Pasadena, California 91125, USA}
\affiliation{Marcus Center for Theoretical Chemistry,
  California Institute of Technology, Pasadena, California 91125, USA}
\author{Meng Zeng}
\affiliation{Max Planck Institute for the Physics of Complex Systems, Nöthnitzer Strasse 38, 01187 Dresden, Germany}

\author{Tao Xiang}
\affiliation{Institute of Physics, Chinese Academy of Sciences, Beijing 100190, China.}
\affiliation{School of Physical Sciences, University of Chinese Academy of Sciences, Beijing 100049, China.}

\author{Yi-Zhuang You}
\email{yzyou@physics.ucsd.edu}
\affiliation{Department of Physics, University of California, San Diego, CA 92093, USA}

\author{Bertrand I. Halperin}
\email{halperin@g.harvard.edu}
\affiliation{Department of Physics, Harvard University, Cambridge, Massachusetts 02138, USA}

\author{Juven Wang}
\email{jw@lims.ac.uk}
\affiliation{Center of Mathematical Sciences and Applications, Harvard University, MA 02138, USA}
\affiliation{London Institute for Mathematical Sciences, Royal Institution, W1S 4BS, UK}

\begin{abstract}

Quantized charge pumping in quantum Hall systems is a bulk topological property, typically
manifesting on gapless boundaries as spectral flow through chiral edge modes upon flux
insertion. When the boundary is anomaly-free, it can be symmetrically gapped by symmetric
mass generation (SMG) interactions. However, the effects on quantized charge pumping at an
interacting gapped boundary are unclear. We address this using density-matrix
renormalization group simulations of a four-leg ladder quasi-(2+1)d realization of the
(1+1)d 3-4-5-0 lattice chiral fermion model with six-fermion interactions at one edge.
We find gapless modes near the interacting edge for both weak and strong
couplings, but the edge can be in a fully gapped SMG phase at intermediate coupling.
We also identify spontaneous breaking of the model's $\mathbb{Z}_2$ symmetry
above a critical interaction strength. Across all three boundary phases, we find
quantized charge pumping with the same pumped charge per flux quantum. The spatial distribution of
the pumped charge varies between phases. Besides charge pumping, we study how a single injected fermion propagates
under open boundary conditions. In the SMG phase, an interaction-induced flavor-mixing
current develops at the endpoints of the gapped edge, which is absent in the weak- and
strong-coupling limits. This current reveals a charge-redistribution mechanism unique to
the SMG phase. Additionally, we propose an analogy between charge pumping in this model
and a topological leptogenesis scenario in the (3+1)d Standard Model. The anomaly-free
chiral U(1) symmetries of the 3-4-5-0 model play a role analogous to the
baryon-minus-lepton symmetry of the Standard Model. In both settings, charge pumping can
be associated with a continuous deformation along a classically defined but
quantum-anomalous direction, while a discrete subgroup remains an exact quantum symmetry.

\end{abstract}

\maketitle

\tableofcontents

\section{Introduction}
Quantized charge pumping is a hallmark transport phenomenon of quantum Hall systems~\cite{Laughlin1981,Thouless1983}. Adiabatic flux insertion induces an integer transfer of charge between boundaries. The response is mediated by gapless chiral edge modes and protected by topology~\cite{Halperin1982,TKNN1982,Niu1985}. %

A natural question is how a quantum Hall system responds to flux insertion when one of its edges is gapped without breaking charge conservation symmetry. In the presence of a chiral anomaly, a single chiral edge cannot be gapped without breaking the symmetry, and therefore remains gapless~\cite{Wang2023}. A prominent mechanism that circumvents this obstruction is symmetric mass generation (SMG)~\cite{Eichten1986,Fidkowski2010, 
Wen:2013ppa-1305.1045, 
Wang2023,
You2018,
Wang2019,
Wang:2018jkc-1809.11171,
Wang:2022ucy2204.14271, You20182,Razamat2021,Tong2022,Butt2021}, in which an anomaly-free set of chiral fermions acquires a many-body gap through interactions while preserving all global symmetries. SMG has been extensively studied from both theoretical and numerical perspectives~\cite{Ayyar2015,He2016}, yet its transport properties, such as charge pumping, remain largely unexplored.

Understanding charge pumping into a gapped edge also bears conceptual relation to leptogenesis scenarios~\cite{Davidson:2008bu0802.2962}, which address the particle-production asymmetry in the early universe.
Instead of the conventional Majorana Leptogenesis~\cite{Fukugita:1986hr}, these  scenarios invoke anomaly, bulk-edge, and topological-phase mechanisms to generate an imbalance between the ordinary-matter and dark-matter sectors, as in Topological Leptogenesis~\cite{Wang:2024auy2501.00607}, with which we will draw
comparisons and analogies 
in a later section.

A concrete two-dimensional lattice realization of SMG is provided by the 3-4-5-0 model~\cite{Chen2013,Wang2023,Wang2019}. Similar models have also been realized through purely one-dimensional lattice constructions~\cite{seifnashri2026,ryan2026,LuShao2026,Zakharov2026,Xu2025staggered}. The Hamiltonian is defined on a two-dimensional spatial lattice of width $L_y$ and length $L_x$,  with continuous time, in the regime $L_x \gg L_y$~\cite{Wang2023, Wang2019}.
The model has four species of fermions, which are noninteracting except on one edge (edge~$B$), where multiparticle interactions interconvert the four species. The model has a topological gap in the (2+1)d bulk for all species, and in the absence of interactions it hosts gapless modes on both (1+1)d edges.
Previous studies of the 3-4-5-0 model on a two-leg ladder ($L_y=2$) showed that, beyond a critical interaction $g_c\simeq 5.7$, edge~$B$ gaps through a Berezinskii--Kosterlitz--Thouless (BKT) transition while edge~A stays gapless~\cite{Zeng2022,Tong2022}. However, the full phase diagram, particularly in the strong-interaction regime, remains unexplored. Moreover, phenomena such as charge pumping can only be studied at larger values of $L_y$. Of particular
interest is the limit of large $L_y$, where there is no coupling between
the $B$ and $A$ edges.
In this paper we perform large-scale density-matrix renormalization group (DMRG) simulations to study the phase diagram of the 3-4-5-0 model on a four-leg ladder and to characterize quantized charge pumping across the distinct boundary phases. We choose our parameters to  minimize the coupling between opposite edges, and we use our  results to discuss implications for systems with larger $L_y$.

\section{Model and method}
\label{sec:model}

The Hamiltonian of the 3-4-5-0 model reads~\cite{Wang2023, Wang2019}
\begin{equation}
H = \sum_{I=1}^{4}\sum_{r,r^{\prime}}\left(t_{I,rr^{\prime}}\, \Psi^{\dagger}_{I,r} \Psi_{I,r^{\prime}} + \text{h.c.}\right) + H_{\mathrm{int}} ,
\label{eq:Ham}
\end{equation}

\begin{figure}[!htbp]
    \centering
    \includegraphics[width=0.95\columnwidth]{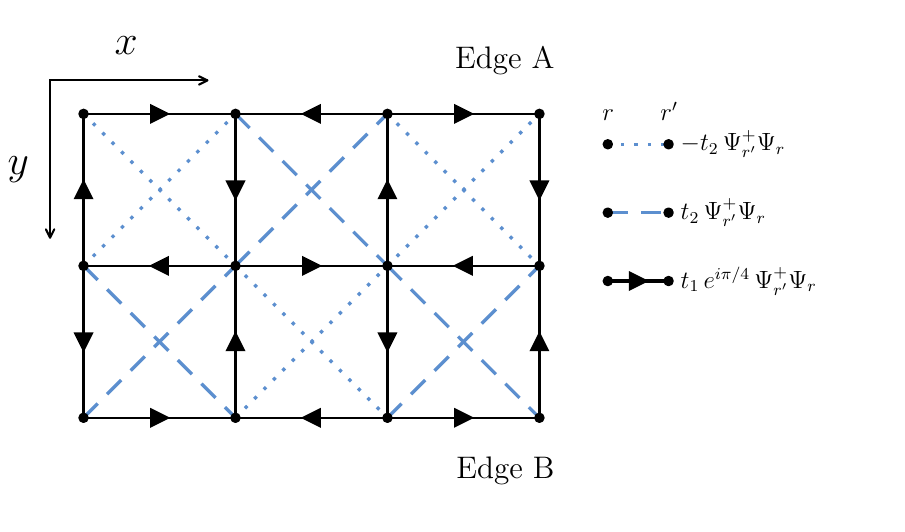}
    \caption{Hopping pattern for fermions of flavors $I=1,2$ on a square lattice in the 3-4-5-0 model. Fermions with flavors $I=3,4$ have complex-conjugate hopping amplitudes. The upper and lower boundaries correspond to edge~A and edge~B, respectively. The hopping amplitudes $t_1$ and $t_2$ are set to $t_1=1.0$ and $t_2=0.5$ throughout this work.
    \label{fig1}} 
\end{figure}
where four flavors of fermions reside on each lattice site, labeled by the flavor index $I=1,2,3,4$. For the first two flavors, the hopping amplitudes $t_{I,rr^{\prime}}$, as illustrated in Fig.~\ref{fig1}, correspond to the Hamiltonian employed in Refs.~\cite{Wang2023, Zeng2022}. We take $t_1=1.0$ as the unit of energy and set $t_2=0.5$. In the noninteracting Chern insulator regime this yields a large bulk gap and sharply boundary-localized edge modes~\cite{Wang2023}, so that moderate $L_y$ already captures the bulk-dominated two-dimensional physics while keeping edge-A/edge-B hybridization weak.
The hopping terms for flavors $I=3,4$ are given by the complex conjugates of those for $I=1,2$, such that they carry opposite Chern numbers.

The free Hamiltonian admits independent phase rotations of the four
fermion flavors. Following Ref.~\cite{Wang:2022fzc2207.14813}, we organize
them into the four $\mathrm{U}(1)$ charge directions in Table~\ref{tab1},
with $q^{(3)}=Kq^{(1)}$, $q^{(4)}=Kq^{(2)}$ and $K=\mathrm{diag}(1,1,-1,-1)$.
Each pair, $(q^{(1)},q^{(2)})$ and $(q^{(3)},q^{(4)})$, is self- and mixed-anomaly-free, satisfying $q^{(a)\intercal}Kq^{(b)}=0$ within the pair.
The two pairs have nonzero mixed anomalies with each other.
Within each pair, the two continuous rotations share a $\mathbb{Z}_3$
subgroup, as their actions on each flavor $I$ obey
\begin{equation}
\begin{aligned}
e^{2\pi\ii q^{(1)}_I/3}&=e^{-2\pi\ii q^{(2)}_I/3},\\
e^{2\pi\ii q^{(3)}_I/3}&=e^{-2\pi\ii q^{(4)}_I/3}.
\end{aligned}
\label{eq:Z3overlap}
\end{equation}
Identifying these shared rotations gives
\begin{equation}
\begin{aligned}
G_{12}&=[\mathrm{U}(1)^{(1)}\times\mathrm{U}(1)^{(2)}]/\mathbb{Z}_3,\\
G_{34}&=[\mathrm{U}(1)^{(3)}\times\mathrm{U}(1)^{(4)}]/\mathbb{Z}_3.
\end{aligned}
\label{eq:G12G34}
\end{equation}

Besides this overlap within each pair, 
the two groups $G_{12}$ and $G_{34}$ share
two order-10 cyclic groups.
Their generators act as $\Psi_I\mapsto e^{2\pi\ii\ell_{a,I}/10}\Psi_I$,
with $\ell_1=(1,-2,1,2)$ and $\ell_2=(2,1,-2,1)$.
Their equivalent representations in the two groups are
\begin{equation}
\begin{aligned}
e^{2\pi\ii\ell_{1,I}/10}&=e^{2\pi\ii(q^{(3)}_I-2q^{(4)}_I)/30}\\
&=e^{2\pi\ii(-3q^{(1)}_I+4q^{(2)}_I)/10},\\
e^{2\pi\ii\ell_{2,I}/10}&=e^{2\pi\ii(2q^{(3)}_I-q^{(4)}_I)/30}\\
&=e^{2\pi\ii(4q^{(1)}_I-3q^{(2)}_I)/10}.
\end{aligned}
\label{eq:Z10reps}
\end{equation}
Thus $G_{12}\cap G_{34}=\mathbb{Z}_{10}^{(1)}\times\mathbb{Z}_{10}^{(2)}$.
The free Hamiltonian therefore has the phase-rotation symmetry
\begin{equation}
G_{\mathrm{free}}\equiv
\frac{G_{12}\times G_{34}}
{\mathbb{Z}_{10}^{(1)}\times\mathbb{Z}_{10}^{(2)}}.
\label{eq:Gfree}
\end{equation}

\begin{table}[htbp]
  \centering
  
  \begin{tabular*}{0.42\textwidth}{@{\extracolsep{\fill}}l cccc}
    \toprule
     & $\Psi_1$ & $\Psi_2$ & $\Psi_3$ & $\Psi_4$ \\
    \midrule
    $\mathrm{U}(1)^{(1)}$ & $3$ & $4$ & $5$ & $0$ \\
    $\mathrm{U}(1)^{(2)}$ & $0$ & $5$ & $4$ & $3$ \\
    $\mathrm{U}(1)^{(3)}$ & $3$ & $4$ & $-5$ & $0$ \\
    $\mathrm{U}(1)^{(4)}$ & $0$ & $5$ & $-4$ & $-3$ \\
    \midrule
    $\mathbb{Z}_{10}^{(1)}$ & $1$ & $-2$ & $1$ & $2$ \\
    $\mathbb{Z}_{10}^{(2)}$ & $2$ & $1$ & $-2$ & $1$ \\
    \bottomrule
  \end{tabular*}
  \caption{Four $\mathrm{U}(1)$ charge assignments $q^{(1)}, q^{(2)}, q^{(3)}, q^{(4)}$ of the free model, following Ref.~\cite{Wang:2022fzc2207.14813}. The common subgroup of $G_{12}$ and $G_{34}$ is $\mathbb{Z}_{10}^{(1)}\times\mathbb{Z}_{10}^{(2)}$, with charge assignments $\ell_1$ and $\ell_2$, respectively.}
  \label{tab1}
\end{table}

In addition to the free-fermion part, the model includes six-body fermion interaction terms acting along the lower boundary (edge~$B$) of the lattice \cite{Wang2023, Wang2019},
\begin{align}
H_{\text{int}} &= H_{\text{int}}^1 + H^{2}_{\text{int}} = \sum_{r\in B} H^{1}_{\text{int},r}+H^{2}_{\text{int},r}
\label{eq:Hint}
\\ \notag 
&= \sum_{r \in B} \Big[
g_1 \left(\Psi_{1,r} \Psi^{\dagger}_{2,r} \Psi^{\dagger}_{2,r+1}
\Psi_{3,r} \Psi_{4,r} \Psi_{4,r+1} + \text{h.c.}\right) \\ \notag 
&\quad +
g_2 \left(\Psi_{1,r} \Psi_{1,r+1} \Psi_{2,r}
\Psi^{\dagger}_{3,r} \Psi^{\dagger}_{3,r+1} \Psi_{4,r} + \text{h.c.}\right)
\Big].
\end{align}

The two interaction channels change the four flavor particle numbers by
$\pm\ell_1$ and $\pm\ell_2$, respectively.
For nonzero $g_1,g_2$, the interactions preserve $G_{12}$ and explicitly break
$G_{34}$ down to the shared $\mathbb{Z}_{10}^{(1)}\times\mathbb{Z}_{10}^{(2)}$ subgroup.
This subgroup already lies in $G_{12}$, so the remaining continuous
symmetry is $G_{12}$.
At $g_1=g_2$, the additional $D_4$ symmetry enlarges the symmetry group to

\begin{equation}
\label{eq:G}
G_{\text{full}} \equiv
\bigl[(\mathrm{U}(1)^{(1)}\times\mathrm{U}(1)^{(2)})/\mathbb{Z}_3\bigr]
\rtimes D_4.
\end{equation}

The symmetry $D_4$, present at $g_1=g_2=g$, is generated by two $\mathbb{Z}_2$ antiunitary symmetries,
$U^{R}_{(1,4)} = \mathcal{R}_{(1,4)}\mathcal{K}$ and $U^{R}_{(1,3)} = \mathcal{R}_{(1,3)}\mathcal{K}$,
where $\mathcal{K}$ denotes complex conjugation in the site occupation basis and $\mathcal{R}_{(1,4)}$ is the flavor relabeling
\begin{equation}
\Psi_1 \rightarrow \ii \Psi_4,\; \Psi_4 \rightarrow \ii \Psi_1,\;
\Psi_2 \rightarrow \ii \Psi_3,\; \Psi_3 \rightarrow \ii \Psi_2 ,
\label{Z_2}
\end{equation}
while $\mathcal{R}_{(1,3)}$ combines a relabeling with a particle--hole conjugation,
\begin{equation}
\Psi_1 \leftrightarrow \Psi_3, \;
\Psi_2 \rightarrow \ii \lambda(r)\, \Psi_4^{\dagger} ,\; \Psi_4 \rightarrow -\ii \lambda^{*}(r)\, \Psi_2^{\dagger} ,
\label{eq:R13}
\end{equation}
with the staggered phase
\begin{equation}
\lambda(r) \;=\; (-1)^{r_x-1}\, (-\ii)^{\,(r_x+r_y)\bmod 2} ,
\label{eq:lambda}
\end{equation}
and $r=(r_x,r_y)$ denotes the integer site coordinates of Fig.~\ref{fig1}, with $r_x=1,\ldots,L_x$ and $r_y=1,\ldots,L_y$. Edge~A is at $r_y=1$, and the leftmost column is at $r_x=1$.

The two generators multiply to a unitary particle--hole symmetry
$U_{\mathcal{C}} = U^{R}_{(1,4)} U^{R}_{(1,3)} U^{R}_{(1,4)} U^{R}_{(1,3)}$, which acts as
\begin{gather}
U_{\mathcal{C}}\, \Psi_{I,r}\, U_{\mathcal{C}}^{\dagger} \;=\; \ii\, \alpha_I\, \lambda_I(r)\, \Psi^{\dagger}_{I,r}, \\ \notag
\lambda_{1,2}(r) \;=\; \lambda^{*}(r) ,
\qquad
\lambda_{3,4}(r) \;=\; \lambda(r) ,
\end{gather}
with $(\alpha_1,\alpha_2,\alpha_3,\alpha_4)=(-1,1,1,-1)$.
The group may also be presented as
$D_4=(\mathbb{Z}_2^{R} \times \mathbb{Z}_2^{\mathcal{C}}) \rtimes \mathbb{Z}_2^{(1,3)}$, with the three factors generated by $U^{R}_{(1,4)}$, $U_{\mathcal{C}}$ and $U^{R}_{(1,3)}$ respectively.

We have investigated the ground-state phase diagram of the Hamiltonian~\eqref{eq:Ham} using DMRG~\cite{White1992}, as implemented in the ITensor library~\cite{ITensor,ITensor-r0.3}. We employ open boundary conditions along the transverse $y$-direction and consider both open and twisted boundary conditions along the longitudinal $x$-direction.

DMRG is formulated in terms of one-dimensional matrix product states, so the two-dimensional lattice must be mapped onto a one-dimensional chain. To reduce the local Hilbert-space dimension, we rearrange the four fermion flavors at each site into separate chains: flavors $I=1,2$ are arranged alternately row by row, while layers $I=3,4$ are unfolded to carry the same effective Chern number as $I=1,2$, with edge~B of layers $I=3,4$ positioned adjacent to edge~B of layers $I=1,2$. This maps the $L_x \times L_y$ lattice with four fermions per site into $4 L_y$ coupled chains of length $L_x$,  with a single fermion per site. We focus on the $\mathrm{U}(1)^{(1)}
\times 
\mathrm{U}(1)^{(2)}$ charge sector   $(Q^{(1)},Q^{(2)})=(6L_xL_y, 6L_xL_y)$ corresponding to half-filling in the noninteracting limit, and also the sector with one additional fermion. The mapping places the edge~B interactions near the center of the chain to minimize entanglement growth. We retain bond dimensions up to $D=17500$, with truncation errors below $5\times10^{-4}$ for all parameters studied.

\section{Weak and strong interaction limits}

We first analyze the two limiting cases of the model.
In the weak-interaction limit, the six-body interactions are irrelevant under renormalization, so that the system reduces to decoupled Chern insulators with both edges gapless.

In the strong-interaction limit, the ground state factorizes into an edge-B state set by $H^{1}_{\mathrm{int}}+H^{2}_{\mathrm{int}}$ and a bulk state of four decoupled Chern insulators on the remaining $L_x\times(L_y-1)$ sites, with emergent gapless edge modes on the row adjacent to edge~B. The nontrivial task is therefore to determine the edge~B state.

On edge~B, $H^{1}_{\mathrm{int}}$ and $H^{2}_{\mathrm{int}}$ commute and are related by the antiunitary $\mathbb{Z}_2^{R}$ symmetry. The two interactions therefore share a common eigenbasis, and we shall first study $H^{1}_{\mathrm{int}}$ alone. We rewrite   $H^{1}_{\mathrm{int}}=g_1\sum_r \left(M_r^{\dagger} M_{r+1}^{\dagger} D_r + \mathrm{h.c.}\right)$ , where the monomer $M_r=\Psi^{B}_{2,r}$ is a single edge-B fermion and the dimer $D_r=\Psi^{B}_{1,r}\Psi^{B}_{3,r}\Psi^{B}_{4,r}\Psi^{B}_{4,r+1}$ is a four-fermion composite on the bond $(r,r{+}1)$. Since this term locally exchanges a dimer for two adjacent monomers, we conjecture a ground-state manifold spanned by non-overlapping monomer/dimer configurations, which a dual transformation maps onto the PXP model~\cite{Fendley2004,Lesanovsky2012}.

As discussed in Appendix~\ref{app:A-ground},
the ground state of $H^{1}_{\mathrm{int}}$ falls within the zero energy  sector of $H^{2}_{\mathrm{int}}$. Therefore, it is an energy eigenstate of the decoupled edge B 
Hamiltonian $H^{1}_{\mathrm{int}}+H^{2}_{\mathrm{int}}$,
with total energy equal to the ground state energy of $H^{1}_{\mathrm{int}}$. Moreover, we find that it is a ground state of 
$H^{1}_{\mathrm{int}}+H^{2}_{\mathrm{int}}$.

The $\mathbb{Z}_2^{R}$ symmetry maps this monomer--dimer manifold onto another manifold at the same energy but in a distinct $\mathrm{U}(1)^{(1)}\times\mathrm{U}(1)^{(2)}$ charge sector, with the roles of $H^{1}_{\mathrm{int}}$ and $H^{2}_{\mathrm{int}}$ interchanged.  These states are orthogonal to each other.
The particle--hole symmetry $U_{\mathcal{C}}$ further maps these two states to two other states in the
complementary charge sectors. Each of these four states is stabilized by a single antiunitary $\mathbb{Z}_2$ symmetry, generated by either $U^{R}_{(1,3)}$ or $U^{R}_{(1,3)}U_{\mathcal{C}}$, and the two stabilizers are exchanged by the broken $\mathbb{Z}_2^{R}$ symmetry (see Appendix~\ref{app:A-ground} for the group analysis). At half filling, the four decoupled Chern insulators accommodate the four
edge-B charge assignments at equal energy, so the fourfold degeneracy carries over to the
full system, which thus spontaneously breaks $D_4$ down to a single antiunitary
$\mathbb{Z}_2$ in the strong-interaction limit.

As we discuss below, the situation for strong but finite coupling should be similar to the case of $g=\infty$. Edge B will be effectively decoupled from the bulk, and there will be four gapless chiral modes centered on the lattice row adjacent to edge B.
Virtual hopping through edge~B generates weak effective interactions, suppressed by its excitation gap of order $g$. The
 perturbed Hamiltonian preserves $\mathrm{U}(1)^{(1)} \times \mathrm{U}(1)^{(2)}$. Under these symmetries no two- or four-fermion gap-opening backscattering is allowed, and the two leading six-fermion backscattering operators remain irrelevant for sufficiently large $g$. The emergent edge therefore remains gapless when hybridization with edge~A is negligible.

\begin{figure}[htbp]
    \centering
    \includegraphics[width=0.82\columnwidth]{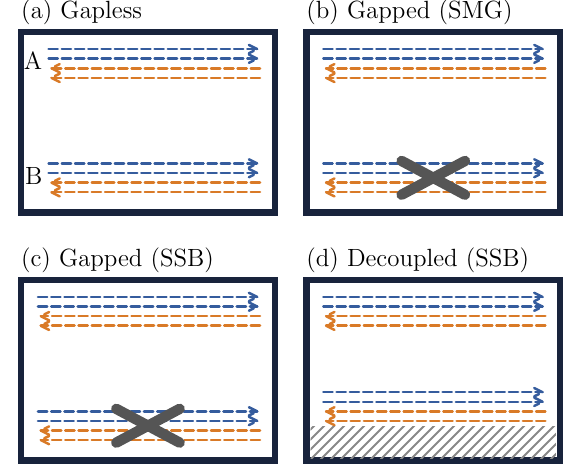}
    \caption{Schematic of the possible phases in the large-$L_y$ limit. (a) Weak-coupling phase with gapless modes on both  edges. (b) Intermediate phase with a fully gapped edge~B and unbroken $D_4$ symmetry (SMG). (c) Intermediate phase with a fully gapped edge~B and spontaneously broken symmetry (SSB). (d) Decoupled phase in which edge~B breaks the symmetry and decouples from the bulk, producing an emergent gapless mode on the adjacent bulk row. A further SSB phase, with a single pair of gapless modes on this emergent edge, is not shown. Different kinds of SSB are possible, with the $\mathbb{Z}_2^{R}$ symmetry alone or with both $\mathbb{Z}_2^{R}$ and $\mathbb{Z}_2^{\mathcal{C}}$ broken.} \label{fig2}
\end{figure}

\section{Phase diagram}

We have seen that there should be gapless propagating modes on or near edge B in the limits of weak or strong interactions.  However, based on our DMRG analysis, we believe there is an intermediate range of coupling constants where these gapless modes are absent.

In the fully gapless phases at weak and strong couplings, we expect that there will be four gapless chiral modes at or near edge B, two left movers and two right movers. If one enters the SMG phase from the weak coupling side, we would expect all four modes to be simultaneously gapped. On the strong coupling side, however, due to the $\mathbb{Z}_2^{R}$ symmetry breaking, there could be a region of coupling constants where two modes are gapped but two others remain gapless. Classified by the gap structure near edge~B alone, the line $g_1=g_2$ could therefore host four phases: gapless, fully gapped, partially gapped, and decoupled. The phases can be further distinguished by which subgroup of $D_4$ survives. A fully gapped edge~B may either preserve the full $D_4$, or break it spontaneously, while a partially gapped edge~B necessarily breaks $\mathbb{Z}_2^{R}$. The decoupled phase also breaks the $D_4$ symmetry, as the large-interaction analysis above shows. Combining the two descriptions, the phase diagram is enriched to five phases: gapless, SMG, SSB fully gapped, SSB partially gapped, and decoupled.

Four of these phases are sketched in Fig.~\ref{fig2}. The SSB partially gapped phase is omitted. The SSB phases break at least the $\mathbb{Z}_2^{R}$ symmetry. The group analysis in \hypersetup{linkcolor=black}Appendix~\ref{app:A-finite} allows the particle-hole symmetry $\mathbb{Z}_2^{\mathcal{C}}$ to break either together with $\mathbb{Z}_2^{R}$ or at a later transition. Each SSB phase can moreover appear in a variant where $\mathbb{Z}_2^{\mathcal{C}}$ is broken as well.

According to our calculations, when $g_1=g_2 $ there is a phase transition at $g \approx 6.0 $ above which the $\mathbb{Z}_2^{R}$ symmetry is broken. A gapped SMG phase appears to set in at a smaller value of $g$, perhaps $g \approx 5, $ which persists to a slightly larger value of $g$, perhaps $g \approx 6$. It is not clear to what extent  the relatively rapid disappearance of the gapped phase on the large $g$ side is simply a result of a decoupling of edge B at large coupling or is strongly enhanced by the $\mathbb{Z}_2^{R}$ symmetry breaking. 

On the weak-coupling side, where the system retains its $\mathbb{Z}_2^{R}$ symmetry, the transitions in both channels should  occur simultaneously. On the strong coupling side, the two transitions may be separated. Transitions are expected to be of the BKT type, where a perturbing operator becomes relevant at a critical value of the coupling constant, as was found consistent with the two-leg ladder simulations  in  
Ref.~\cite{Zeng2022}.

\begin{figure}[!t]
\centering
\includegraphics[width=0.92\columnwidth]{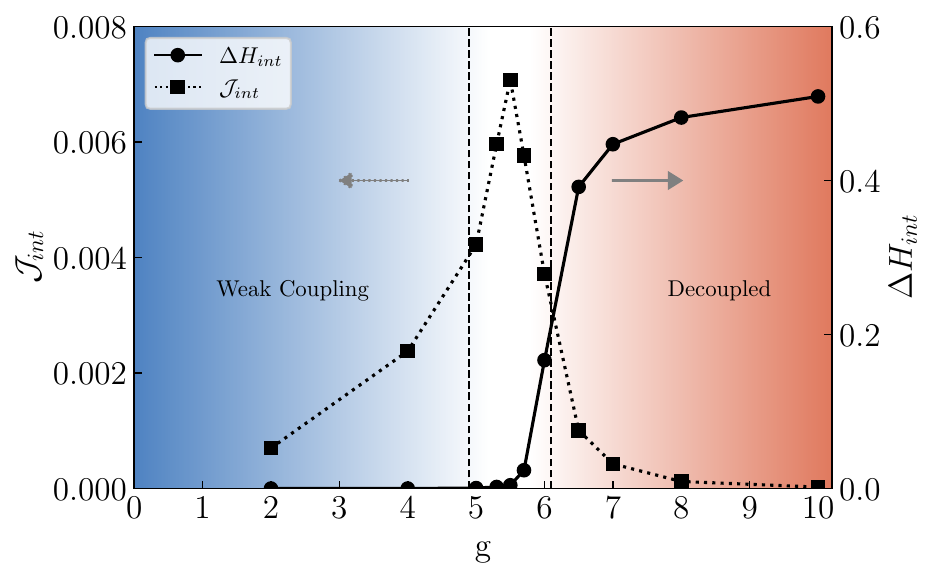}
\caption{Phase diagram for the four-leg ladder as a function of interaction strength $g$. The solid line denotes the symmetry-breaking order parameter $\Delta H_{\mathrm{int}}$ (antiperiodic boundary conditions). The dotted line denotes the  interaction-induced current $\mathcal{J}_{\mathrm{int}}$ (open boundary conditions). The dashed vertical lines mark the tentative positions of the BKT transition from the weak-coupling gapless phase into the SMG regime, and of the $\mathbb{Z}^R_2$ symmetry-breaking transition.
\label{fig3}}
\end{figure}

We use numerics to examine the proposed phases. To characterize the symmetry breaking at finite interaction strength, we introduce the order parameter
\begin{equation}
\Delta H_{\text{int}} = \frac{1}{gL_x}\left|\sum_{r\in B}^{L_x}(\langle H^{1}_{\text{int},r}\rangle-\langle H^{2}_{\text{int},r}\rangle)\right|. 
\end{equation}
The solid line in Fig.~\ref{fig3} shows $\Delta H_{\text{int}}$ under antiperiodic boundary conditions along the $x$-direction for a cylinder with $L_x=8$, $L_y=4$. We start from the sector   $\bigl|\langle H^1_{\text{int}}\rangle\bigr| > \bigl|\langle H^2_{\text{int}}\rangle\bigr|$  at large $g$ and adiabatically vary $g$ to obtain a smooth $\Delta H_{\text{int}}(g)$. The same adiabatic branch is used below in the discussion of the real-space density and current distribution under open boundary conditions. The order parameter exhibits Ising-like behavior, with the transition near $g \approx 6.0$.

Previous DMRG studies~\cite{Zeng2022} showed that on a two-leg ladder, a finite critical interaction strength $g_c$ is required for edge~B to develop a spectral gap $\Delta_B^{\mathrm{gap}}$. For a ladder with four or more legs, however, as $g$ increases further, the boundary gap may eventually exceed the bulk gap $\Delta_{\mathrm{bulk}}^{\mathrm{gap}}$, leading to an essentially decoupled gapped edge B, with gapless  modes on the next leg (see Appendix~\ref{app:A-excite} for the edge-B excitation gaps in the strong-coupling limit).

\begin{figure*}[htbp]
    \centering
    \includegraphics[width=0.92\textwidth]{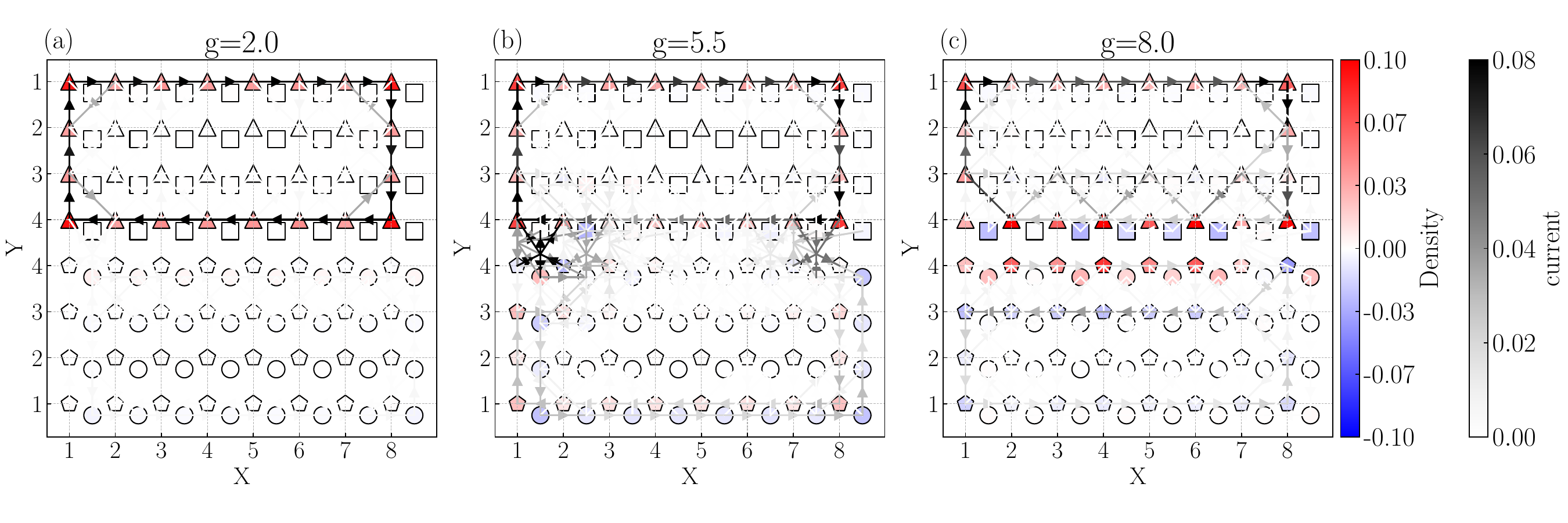}
    \caption{Deviations in the fermion densities and currents for the separate species for $g=2.0$, $5.5$, and $8.0$ under open boundary conditions on the X-Y rectangle, after  a fermion of flavor $I=1$ has been added to the  ground state at half filling. Fermion flavors $I=1,2,3,4$ ($\mathrm{U}(1)^{(1)}$ charges $3,4,5,0$) are represented by triangles $\triangle$, squares $\square$, pentagons $\pentagon$, and circles $\circ$. The blue-red color scale encodes the density deviation from the half-filling value $0.5$. The currents are shown without any background subtraction. For clarity, the current amplitude from the nonlocal interaction term is scaled by a factor of $5$ relative to the hopping-induced current. To visualize all four fermion flavors in a single panel, the upper half-plane shows flavors $I=1,2$ (triangles and squares), the lower half-plane shows flavors $I=3,4$ (pentagons and circles), reflected along the $y$-direction. With this arrangement, all flavors carry the same Chern number, while the interaction on edge~B acts on the central 
    two rows.
    \label{fig4}}
\end{figure*}

The spatial distribution of a low-energy injected charge directly probes this gap hierarchy: at intermediate interaction strengths, where $0 < \Delta_B^{\mathrm{gap}} \ll \Delta_{\mathrm{bulk}}^{\mathrm{gap}}$, the added fermion can only travel along the gapless edge. When $\Delta_B^{\mathrm{gap}} \gg \Delta_{\mathrm{bulk}}^{\mathrm{gap}}$, the gapped edge decouples and { a portion of the  charge can redistribute to the emergent gapless edge adjacent to the gapped edge.   To determine whether the  intermediate regime with fully gapped edge B exists,} we examine the deviations in the fermion densities and currents for the separate species, in the ground state with one additional fermion of flavor $I=1$ carrying charge $(Q^{(1)},Q^{(2)})=(3,0)$, under open boundary conditions (Fig.~\ref{fig4}). The density is displayed as the deviation from the half-filling background (density $0.5$ at every site), while the fermion currents are shown as measured, as the currents vanish in the absence of the additional fermion.

In the weak-interaction regime ($g=2.0$), edge~B remains gapless. The additional fermion is predominantly localized at the boundary of the $I=1$ Chern insulator and forms a chiral current loop characteristic of a free chiral edge mode.

 At intermediate interaction strength ($g=5.5$), a qualitatively different behavior emerges. Part of the charge redistributes among different flavor sectors: the density deviation becomes positive in the $I=3$ sector and negative in the $I=4$ sector. Simultaneously, pronounced interaction-induced flavor-mixing currents develop at the two ends of edge~B, which mediate charge transfer between the $I=1$ sector and the other sectors, and are indicated by six-leg vertices in Fig.~\ref{fig4}.
 To demonstrate that this pattern is not a metastable state inherited from one of the strong-coupling branches, we adiabatically evolve the DMRG calculation from each of the four strong-coupling charge sectors of the large-$g$ limit and recover the same density and current distributions (Fig.~\ref{figC4} in \hypersetup{linkcolor=black}Appendix~\ref{app:C-sectors}). This initialization independence places $g=5.5$ on the symmetric side of the $\mathbb{Z}_2^R$ transition.

To quantify the interaction-induced transport, we define the interaction-induced flavor-mixing current as the commutator of the density operator for each flavor $I$ with the interaction term $H_{\mathrm{int},r}^{\alpha}$, which acts on sites $r$ and $r+1$ ($I=1,2,3,4$, $\alpha=1,2$):
\begin{equation}
J^{\alpha}_{(I,r^{\prime}),r} = i \left[\Psi_{I,r^{\prime}}^{\dagger}\Psi_{I,r^{\prime}}, H_{\text{int},r}^{\alpha} \right].
\end{equation}

These operators can be written as
\begin{equation}
J^{\alpha}_{(I,r^{\prime}),r} = \sigma^{\alpha, I}_{r^{\prime}, r}\, K^{\alpha}_r , 
\end{equation}
where $K^{\alpha}_r$ is independent of $(I,r^{\prime})$. For $\alpha=1$ we have
$\sigma^{1, I}_{r^{\prime}, r} = 1$ if $I=1$ or 3 and $r^{\prime}=r$, and also if $I=4$ and
$r^{\prime}=r$ or $r^{\prime}=r+1$, while $\sigma^{1, I}_{r^{\prime}, r} = -1$ if $I=2$ and
$r^{\prime}=r$ or $r^{\prime}=r+1$. For $\alpha=2$ we have
$\sigma^{2, I}_{r^{\prime}, r} = 1$ if $I=2$ or 4 and $r^{\prime}=r$, and also if $I=1$ and
$r^{\prime}=r$ or $r^{\prime}=r+1$, while $\sigma^{2, I}_{r^{\prime}, r} = -1$ if $I=3$ and
$r^{\prime}=r$ or $r^{\prime}=r+1$. In all other cases $\sigma^{\alpha, I}_{r^{\prime}, r} = 0$.
Thus, there are six non-zero legs for any choice of $r$ and $\alpha$, and because
$|\sigma^{\alpha, I}_{r^{\prime}, r}| = 1$ on each of them, all six carry the same magnitude
$|\langle K^{\alpha}_r \rangle|$, with $\sigma^{\alpha, I}_{r^{\prime}, r}$ fixing their relative signs. 

In Fig.~\ref{fig4}, expectation values of the non-zero flavor-mixing currents are represented by
six-leg tensors \tensorSixLadder and \tensorSix for $\alpha =1$ and 2, respectively.
The gray scale encodes the current magnitude, while an arrow away from or into a site $(I, r^{\prime})$ signifies a positive or negative value of $\langle J^{\alpha}_{(I,r^{\prime}),r} \rangle$, respectively.

We define $\mathcal{J}_{\mathrm{int}}$ from the interaction-induced currents at the two endpoints of edge~B, taking the common magnitude $|\langle K^{\alpha}_r \rangle|$ shared by their six legs,
\begin{equation}
\mathcal{J}_{\mathrm{int}} = \frac{1}{g} \sum_{r \in \{1,L_x-1\}} \sum_{\alpha} \left| \langle K^{\alpha}_r \rangle \right|.
\label{intercurrent}
\end{equation}
The dotted line in Fig.~\ref{fig3} shows an enhancement of $\mathcal{J}_{\mathrm{int}}$ in the intermediate-coupling regime on the four-leg ladder. We expect $\mathcal{J}_{\mathrm{int}}$ to remain nonzero in the two-dimensional lattice limit, $L_y \to \infty$, where the endpoint of edge~B mediates scattering between flavors.

For stronger interactions ($g=8.0$), edge~B becomes decoupled from the rest of the system, and the interaction-induced current along edge~B vanishes. With total charge conserved and one additional fermion at half-filling, part of the charge localizes near edge~B, while the remainder redistributes in the bulk to minimize the total energy. Additional real-space data and further discussion are presented in Appendix~\ref{app:C-profiles}. The density and current patterns across the three regimes are similar at the smaller sizes $L_x=4$ and $6$ (Appendix~\ref{app:H-profiles}).

Fig.~\ref{fig3} overlays $\mathcal{J}_{\mathrm{int}}$ and the $\mathbb{Z}^R_2$ order parameter $\Delta H_{\mathrm{int}}$ along the symmetric line $g_1=g_2$. Two vertical dashed lines mark the tentative boundaries of the three phases that are resolved at the finite $L_x$ and $L_y$ accessible here. They correspond to the weak-side merged BKT separating the gapless phase from the SMG regime, and to the symmtry-breaking transition into the $\mathbb{Z}^R_2$-SSB region. The two strong-side BKTs cannot be resolved at this $L_x$, and the apparent positions of all boundaries may additionally be shifted by edge-A/edge-B hybridization, which is suppressed as $\exp(-L_y)$ but can still induce a small gap at finite~$L_y$. As a first step toward resolving the strong-side transitions, Appendix~\ref{app:E} reports the row-resolved   single-particle  correlation functions for $L_x=8$ at representative $g$ in each of the three resolved regimes, whose distinct decay behaviors provide a diagnostic for phase identification.

\section{Charge pumping}

Having established the phase diagram, we now turn to charge pumping.
Without loss of generality, we consider the $\mathbb{Z}_{10}^{(2)}$
symmetry twist introduced in Sec.~\ref{sec:model}.
Charge pumping can arise by continuously interpolating this anomaly-free twist through a
$\mathrm{U}(1)$ with nonzero mixed anomalies with the conserved symmetries.
We implement this interpolation along the $\ell_2=(2,1,-2,1)$ direction
through the boundary twist
\begin{equation*}
\Psi_{I,L_x+1}=e^{\ii\theta\ell_{2,I}}\Psi_{I,1}.
\end{equation*}
This $\mathrm{U}(1)$ has nonzero mixed anomalies with $\mathrm{U}(1)^{(1)}$
and $\mathrm{U}(1)^{(2)}$, and the associated bulk Hall response
predicts an edge-A charge change of $(-2,-1)$ when one
$\mathbb{Z}_{10}^{(2)}$ flux quantum is removed (Appendix~\ref{app:D-bulk}).
In the weak-interaction limit, edge~B is gapless and flux insertion pumps
charge between the two gapless edges as in the conventional Laughlin argument.
However, the behaviors at strong coupling and in the SMG phase are less clear.

\begin{figure}[htbp]
\centering
\includegraphics[width=1.0\columnwidth]{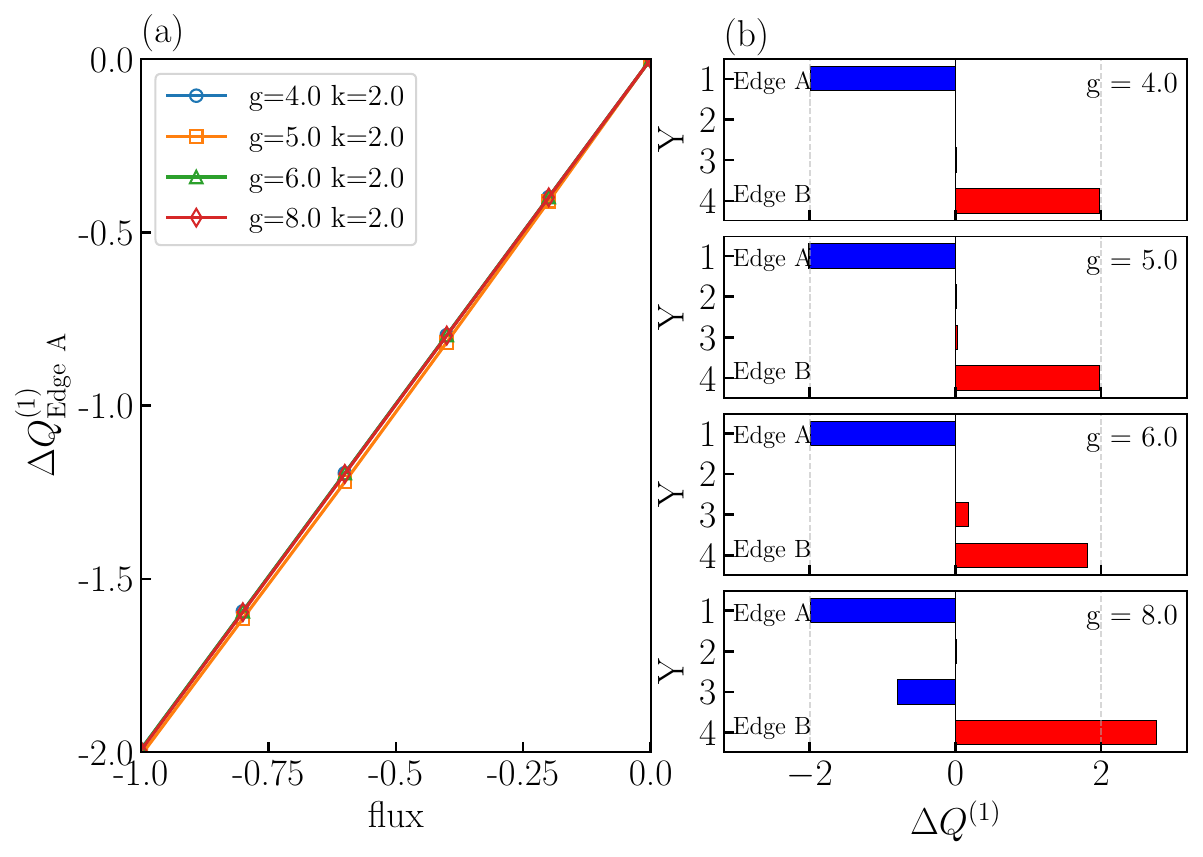}
\caption{Charge pumping under adiabatic removal of $\mathbb{Z}_{10}^{(2)}$ flux at $g=4.0$, $5.0$, $6.0$, and $8.0$ ($L_x=8$, $L_y=4$). (a) Charge deviation at edge~A, $\Delta Q^{(1)}_{\text{edge A}}$, as a function of the threaded flux. The slope $k \approx 2$ reflects the quantized Hall response. (b) Row-resolved charge deviation $\Delta Q^{(1)}_Y$ for each row $Y$ after removal of one $\mathbb{Z}_{10}^{(2)}$ flux quantum.
    \label{fig5}}
\end{figure}

To address these questions, we perform DMRG simulations with $L_y=4$ and $L_x=8$, starting from antiperiodic boundary conditions to lift the free-fermion degeneracies at half-filling. We adiabatically remove one $\mathbb{Z}_{10}^{(2)}$ flux quantum and measure the resulting per-row charge deviation $\Delta Q^{(1)}_{Y}$ from the half-filled background (see \hypersetup{linkcolor=black}Appendix~\ref{app:D-bulk} for the detailed numerical setup). Along this insertion path, only the antiunitary $U^{R}_{(1,3)}$ among the $D_4$ generators remains exact at every flux. As shown in Fig.~\ref{fig5}(a), the pumped charge from edge~A scales linearly with the threaded flux for all interaction strengths considered ($g=4.0$, $5.0$, $6.0$, and $8.0$): in all cases,   the results are consistent with  two $\mathrm{U}(1)^{(1)}$ charges   being  pumped from edge~A to edge~B when one flux quantum is removed. A similar quantized pumping is found at the smaller sizes $L_x=4$ and $6$, consistent with an $L_x$-independent response (\hypersetup{linkcolor=black}Appendix~\ref{app:H-pumping}).

The insensitivity of the quantized pumping to the state of edge~B indicates that the response is a bulk property.

Figure~\ref{fig5}(b) shows the distribution of the pumped charge along the $Y$ direction after removal of one $\mathbb{Z}_{10}^{(2)}$ flux quantum. Under integer flux insertion the charge density is uniform along each row, enforced by a modified translation symmetry $\tilde{T}$, so we report the total charge deviation per row relative to half-filling. Densities at intermediate, non-integer flux values are given in \hypersetup{linkcolor=black}Appendix~\ref{app:D-uniformity}.

While the charge depletion at edge~A is independent of $g$, the density pattern near edge~B depends strongly on the boundary phase. In the weak-interaction limit ($g=4.0$), the pumped charge sits almost entirely on edge~B, as expected for a gapless chiral mode. At intermediate coupling it remains concentrated near edge~B but partially spills into the adjacent bulk row, reflecting a gapped but still hybridized boundary. At strong coupling ($g=8.0$), the edge~B mode no longer carries the full response, and the local edge-B charge is set by a flux-biased superposition of two charge sectors rather than the quantized pumped charge alone, with the neighboring bulk row compensating so that their combined charge stays quantized (see \hypersetup{linkcolor=black}Appendix~\ref{app:D-accumulation} for details).

On the other hand, the same $\mathbb{Z}_{10}^{(2)}$ symmetry twist can also be interpolated along a different continuous path through the boundary condition
\begin{equation*}
\Psi_{I,L_x+1}=e^{\ii\theta(4q^{(1)}_I-3q^{(2)}_I)}\Psi_{I,1}.
\end{equation*}
Here $4q^{(1)}-3q^{(2)}=(12,1,8,-9)$. The twist is continuously interpolated within the preserved $\mathrm{U}(1)^{(1)}\times\mathrm{U}(1)^{(2)}$ symmetry. The continuous $\mathrm{U}(1)$ used for this interpolation has zero mixed anomaly with each of $\mathrm{U}(1)^{(1)}$ and $\mathrm{U}(1)^{(2)}$. The bulk Hall response therefore predicts zero pumped charge along this path. Appendix~\ref{app:D-embeddings} compares the charge response and energy evolution along the two paths and shows the species-resolved densities along the second. The two paths give identical Hamiltonians after removal of one flux quantum but differ at intermediate flux and have different bulk Hall responses, so the transported charge depends on the continuous insertion path.

We note that in all cases, a large fraction of the charge ends up on the final leg of edge B.    In the SMG and strong coupling phases, the existence of an energy gap means that
the resulting final state has an energy larger than the initial state by a finite amount.  This is allowed because the action of flux insertion does finite work on the system when $g \neq 0$.

\section{Analogy with the scenarios of leptogenesis} 

{Let us make some final comments on analogies with different leptogenesis scenarios~\cite{Davidson:2008bu0802.2962, Fukugita:1986hr, Wang:2024auy2501.00607}, which address different mechanisms of the particle versus anti-particle production asymmetry in the early universe.} 
First note that our lattice model is (1+1)d with a few ladder-width  (2+1)d bulk,
while 
our realistic Standard Model world \cite{Weinberg:1996kr-book} is (3+1)d
with dynamical gauge field sector of Lie algebra
$$su(3) \times su(2) \times u(1),$$ 
with quarks and leptons written as 16 Weyl fermions in the following representation:
\begin{multline}
\label{eq:psi-SM}
 \psi_{\rm SM}\equiv \bar{d}_R \oplus {l}_L  \oplus q_L  \oplus \bar{u}_R \oplus   \bar{e}_R  \oplus  {\bar{\nu}_{R}} \sim \\ 
(\overline{\bf 3},{\bf 1})_{2} \oplus ({\bf 1},{\bf 2})_{-3} \oplus ({\bf 3},{\bf 2})_{1} \\
\oplus (\overline{\bf 3},{\bf 1})_{-4} \oplus ({\bf 1},{\bf 1})_{6}  \oplus  ({\bf 1},{\bf 1})_{0}.
\end{multline}
This pattern repeats for $N_f=3$ families.
The subscript in \eqref{eq:psi-SM} has the properly quantized $u(1)= u(1)_{\tilde Y}$
hypercharge ${\tilde Y}$. 
Here we include the sterile neutrino.
We will focus on a single $N_f=1$ family Standard Model without losing the generality of our main observation, and we can 
triplicate the effect to $N_f=3$ families.

\begin{figure}[!htbp]
    \centering
    \includegraphics[width=0.44\textwidth]{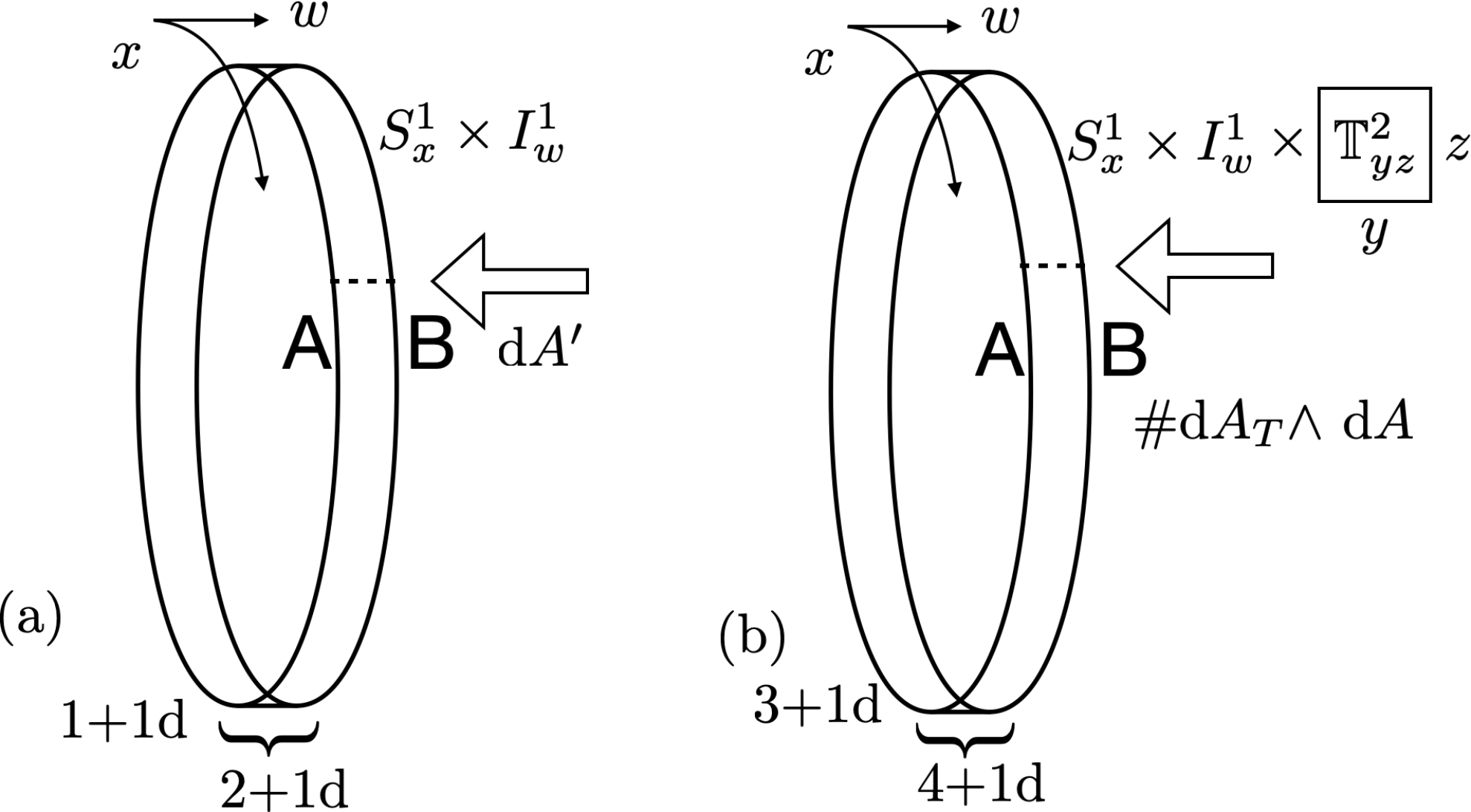}
    \caption{Annular or strip geometries with
    two boundaries, A and B.
     The dashed lines
     extending along
     $I^1_w$ schematically 
     represent codimension-1 symmetry-twist defects, projected onto the two-dimensional strip $S^1_{x} \times I^1_{w}$.
    (a) Charge pumping on the
    (1+1)d edge world coupled to a (2+1)d bulk,
     with the anomalous current given by \eqref{eq:1+1d-edge-current-mutual}, similar to the setting of \cite{Laughlin1981}.
     The symmetry twist defect 
is associated with the background $A'$. 
At fixed time, a twist defect localized at $x=x_0$ extends along $I_w^1$.
    (b) Schematic analogy to the charge-pumping bulk-edge correspondence version of Topological Leptogenesis for 
    the (3+1)d Standard Model 
    world coupled to a (4+1)d bulk, 
     with the anomalous current given by
     \eqref{eq:d*j-current-Tbb}.
     Here the (4+1)d {spacetime is} $\mathbb{T}^3_{xyz} \times I^1_{w} \times I^1_{t}$, where
$I^1_{t} \subset \mathbb{R}^1_t$ is a time interval and
$\mathbb{T}^3_{xyz}=
S^1_{x} \times S^1_{y} \times S^1_{z}$ is the spatial torus. There are several possible configurations for inducing the charge pumping; 
as an example, we can arrange
that the time-dependent $\dd A_T$ flux pumps through $S^1_{x} \times I^1_{t}$
and
the $\dd A$ flux threads through the 2-torus
$\mathbb{T}^2_{yz}$.
The symmetry twist defect 
is associated with the background $A_T$, particularly the discrete $\mathbb Z_{4,T}^{F}$ symmetry. At fixed time, a twist defect localized at $x=x_0$ extends along $T_{yz}^2\times I_w^1$, forming a 3d spatial hypersurface, or a 4d defect worldvolume $T_{yz}^2\times I_w^1 \times I_t$ including time.
    \label{fig6}}
\end{figure}

We now compare the two models.
One is the (1+1)d 3-4-5-0 chiral fermion
edge world
coupled to a (2+1)d bulk 
and the other is the (3+1)d Standard Model world coupled to a (4+1)d bulk that we will determine later (see Fig.~\ref{fig6}).
Table~\ref{tab:twist-lift-comparison} summarizes the comparison.

\begin{table}[!htbp]
\centering
\footnotesize
\setlength{\tabcolsep}{4pt}
\renewcommand{\arraystretch}{1.15}
\begin{tabular*}{\columnwidth}{@{\extracolsep{\fill}}lcc@{}}
\toprule
 & 3-4-5-0 model & 
 \makecell{16-Weyl fermion \\ Standard Model} \\
\midrule
Boundary dimension & $(1+1)$d & $(3+1)$d \\
\midrule
Anomaly-free symmetry & $\mathrm{U}(1)^{(1)} \times \mathrm{U}(1)^{(2)}$ & $\U(1)_{\bf B - L}$ or  $\U(1)_X$ \\
\midrule
Discrete subgroup &  \makecell{$\mathbb{Z}_{10}^{(1)} \times \mathbb{Z}_{10}^{(2)}$
} & $\mathbb{Z}_4^F = 
\mathbb{Z}_{4,X}^F=
\mathbb{Z}_{4,T}^F$ \\
\midrule
\makecell{Anomalous \\
continuous 
deformation\\
and its gauge field}
& \makecell{ $\mathrm{U}(1)^{(3)} \times \mathrm{U}(1)^{(4)}$\\[2mm] 
$A'$
}& 
\makecell{ $\U(1)_T$ \\[2mm] 
$A_T$
}\\
\midrule
Anomalous current & Eq.~\eqref{eq:1+1d-edge-current-mutual} & Eq.~\eqref{eq:d*j-current-Tbb} \\
\midrule
Transported charge 
& Eq.~\eqref{eq:DeltaQ} & Eq.~\eqref{eq:DeltaB-L}\\
\midrule
Inserted flux & $\theta/(2\pi)=1/10$ & \makecell{$\theta_T/(2\pi)=1/4$\\[2mm]$\int_{\mathbb{T}^2_{yz}}\frac{\dd A}{2\pi}=3$} \\
\midrule
Pumped charge & $(2,1)$ & $4$ \\
\bottomrule
\end{tabular*}
\caption{{Comparison of the two models and their bulk Hall contributions to the pumped charge:
$(\Delta Q^{(1)},\Delta Q^{(2)})$ for the 3-4-5-0 model and
$\Delta({\bf B}-{\bf L})$ for the sixteen-Weyl analogue.
Assumptions and sign conventions are specified in the text.}}
\label{tab:twist-lift-comparison}
\end{table}

\subsection{(1+1)d 3-4-5-0 chiral fermion model}
In our present work, the (2+1)d bulk is described by an invertible topological phase written as
the pure-gauge Chern-Simons 3-form 
$$A \wedge \dd A,$$
with some U(1) symmetry's 1-form background field $A$.
If the (1+1)d edge mode corresponds to a U(1) symmetric charge-$q$ chiral Weyl fermion,
the full (2+1)d invertible field theory  partition function is a $\U(1)_{k=1}$ Chern-Simons theory
at the level $k=1$,
\begin{equation}
    \exp(\ii  S_3)
\equiv \exp\Bigg[\ii q^2    \int_{M^3}  \,A \wedge \frac{\dd A}{4 \pi}  \Bigg].
\end{equation}
The current $j$ non-conservation of the (1+1)d edge due to the flux insertion into the
(2+1)d annulus or cylinder shape 
 $S^1_{x} \times I^1_{w} \times I^1_{t}$
(of ladders in our lattice model) is governed by
$$
\dd \star j 
=  \sigma_{xy} {\dd A}
=  \frac{q^2}{k} \frac{\dd A}{2 \pi} = \star J,
$$
where the quantum Hall conductance
is given by $\sigma_{xy}=\frac{q^2}{{2 \pi} k}$.
The $\star$ is the Hodge star operator.
On the left-hand side equality, 
the edge current non-conservation 
$\dd \star j \equiv \dd \star_2 j$
comes from 
the inflow of the bulk Hall current $J$ along the $w$ direction on the right-hand side equality; 
such that its Hodge dual 
$\star J \equiv \star_3 J$ 
would be codimension-1 to 
the $w$ direction respect to the (2+1)d bulk, namely the full spacetime dimensions of the (1+1)d edge theory in the $t$-$x$ plane.
Here the $w$ direction is the same as the conventional $y$ direction in terms of the Hall conductance
$\sigma_{xy}$.

If we consider the model 
with four flavors of fermions, we have four
components in the current $j$ and in the gauge field $A$,  with corresponding charges denoted by a four-component vector ${\bf q}$. 
Thus, the charge $q$ becomes a
charge vector ${\bf q}$ and the level $k$ becomes the rank-4 $K$-matrix 
$K=\text{diag}(1,1,-1,-1)$,
so quantum Hall conductance becomes
$\sigma_{xy}=
\frac{{\bf q}^{\intercal} K^{-1} {\bf q}}{2 \pi}$ \cite{Wang2023}.
The anomalous current $j$ non-conservation of the (1+1)d edge becomes
\begin{equation}
\label{eq:1+1d-edge-current}
\dd \star j 
=  \sigma_{xy} {\dd A}
=  ({{\bf q}^{\intercal} K^{-1} {\bf q}}) \frac{\dd A}{2 \pi}.
\end{equation}

Note that the Hall conductance is in the unit of $e^2/h$. Since we set $e=c=\hbar \equiv h/(2 \pi)=1$ in field theory, 
so our Hall conductance has a unit of $e^2/h= e^2/(2 \pi \hbar) 
= 1/(2 \pi)$.

For the 3-4-5-0 model, 
to transport particles of 
charge ${\bf q}$
of 
$\mathrm{U}(1)^{(1)} \times \mathrm{U}(1)^{(2)}$ symmetry, we need to implement
the  $\mathbb{Z}_{10}^{(1)} \times \mathbb{Z}_{10}^{(2)}$ symmetry twist with a corresponding charge ${\bf q}'$  
and its gauge field $A'$
along the
$\mathrm{U}(1)^{(3)} \times \mathrm{U}(1)^{(4)}$ continuous deformation
direction, 
through their mixed anomaly captured by
the mutual Hall conductance
$\sigma_{xy}'=
\frac{{\bf q}^{\intercal} K^{-1} {\bf q}'}{2 \pi}$ \cite{Wang2023, Santos:2013uda1310.8291}, so \eqref{eq:1+1d-edge-current} becomes
\begin{equation}
\label{eq:1+1d-edge-current-mutual}
\dd \star j 
=  \sigma_{xy}' {\dd A'}
=  ({{\bf q}^{\intercal} K^{-1} {\bf q}'}) \frac{\dd A'}{2 \pi}.
\end{equation}

{\bf {\emph{Leptogenesis on the edge A from the edge B}}:}
We consider the regime in which edge A remains
gapless while edge B is gapped by SMG, with the anomaly-free
$\mathrm{U}(1)^{(1)} \times \mathrm{U}(1)^{(2)}$
symmetry preserved.
The SMG-gapped 
mirror fermions in the 
{mirror world} at edge B can be transported to the {original world} on the gapless edge A by the 
discrete $$\mathbb{Z}_{10}^{(1)} \times \mathbb{Z}_{10}^{(2)}
\subset
\mathrm{U}(1)^{(3)} \times \mathrm{U}(1)^{(4)}$$ 
flux insertion, which is demonstrated by our numerical simulation. Conceptually, the inserted
$\mathrm{U}(1)^{(3)} \times \mathrm{U}(1)^{(4)}$ flux $\dd A'$ lies in the $t$-$x$ plane
$S^1_x\times I^1_{t}$,
with the flux insertion characterized by the surface integral 
$\iint_{S^1_x\times I^1_{t}} \dd A'$,
while the dashed line in Fig.~\ref{fig6} 
represents a symmetry-twist defect 
 associated with the background $A'$. 
The symmetry-twist defect is a  codimension-1 
defect transverse to the $x$-direction, along which the holonomy $\oint_{S_x^1}A'$ is defined and the boundary current flows.
Namely, at fixed time, a twist defect extending along $I_w^1$, is localized at $x=x_0$, forming a 1d spatial line, or a 2d defect worldvolume within $I_w^1 \times I_t$ including time.

The transported charge $\Delta Q$ on the edge A from the edge B for a twist angle
$\theta = \frac{2 \pi}{10} {n}$, with 
$n \in \mathbb{Z}$
and 
$(n \mod 10)$ belonging to a subgroup of
$\mathbb{Z}_{10}^{(1)} \times \mathbb{Z}_{10}^{(2)}$, 
is given by:
\begin{equation}
\label{eq:DeltaQ}
\Delta Q=\frac{\theta}{2\pi}\,({{\bf q}^{\intercal} K^{-1} {\bf q}'}).
\end{equation} 

In our simulations, for example, we choose ${\bf q}'=\ell_2=(2,1,-2,1)$ and remove one $\mathbb{Z}_{10}^{(2)}$ flux quantum, $\theta=-2\pi/10$. Equation (\ref{eq:DeltaQ}) then gives $(\Delta Q_A^{(1)},\Delta Q_A^{(2)})=(-2,-1)$, corresponding to a transfer of charge $(2,1)$ from edge A to edge B.

\subsection{(3+1)d \texorpdfstring{$su(3) \times su(2) \times u(1)$}{su(3) x su(2) x u(1)} Standard Model}

In analogy to our (3+1)d physical Standard Model world, each Weyl fermion in (3+1)d can be attached to a (4+1)d bulk described by an invertible topological phase written as
a combination of the pure-gauge Chern-Simons 5-form 
$$A \wedge \dd A \wedge \dd A$$ and
the mixed gauge-gravity 5-form 
$$A \wedge \Tr[R \wedge R].$$
These two 5d terms come from the
anomaly inflow of two types of Feynman triangle diagram anomalies in (3+1)d: $\U(1)^3$
and $\U(1)$-gravity-gravity.
Here
$A= A_{{\bf B} - {\bf L} } \equiv A_b$ is the background gauge field,
and $R$ is the curvature 2-form
over the spacetime manifold.
We shorthand the 
baryon minus lepton
${{\bf B} - {\bf L} } \equiv b$
and the ${{\bf B} - {\bf L} }$ number as 
the $q_b$ charge.
The 
$\psi_{\rm SM}$ multiplet \eqref{eq:psi-SM}
has the $q_b$ charge
$$
q_b=(-\tfrac{1}{3},-1,\tfrac{1}{3},-\tfrac{1}{3},1,1),
$$
or the properly normalized 
$\tilde{b} \equiv 3b = 3 ({{\bf B} - {\bf L} })$, so 
$$q_{\tilde{b}}=3 q_b
= (-1,-3, 1,-1,3,3).$$ 

The trace Tr is 
over the tangent-bundle vector indices, in the vector representation of the Lorentz group with the Lie algebra $so(1,3)$, or its Euclidean version $so(4)$.
The partition function of the (4+1)d invertible field theory on a 5-manifold $M^5$ contains
the exponentiation of the action 
\begin{multline}
    \label{eq:bulk}
\exp\Bigg[\ii   \int_{M^5}  \mathcal{A}_{bbb} \, \frac{1}{6} A \wedge    \frac{\dd A}{2 \pi} \wedge  \frac{\dd A}{2 \pi} \\
+ \mathcal{A}_{b}^{\rm Grav} \frac{1}{24} \frac{1}{8 \pi^2}  A \wedge \text{Tr}[ R \wedge R] \Bigg]
.
\end{multline}
{A (3+1)d world at the boundary of the (4+1)d world can then have} an anomalous current 
$j \equiv j_{{\bf B}-{\bf L}}$,
not conserved with appropriate background fields:
\begin{multline}
\label{eq:d*j-current}
\dd \star j 
=  
 \mathcal{A}_{bbb}
 \frac{1}{2}
 \Big(\frac{\dd A}{2 \pi}\Big)^2
+ \mathcal{A}_{b}^{\rm Grav}
\frac{1}{24}  \frac{1}{8 \pi^2}   \text{Tr}[ R \wedge R]
.
\end{multline}
Here $j$ is the covariant current rather than the consistent current \cite{BardeenZumino1984}, which would introduce an extra factor of 1/3 in the first part of \eqref{eq:d*j-current}. 

However, it turns out that these anomaly coefficients
of the 16-Weyl fermion Standard Model vanish:
$$
\mathcal{A}_{bbb}=\sum_I q_b^3 =0, \quad \mathcal{A}_{b}^{\rm Grav}=
\sum_I q_b =0
$$
where $I$ denotes each of the 16 components of the Weyl fermion multiplet.

The current $j$ is therefore
conserved, $\dd \star j
=  0$, within the (3+1)d world, while the
{gauge instanton} 
(namely ${(\dd A)}^2$ term)
and {gravitational instanton}
(namely $R \wedge R$ term)
insertion cannot generate any leptogenesis.

Nonetheless, we can consider a $\U(1)_T$ deformation where the $\psi_{\rm SM}$ multiplet \eqref{eq:psi-SM} is assigned the uniform $q_T$ charge
\begin{eqnarray}
\label{eq:qT}
&&q_T=(1,1,1,1,1,1), \text{ so } \nonumber \\
&&\mathrm{U}(1)_T:\quad\Psi_I\mapsto e^{\ii\theta_T q_{T_I}}\Psi_I,
\quad \theta_T \in [0, 2\pi).
\end{eqnarray}
Note that $\mathrm{U}(1)_T$ is a classical symmetry but \emph{not} a quantum symmetry of the Standard Model, because $\mathrm{U}(1)_T$ is
incompatible with
$su(3) \times su(2) \times u(1)$
gauge structure due to the Adler-Bell-Jackiw anomaly.

In contrast, there is indeed a quantum symmetry
$X$ as a particular linear combination of
${\bf B} - {\bf L}=b$ and hypercharge $\tilde Y$
\cite{Wilczek1979hcZee, WilczekZeePLB1979}
compatible with the 16-Weyl fermion Standard Model,
where the $\psi_{\rm SM}$ multiplet \eqref{eq:psi-SM} is assigned the $q_X$ charge
\begin{eqnarray}
\label{eq:qX}
&&q_X=(-3,-3,1,1,1,5), \quad {X \equiv 5({{\bf B} - {\bf L} })-\frac{2}{3} \tilde Y},
\text{ so } \nonumber \\
&&\mathrm{U}(1)_{X}:
\quad\Psi_I\mapsto e^{\ii\theta_X X_I}\Psi_I,
\quad 
\theta_X \in [0, 2\pi).
\end{eqnarray}
Importantly and surprisingly, although
the full continuous $\mathrm{U}(1)_T$ is \emph{not} a quantum symmetry,
the $\mathrm{U}(1)_T$ has a discrete 
$\mathbb{Z}_{4,T}^F$ that overlaps
with $\mathbb{Z}_{4,X}^F$ which is a good quantum symmetry, such that the common subgroup
$\mathbb{Z}_{4,T}^F = \mathbb{Z}_{4,X}^F$ 
contains the fermion parity normal subgroup
$\mathbb{Z}_2^F$.

The (3+1)d case here is analogous to the (1+1)d relation following Eq.~\eqref{eq:Z10reps},
\begin{eqnarray}
\mathrm{U}(1)_X \cap \mathrm{U}(1)_T \equiv
\mathbb{Z}_{4}^F =
\mathbb{Z}_{4,T}^F = \mathbb{Z}_{4,X}^F \supset \mathbb{Z}_2^F.
\end{eqnarray}
Since $q_T = q_X \mod 4$, the transformations at
$\theta_T=\theta_X = \pi/2$ define the same generator:
$$
g=e^{\ii\pi q_T/2}=e^{\ii\pi q_X/2},
\qquad g^2=(-1)^F.
$$

For the 16-Weyl fermion Standard Model, 
to transport the particles of 
a charge $q_b$
of 
$\U(1)_{\bf B - L} \equiv \mathrm{U}(1)_b$ 
symmetry (or $q_X$ of $\mathrm{U}(1)_X$ symmetry up to a gauge-equivalent transformation),
we can implement
the  discrete $\mathbb{Z}_{4}^F$ symmetry twist with a corresponding charge $q_T$  
and its gauge field $A_T$
along the
$\mathrm{U}(1)_T$ continuous deformation
direction, 
through their mixed anomaly captured by
the Feynman triangle diagram
$\mathrm{U}(1)_T$-$\U(1)_b$-$\U(1)_b$ and the current non-conservation; see Fig.~\ref{fig6} (b):
\begin{equation}
\label{eq:d*j-current-Tbb}
\begin{aligned}
\dd \star j &=
2 \mathcal{A}_{Tbb}
 \frac{1}{2}
\frac{\dd A_T}{2 \pi}
\wedge
\frac{\dd A}{2 \pi}
,\\
\mathcal{A}_{Tbb}&=\sum_I q_b^2 =\frac{16}{3}.
\end{aligned}
\end{equation}
Here $j$ is the covariant current instead of the consistent current \cite{BardeenZumino1984}, which would introduce an extra 1/3 factor. 

{\bf {\emph{Leptogenesis on the boundary A from the boundary B}}:}
The (3+1)d boundary A is
gapless, while the (3+1)d boundary B is gapped by SMG
because the anomaly-free conditions among the gauge sector $su(3) \times su(2) \times u(1)$
and the global symmetry $\U(1)_{\bf B - L} \equiv  \U(1)_b$ (thus also $\U(1)_X$)
hold.
The SMG-gapped 
mirror fermions in the 
{mirror world} at boundary B can be transported to the {original world} on the gapless boundary A by the 
discrete $\mathbb{Z}_{4,T}^F  \subset \mathrm{U}(1)_T$
flux insertion, due to the mixed anomaly between 
$\mathrm{U}(1)_T$ and 
$\U(1)_{\bf B - L} \equiv \mathrm{U}(1)_b$
(thus also $\U(1)_X$) given in \eqref{eq:d*j-current-Tbb}. 
Conceptually, the continuous
insertion of $\mathrm{U}(1)_T$ flux $(\dd A_T)_{tx}$
in the $t$-$x$ plane
$S^1_x\times I^1_{t}$
varies
the boundary twist around the
periodic $x$ direction from $\theta_T =0$ to $\theta_T=\pi/2$, when the time evolves from
$t =0$ to $t = I^1_t$ for $t \in [0, I^1_t]$,
 \begin{equation}
 \Psi_I(x+L_x)=e^{\ii\theta_T T_I}\Psi_I(x).
 \end{equation}

 In addition, we introduce a $\mathrm{U}(1)_b$ background flux
$(\mathrm{d}A)_{yz}$ through the two-torus $\mathbb{T}_{yz}^{2}$.
The dashed line in Fig.~\ref{fig6} schematically represents the projection
of a $T$-symmetry twist defect hypersurface, localized at $x=x_0$ and extending
along $\mathbb{T}_{yz}^{2}\times I_w^{1}$ at fixed time.
The mixed background flux
$(\mathrm{d}A_T)_{tx}\wedge(\mathrm{d}A)_{yz}$ occupies the
$t$--$x$--$y$--$z$ directions, whose 5-dimensional Hodge dual points
along $I_w^{1}$. Accordingly, the associated bulk inflow current is
directed between the two boundaries along the $w$ direction.

The transported charge $\Delta ({\bf B} - {\bf L}) \big\vert_T$ to the boundary A from the boundary B in the configuration of Fig.~\ref{fig6}(b)
along the $\mathrm{U}(1)_T$ continuous deformation  
for a twist angle
$\theta_T = \frac{2 \pi}{4} {n_T}$
with
$n_T \in \mathbb{Z}$ 
labeling the deformation path along the $\U(1)_T$ direction
and
$(n_T \mod4) \in \mathbb{Z}_{4,T}^F$
is given by:
\begin{equation}
\begin{split}
\label{eq:DeltaB-L}
\Delta ({\bf B} - {\bf L}) \big\vert_T &=
\mathcal{A}_{Tbb} \cdot
\frac{\theta_T}{2\pi} \cdot
\int_{\mathbb{T}^2_{yz}} \frac{\dd A}{2 \pi}\\
&=\frac{16}{3} \cdot \frac{n_T}{4} \cdot n_b 
= 4 n_T n_{\bf Q} \in 4 \mathbb{Z}.
\end{split}
\end{equation} 
Here the $\mathrm{U}(1)_b$ flux
is quantized in multiples of three:
$$\frac{1}{2 \pi} 
\int {\dd A} =n_b = 3 n_{\bf Q},$$ 
with $n_{\bf Q} \in \mathbb{Z}$,
where ${\bf Q}$ stands for the proper normalization in the quark basis $({\bf Q} - N_c {\bf L}) =  N_c ({\bf B} - {\bf L})$ with the color number $N_c=3$. So we derive that
\begin{equation}
\label{eq:B-L4}
\Delta ({\bf B} - {\bf L}) \big\vert_T = 0 \mod 4 ,
\end{equation}
which also preserves the 
$\mathbb{Z}_{4,X}^F$ symmetry with
$\Delta X = 0 \mod 4$, because the gauge charge
$\frac{2}{3}\Delta \tilde Y =0 \mod 4$ in \eqref{eq:qX}.
This contrasts with the pumping along the $\mathrm{U}(1)_X$ deformation that produces
$$
\Delta ({\bf B} - {\bf L}) \big\vert_X = 0 ,
$$ 
because
$\mathcal{A}_{Xbb} = 
\mathcal{A}_{XXb} = \mathcal{A}_{bbb} = 0$.

In this hypothetical picture, our Universe sits on boundary~A, while the
mirror fermions on boundary~B are gapped by SMG with the preserved
symmetries unbroken. An early-Universe event could supply the energy for a
$\mathbb{Z}_4^F$ flux insertion along the $\mathrm{U}(1)_T$ path, pumping
${\bf B} - {\bf L}$ from the gapped mirror side through the $(4+1)$d bulk to our boundary
in the background specified above. If this pumping direction is favored,
the resulting asymmetry in our Universe could survive after charge exchange
freezes out, while the total ${\bf B} - {\bf L}$ of the full system remains conserved. Namely,
 the boundary theories
 can have an imbalance
 $\Delta ({\bf B} - {\bf L}) \big\vert_T \in 4 \mathbb{Z}$
 on edge A and on edge B,
 but $\Delta ({\bf B} - {\bf L}) \big\vert_T = 0$ for the full system.

Moreover, since the (4+1)d 
bulk without SPTs can be detached,
if the 16th Weyl fermion $\nu_R$ also has obtained a Majorana mass,
the (3+1)d world itself can induce the familiar
\emph{Majorana Leptogenesis} ~\cite{Fukugita:1986hr},
which is a rather different mechanism because
the Majorana mass breaks the continuous symmetry
$\U(1)_{\bf B - L} \equiv \mathrm{U}(1)_b$
down to ${\mathbb{Z}_2^{\rm F}}$,
so it is expected to have a different conservation law:
\begin{eqnarray}
\label{eq:B-L2}
\Delta ({\bf B} - {\bf L}) = 0 \mod 2.
\end{eqnarray}
In summary, because the
underlying symmetries are different in \eqref{eq:B-L4} and 
\eqref{eq:B-L2},
our leptogenesis scenario is fundamentally different from
\emph{Majorana Leptogenesis}.

Typically, 
we expect that leptogenesis happens at a higher energy scale $E_{\text{lepto}}$, while
 baryogenesis
happens at an intermediate energy scale
$E_{\text{baryo}}$ when the sphaleron process occurs \cite{Kuzmin:1985mmShaposhnikovBaryon},
and the Standard Model electroweak or Higgs scale
$\Lambda_{\rm H}$ is 
at a lower energy scale, such that
\begin{equation}
E_{\text{lepto}} >  E_{\text{baryo}}
> \Lambda_{\rm H}.
\end{equation}
In fact, we expect that
$$E_{\text{lepto}} \sim \Lambda_{\rm SMG}$$
when SMG is involved in the 
\emph{bulk-edge Topological Leptogenesis},
while in conventional \emph{Majorana Leptogenesis}, one expects
$$E_{\text{lepto}} \sim M_{\rm Majorana}$$
with $M_{\rm Majorana}$ the Majorana mass of the right-handed neutrino.

Note that our present work on the
(1+1)d-(2+1)d-coupled system is most closely related to
the \emph{bulk-edge correspondence version of
Topological
Leptogenesis} \cite{Wang:2024auy2501.00607},
which is also illustrated above in the
(3+1)d-(4+1)d-coupled system for the Standard Model. 
However, in Ref.~\cite{Wang:2024auy2501.00607},
there is also 
\emph{Topological
Leptogenesis} within a (3+1)d world
alone (without the need for a (4+1)d bulk)
 that involves the replacement
of $\nu_R$ by a gapped
(3+1)d long-range entangled topological order. This will be a different leptogenesis 
scenario that replaces the conventional
\emph{Majorana Leptogenesis}. 
We do not have that analogy in the (1+1)d edge
world, since
stable topological order 
only occurs in spacetime dimensions higher than (1+1)d.

\section{Conclusion and discussion}

\begingroup\sloppy\hbadness=10000\relax
Combining DMRG with a strong-coupling analysis, we have explored the ground-state phase diagram of the 3-4-5-0 model on a four-leg ladder. In addition to the weak-coupling regime and the SMG phase found for a two-leg ladder in Ref.~\cite{Zeng2022}, we find a strong-coupling $\mathbb{Z}_2^{R}$-symmetry-broken regime in which edge~B becomes interaction-dominated. When the interaction is sufficiently strong, the interacting leg at edge~B effectively decouples from the bulk of the system, and new gapless modes are formed on the leg adjacent to the interacting leg. Consequently, a truly gapped edge~B is only present within an intermediate range of interaction strengths. We remark that in the limit of large interaction strength, edge~B can be mapped into a one-dimensional model of dimers and monomers with four degenerate ground states, which can be described exactly. We expect that the aforementioned features will persist for arbitrarily large values of $L_y$.

A central result of our analysis is that adiabatic pumping of $\mathrm{U}(1)$ charge between edges~A and~B under the insertion of one unit of $\mathbb{Z}_{10}^{(2)}$ flux along the interpolation with flavor charges $(2,1,-2,1)$ remains quantized across all regimes. The quantized value is fixed by the bulk mixed Hall response for this interpolation, and therefore does not depend on whether edge~B is gapless, SMG-gapped, or symmetry-broken. In the phases with gapless modes on or adjacent to edge~B, the added charge is uniformly distributed along the length of the edge. In the gapped SMG phase, the pumped charge is uniformly distributed after the insertion of integer $\mathbb{Z}_{10}^{(2)}$ flux
but it is non-uniform, concentrated near the flux insertion point, when only a fractional flux has been inserted. The energy necessary to add charge to the gapped edge is presumably supplied by the work done during the flux insertion.
In the SMG regime, this bulk-mediated transfer creates a charge imbalance between the gapless sector and the gapped mirror sector while conserving the total charge. This offers a lower-dimensional analogy for understanding how a matter--antimatter imbalance could arise in $(3+1)$ dimensions through charge transfer between ordinary and gapped mirror sectors.

Studies of the system with open boundary conditions reveal additional phase-dependent real-space signatures. In the ground state with one added charge-$(3,0)$ fermion inserted in the $I{=}1$ layer, we find in the intermediate regime with fully gapped edge~B, and only in that regime, interaction-induced local ``currents'' at the end points of the gapped region. These currents describe conversion between different fermion species near the end points.
\par\endgroup

The existence of a fully gapped edge~B for intermediate coupling constants raises an important dynamical question: what happens if, say, a low-energy $I{=}1$ particle, carrying charge $(3,0)$, is incident from the gapless side on the boundary between the gapless edge and the gapped edge? (We  consider here a system where { the six-body interaction is present on a portion of edge B and that portion is in the gapped SMG phase, while the remaining portion is gapless. We assume that  
$L_x = \infty$ and $L_y$ is sufficiently large that interaction between the two edges can be neglected.)}
The incident particle cannot be simply reflected into an $I{=}3$ or $I{=}4$ particle, as that would violate $\mathrm{U}(1)$ charge conservation. In fact, there is no combination of a finite number of $I{=}3$ and $I{=}4$ particles or holes that can give a total charge $(3,0)$. Therefore, the only way that the final state could result in a finite number of fermions emitted from the interface is if a finite amount of charge remains trapped at the interface. This requires that there be a localized many-body state at the interface, with well-defined energy, that can store the missing $\mathrm{U}(1)$ charges. As an example, if the emitted charges consist of one $I{=}3$ fermion and one $I{=}4$ hole, the stored charges would have to be $(-2,-1)$. Since the incident $I{=}1$ particle could have arbitrarily low energy, the localized many-body state would need to have precisely zero energy. It is not clear what would protect this energy.

Repeated incidence of $I{=}1$ particles would necessarily produce additional changes in the charge of the localized state. Charge conservation will allow the localized state to return to its original charge state only after the fifth incident $I{=}1$ particle. This implies that the localized mode must therefore have at least five possible zero-energy states. 
Such a localized mode could also return to its original state after the incidence of five $I{=}2$ fermions by emitting a total of four $I{=}3$ and three $I{=}4$ fermions.  

The alternative scenario, which seems to us more likely, is that an incident $I{=}1$ fermion will result in an infinite cascade of emitted fermions and holes with ever-decreasing energy, such that the total energy of the emitted particles equals the energy of the incident fermion. By the uncertainty principle, particles with arbitrarily low energy must take an arbitrarily long time to exit the interface, so there will never be a time when the process is complete, and there is no need for a strictly zero-energy { localized} state at the interface.

Our discussion of the first scenario tacitly assumed that there exist integer-valued commuting operators that can measure the numbers $N_I^{\pm}$ of particles and holes of the various species on the gapless edge, at least at times long after the collision. This should be possible, as the system far from the SMG interface consists of four independent Chern insulators with non-interacting fermions. 
We note that the expectation values of these occupation numbers need not be integers, as the outgoing wave function may contain a linear superposition of several states with different values of the occupations. However, measurement of the occupations will collapse the wave function into a state with integer values of the occupations, and these occupation numbers will then persist for arbitrarily long times. Occupation numbers in the collapsed state must satisfy the charge-conservation laws.

Charge conservation also imposes restrictions on the outgoing currents in the case where there is a steady incident current of $I{=}1$ and/or $I{=}2$ fermions. If the incident currents are denoted $J_1$ and $J_2$, the time-averaged outgoing currents must be given by

\begin{equation}
J_3 = \frac{3J_1 + 4J_2}{5}, \qquad J_4 = \frac{-4J_1+3J_2}{5}.
\end{equation}

Issues related to the interface between a gapped and a gapless edge in a 3-4-5-0 model have been discussed previously in the literature.  In the language of conformal field theory, if the interacting region flows to a conformal boundary, the incident fermion is reflected into a twisted sector with outgoing fermions attached to a topological line~\cite{Beest2023,Beest2023b}. Similar issues have also been discussed in the fermion--rotor model~\cite{Polchinski1984,Loladze2024,Loladze2025}, introduced for fermion scattering off a magnetic monopole~\cite{Rubakov1982,Callan1982}.
Related scattering problems have been studied for fermions incident on conformal boundaries~ \cite{Affleck1994,Maldacena1995,Wei2026,Tachikawa2026} , alongside lattice constructions of duality defects~\cite{Aasen2016,Ueda2026}.

{
In most cases, the above references do not directly address the question of which scenario correctly describes the outcome after a low-energy particle is incident on the barrier. Moreover, the various models considered might lead to different answers. 
For our lattice model, questions include whether the junction flows at low energies to the same boundary fixed point as one or more of the previous studied models, and if so, 
 through which microscopic mechanism. We leave these questions for future investigation.} %

\begin{acknowledgments}
R.Z.C. acknowledges support from Garnet Kin-Lic Chan.
M.Z. is supported in part
by the Alexander von Humboldt Foundation through a Humboldt Research Fellowship.
Y.Z.Y. is supported by the NSF Grant No. DMR-2238360.
J.W. is supported by LIMS and Ben Delo Fellowship.
This idea and work started while J.W. was affiliated with Harvard CMSA in 2023. J.W. thanks Shinsei Ryu and Biao Lian for related discussions and feedback in 2025.
We used OpenAI Codex and Anthropic Claude Code for assistance with parts of the code, data analysis, visualization, and scientific discussions. The authors reviewed and verified the AI-assisted results and take full responsibility for this work.

\end{acknowledgments}

\bibliography{Ref}

@article{BardeenZumino1984,
    author = "Bardeen, William A. and Zumino, Bruno",
    title = "{Consistent and Covariant Anomalies in Gauge and Gravitational Theories}",
    reportNumber = "FERMILAB-PUB-84-038-T, LBL-17639, UCB-PTH-84-12, LBL-17639-REV, UCB-PTH-84-12-REV",
    doi = "10.1016/0550-3213(84)90322-5",
    journal = "Nucl. Phys. B",
    volume = "244",
    pages = "421--453",
    year = "1984"
}

@misc{Mark2025O2510.11679,
    author = {Mark, Daniel K. and Surace, Federica M. and Schuster, Thomas and Shaw, Adam L. and Gong, Wenjie and Choi, Soonwon and Endres, Manuel},
    title = {Observation of ballistic plasma and memory in high-energy gauge theory dynamics},
    eprint = {2510.11679},
    archivePrefix = {arXiv},
    primaryClass = {quant-ph},
    year = {2025}
}

@article{Wang:2022ucy2204.14271,
    author = "Wang, Juven and You, Yi-Zhuang",
    title = "{Symmetric Mass Generation}",
    eprint = "2204.14271",
    archivePrefix = "arXiv",
    primaryClass = "cond-mat.str-el",
    doi = "10.3390/sym14071475",
    journal = "Symmetry",
    volume = "14",
    number = "7",
    pages = "1475",
    year = "2022"
}

@article{Santos:2013uda1310.8291,
    author = "Santos, Luiz H. and Wang, Juven",
    title = "{Symmetry-protected many-body Aharonov-Bohm effect}",
    eprint = "1310.8291",
    archivePrefix = "arXiv",
    primaryClass = "quant-ph",
    doi = "10.1103/PhysRevB.89.195122",
    journal = "Phys. Rev. B",
    volume = "89",
    number = "19",
    pages = "195122",
    year = "2014"
}

@article{Wang2023,
 author = "Wang, Juven and Wen, Xiao-Gang",
    title = "{Nonperturbative regularization of (1+1)-dimensional anomaly-free chiral fermions and bosons: On the equivalence of anomaly matching conditions and boundary gapping rules}",
    eprint = "1307.7480",
    archivePrefix = "arXiv",
    primaryClass = "hep-lat",
    doi = "10.1103/PhysRevB.107.014311",
    journal = "Phys. Rev. B",
    volume = "107",
    number = "1",
    pages = "014311",
    year = "2023"
}

@article{Wang2019,
 author = "Wang, Juven and Wen, Xiao-Gang",
    title = "{A Solution to the 1+1D Gauged Chiral Fermion Problem}",
    eprint = "1807.05998",
    archivePrefix = "arXiv",
    primaryClass = "hep-lat",
    doi = "10.1103/PhysRevD.99.111501",
    journal = "Phys. Rev. D",
    volume = "99",
    number = "11",
    pages = "111501",
    month = "7",
    year = "2019"
}

@article{Wang:2022fzc2207.14813,
    author = "Wang, Juven",
    title = "{CT or P problem and symmetric gapped fermion solution}",
    eprint = "2207.14813",
    archivePrefix = "arXiv",
    primaryClass = "hep-th",
    doi = "10.1103/PhysRevD.106.125007",
    journal = "Phys. Rev. D",
    volume = "106",
    number = "12",
    pages = "125007",
    year = "2022"
}

@article{Davidson:2008bu0802.2962,
    author = "Davidson, Sacha and Nardi, Enrico and Nir, Yosef",
    title = "{Leptogenesis}",
    eprint = "0802.2962",
    archivePrefix = "arXiv",
    primaryClass = "hep-ph",
    doi = "10.1016/j.physrep.2008.06.002",
    journal = "Phys. Rept.",
    volume = "466",
    pages = "105--177",
    year = "2008"
}

@article{Fukugita:1986hr,
    author = "Fukugita, M. and Yanagida, T.",
    title = "{Baryogenesis Without Grand Unification}",
    reportNumber = "RIFP-641",
    doi = "10.1016/0370-2693(86)91126-3",
    journal = "Phys. Lett. B",
    volume = "174",
    pages = "45--47",
    year = "1986"
}

@article{Wang:2024auy2501.00607,
    author = "Wang, Juven",
    title = "{Topological leptogenesis}",
    eprint = "2501.00607",
    archivePrefix = "arXiv",
    primaryClass = "hep-ph",
    doi = "10.1103/r763-lq1b",
    journal = "Phys. Rev. Res.",
    volume = "8",
    number = "2",
    pages = "L022024",
    year = "2026"
}

@article{Kuzmin:1985mmShaposhnikovBaryon,
    author = "Kuzmin, V. A. and Rubakov, V. A. and Shaposhnikov, M. E.",
    title = "{On the Anomalous Electroweak Baryon Number Nonconservation in the Early Universe}",
    reportNumber = "IC/85/8",
    doi = "10.1016/0370-2693(85)91028-7",
    journal = "Phys. Lett. B",
    volume = "155",
    pages = "36",
    year = "1985"
}

@article{Tong2022,
  title={Comments on symmetric mass generation in 2d and 4d},
  author={Tong, David},
  journal={JHEP},
  volume={2022},
  number={07},
  pages={001},
  year={2022},
  doi={10.1007/JHEP07(2022)001},
  eprint={2104.03997},
  archivePrefix={arXiv},
  primaryClass={hep-th}
}

@article{Zeng2022,
 author = "Zeng, Meng and Zhu, Zheng and Wang, Juven and You, Yi-Zhuang",
    title = "{Symmetric Mass Generation in the 1+1 Dimensional Chiral Fermion 3-4-5-0 Model}",
    eprint = "2202.12355",
    archivePrefix = "arXiv",
    primaryClass = "cond-mat.str-el",
    doi = "10.1103/PhysRevLett.128.185301",
    journal = "Phys. Rev. Lett.",
    volume = "128",
    number = "18",
    pages = "185301",
    year = "2022"
}

@article{Beest2023,
	archiveprefix = {arXiv},
	author = {Marieke van Beest and Philip Boyle Smith and Diego Delmastro and Zohar Komargodski and David Tong},
	eprint = {2306.07318},
	primaryclass = {hep-th},
	title = {Monopoles, Scattering, and Generalized Symmetries},
	journal = {JHEP},
	volume = {2025},
	number = {03},
	pages = {014},
	doi = {10.1007/JHEP03(2025)014},
	year = {2025}}

@article{Turner2018,
	author = {Turner, C. J. and Michailidis, A. A. and Abanin, D. A. and Serbyn, M. and Papi\ifmmode \acute{c}\else \'{c}\fi{}, Z.},
	doi = {10.1103/PhysRevB.98.155134},
	eprint = {1806.10933},
	archivePrefix = {arXiv},
	primaryClass = {cond-mat.quant-gas},
	issue = {15},
	journal = {Phys. Rev. B},
	month = {Oct},
	numpages = {23},
	pages = {155134},
	publisher = {American Physical Society},
	title = {Quantum scarred eigenstates in a Rydberg atom chain: Entanglement, breakdown of thermalization, and stability to perturbations},
	url = {https://link.aps.org/doi/10.1103/PhysRevB.98.155134},
	volume = {98},
	year = {2018}}

@article{White1992,
  title = {Density matrix formulation for quantum renormalization groups},
  author = {White, Steven R.},
  journal = {Phys. Rev. Lett.},
  volume = {69},
  issue = {19},
  pages = {2863--2866},
  numpages = {0},
  year = {1992},
  month = {Nov},
  publisher = {American Physical Society},
  doi = {10.1103/PhysRevLett.69.2863}
}

@article{ITensor,
	title={{The ITensor Software Library for Tensor Network Calculations}},
	author={Matthew Fishman and Steven R. White and E. Miles Stoudenmire},
	journal={SciPost Phys. Codebases},
	pages={4},
	year={2022},
	publisher={SciPost},
	doi={10.21468/SciPostPhysCodeb.4},
	eprint={2007.14822},
	archivePrefix={arXiv},
	primaryClass={cs.MS},
	url={https://scipost.org/10.21468/SciPostPhysCodeb.4},
}

@article{ITensor-r0.3,
	title={{Codebase release 0.3 for ITensor}},
	author={Matthew Fishman and Steven R. White and E. Miles Stoudenmire},
	journal={SciPost Phys. Codebases},
	pages={4-r0.3},
	year={2022},
	publisher={SciPost},
	doi={10.21468/SciPostPhysCodeb.4-r0.3},
	url={https://scipost.org/10.21468/SciPostPhysCodeb.4-r0.3},
}

@article{Fendley2004,
  title = {Competing density-wave orders in a one-dimensional hard-boson model},
  author = {Fendley, Paul and Sengupta, K. and Sachdev, Subir},
  journal = {Phys. Rev. B},
  volume = {69},
  issue = {7},
  pages = {075106},
  numpages = {15},
  year = {2004},
  month = {Feb},
  publisher = {American Physical Society},
  doi = {10.1103/PhysRevB.69.075106},
  eprint = {cond-mat/0309438},
  archivePrefix = {arXiv},
  url = {https://link.aps.org/doi/10.1103/PhysRevB.69.075106}
}

@article{Lesanovsky2012,
  title = {Interacting Fibonacci anyons in a Rydberg gas},
  author = {Lesanovsky, Igor and Katsura, Hosho},
  journal = {Phys. Rev. A},
  volume = {86},
  issue = {4},
  pages = {041601},
  numpages = {5},
  year = {2012},
  month = {Oct},
  publisher = {American Physical Society},
  doi = {10.1103/PhysRevA.86.041601},
  eprint = {1204.0903},
  archivePrefix = {arXiv},
  primaryClass = {cond-mat.quant-gas},
  url = {https://link.aps.org/doi/10.1103/PhysRevA.86.041601}
}

@article{Laughlin1981,
  title = {Quantized Hall conductivity in two dimensions},
  author = {Laughlin, R. B.},
  journal = {Phys. Rev. B},
  volume = {23},
  issue = {10},
  pages = {5632--5633},
  numpages = {0},
  year = {1981},
  month = {May},
  publisher = {American Physical Society},
  doi = {10.1103/PhysRevB.23.5632},
  url = {https://link.aps.org/doi/10.1103/PhysRevB.23.5632}
}

@article{Thouless1983,
  title = {Quantization of particle transport},
  author = {Thouless, D. J.},
  journal = {Phys. Rev. B},
  volume = {27},
  issue = {10},
  pages = {6083--6087},
  numpages = {0},
  year = {1983},
  month = {May},
  publisher = {American Physical Society},
  doi = {10.1103/PhysRevB.27.6083},
  url = {https://link.aps.org/doi/10.1103/PhysRevB.27.6083}
}

@article{You2018,
  title = {Symmetric Fermion Mass Generation as Deconfined Quantum Criticality},
  author = {You, Yi-Zhuang and He, Yin-Chen and Xu, Cenke and Vishwanath, Ashvin},
  journal = {Phys. Rev. X},
  volume = {8},
  issue = {1},
  pages = {011026},
  numpages = {14},
  year = {2018},
  month = {Feb},
  publisher = {American Physical Society},
  doi = {10.1103/PhysRevX.8.011026},
  eprint = {1705.09313},
  archivePrefix = {arXiv},
  primaryClass = {cond-mat.str-el},
  url = {https://link.aps.org/doi/10.1103/PhysRevX.8.011026}
}

@article{You20182,
  title = {From bosonic topological transition to symmetric fermion mass generation},
  author = {You, Yi-Zhuang and He, Yin-Chen and Vishwanath, Ashvin and Xu, Cenke},
  journal = {Phys. Rev. B},
  volume = {97},
  issue = {12},
  pages = {125112},
  numpages = {17},
  year = {2018},
  month = {Mar},
  publisher = {American Physical Society},
  doi = {10.1103/PhysRevB.97.125112},
  eprint = {1711.00863},
  archivePrefix = {arXiv},
  primaryClass = {cond-mat.str-el},
  url = {https://link.aps.org/doi/10.1103/PhysRevB.97.125112}
}

@article{Razamat2021,
  title = {Gapped Chiral Fermions},
  author = {Razamat, Shlomo S. and Tong, David},
  journal = {Phys. Rev. X},
  volume = {11},
  issue = {1},
  pages = {011063},
  numpages = {14},
  year = {2021},
  month = {Mar},
  publisher = {American Physical Society},
  doi = {10.1103/PhysRevX.11.011063},
  eprint = {2009.05037},
  archivePrefix = {arXiv},
  primaryClass = {hep-th},
  url = {https://link.aps.org/doi/10.1103/PhysRevX.11.011063}
}

@article{Butt2021,
  title = {Anomalies and symmetric mass generation for K\"ahler-Dirac fermions},
  author = {Butt, Nouman and Catterall, Simon and Pradhan, Arnab and Toga, Goksu Can},
  journal = {Phys. Rev. D},
  volume = {104},
  issue = {9},
  pages = {094504},
  numpages = {8},
  year = {2021},
  month = {Nov},
  publisher = {American Physical Society},
  doi = {10.1103/PhysRevD.104.094504},
  eprint = {2101.01026},
  archivePrefix = {arXiv},
  primaryClass = {hep-th},
  url = {https://link.aps.org/doi/10.1103/PhysRevD.104.094504}
}

@article{ryan2026,
      title={Chiral Lattice Gauge Theories from Symmetry Disentanglers}, 
      author={Ryan Thorngren and John Preskill and Lukasz Fidkowski},
      year={2026},
      journal={JHEP},
      volume={2026},
      number={07},
      pages={271},
      doi={10.1007/JHEP07(2026)271},
      eprint={2601.04304},
      archivePrefix={arXiv},
      primaryClass={hep-th},
      url={https://arxiv.org/abs/2601.04304}, 
}

@article{Rubakov1982,
    author = "Rubakov, V. A.",
    title = "{Adler-Bell-Jackiw Anomaly and Fermion Number Breaking in the Presence of a Magnetic Monopole}",
    doi = "10.1016/0550-3213(82)90034-7",
    journal = "Nucl. Phys. B",
    volume = "203",
    pages = "311--348",
    year = "1982"
}

@article{Callan1982,
    author = "Callan, Jr., Curtis G.",
    title = "{Dyon-Fermion Dynamics}",
    doi = "10.1103/PhysRevD.26.2058",
    journal = "Phys. Rev. D",
    volume = "26",
    pages = "2058--2068",
    year = "1982"
}

@article{Polchinski1984,
    author = "Polchinski, Joseph",
    title = "{Monopole Catalysis: The Fermion-Rotor System}",
    doi = "10.1016/0550-3213(84)90398-5",
    journal = "Nucl. Phys. B",
    volume = "242",
    pages = "345--363",
    year = "1984"
}

@article{Affleck1994,
    author = "Affleck, Ian and Sagi, Jacob",
    title = "{Monopole catalyzed baryon decay: A Boundary conformal field theory approach}",
    eprint = "hep-th/9311056",
    archivePrefix = "arXiv",
    doi = "10.1016/0550-3213(94)90478-2",
    journal = "Nucl. Phys. B",
    volume = "417",
    pages = "374--402",
    year = "1994"
}

@article{Maldacena1995,
    author = "Maldacena, Juan Martin and Ludwig, Andreas W. W.",
    title = "{Majorana fermions, exact mapping between quantum impurity fixed points with four bulk fermion species, and solution of the `unitarity puzzle'}",
    eprint = "cond-mat/9502109",
    archivePrefix = "arXiv",
    doi = "10.1016/S0550-3213(97)00596-8",
    journal = "Nucl. Phys. B",
    volume = "506",
    pages = "565--588",
    year = "1997"
}

@article{Loladze2025,
    author = "Loladze, Vazha and Okui, Takemichi and Tong, David",
    title = "{Dynamics of the Fermion-Rotor System}",
    eprint = "2508.21059",
    archivePrefix = "arXiv",
    primaryClass = "hep-th",
    doi = "10.1007/JHEP01(2026)052",
    journal = "JHEP",
    volume = "2026",
    number = "01",
    pages = "052",
    year = "2026"
}

@article{Loladze2024,
    author = "Loladze, Vazha and Okui, Takemichi",
    title = "{Monopole-Fermion Scattering and the Solution to the Semiton-Unitarity Puzzle}",
    eprint = "2408.04577",
    archivePrefix = "arXiv",
    primaryClass = "hep-th",
    doi = "10.1103/PhysRevLett.134.051602",
    journal = "Phys. Rev. Lett.",
    volume = "134",
    pages = "051602",
    year = "2025"
}

@article{Beest2023b,
    author = "van Beest, Marieke and Boyle Smith, Philip and Delmastro, Diego and Mouland, Rishi and Tong, David",
    title = "{Fermion-monopole scattering in the Standard Model}",
    eprint = "2312.17746",
    archivePrefix = "arXiv",
    primaryClass = "hep-th",
    doi = "10.1007/JHEP08(2024)004",
    journal = "JHEP",
    volume = "2024",
    number = "08",
    pages = "004",
    year = "2024"
}

@article{Tachikawa2026,
    author = "Tachikawa, Yuji and Tsuji, Keita and Watanabe, Masataka",
    title = "{What happens to wavepackets of fermions when scattered by the Maldacena-Ludwig wall?}",
    eprint = "2603.25508",
    archivePrefix = "arXiv",
    primaryClass = "hep-th",
    doi = "10.1007/JHEP10(2026)140",
    journal = "JHEP",
    volume = "2026",
    number = "10",
    pages = "140",
    year = "2026"
}

@article{Aasen2016,
    author = "Aasen, David and Mong, Roger S. K. and Fendley, Paul",
    title = "{Topological Defects on the Lattice I: The Ising model}",
    eprint = "1601.07185",
    archivePrefix = "arXiv",
    primaryClass = "cond-mat.stat-mech",
    doi = "10.1088/1751-8113/49/35/354001",
    journal = "J. Phys. A",
    volume = "49",
    number = "35",
    pages = "354001",
    year = "2016"
}

@misc{Wei2026,
    author = "Wei, Pengcheng and Zheng, Yunqin",
    title = "{Non-invertible Symmetries in Weyl Fermions, and Applications to Fermion-Boundary Scattering Problem}",
    eprint = "2605.19363",
    archivePrefix = "arXiv",
    primaryClass = "hep-th",
    year = "2026"
}

@article{Ueda2026,
    author = "Ueda, Atsushi and Numasawa, Tokiro and De Vos, Boris and Watanabe, Masataka",
    title = "{Emergent Andreev Reflection from a Lattice Duality Defect}",
    eprint = "2606.23684",
    archivePrefix = "arXiv",
    primaryClass = "cond-mat.str-el",
    doi = "10.1103/h1rv-bv9y",
    journal = "Phys. Rev. Lett.",
    volume = "137",
    pages = "146504",
    year = "2026"
}

@article{Eichten1986,
    author = "Eichten, Estia and Preskill, John",
    title = "{Chiral Gauge Theories on the Lattice}",
    doi = "10.1016/0550-3213(86)90207-5",
    journal = "Nucl. Phys. B",
    volume = "268",
    pages = "179--208",
    year = "1986"
}

@article{Fidkowski2010,
    author = "Fidkowski, Lukasz and Kitaev, Alexei",
    title = "{Effects of interactions on the topological classification of free fermion systems}",
    eprint = "0904.2197",
    archivePrefix = "arXiv",
    primaryClass = "cond-mat.str-el",
    doi = "10.1103/PhysRevB.81.134509",
    journal = "Phys. Rev. B",
    volume = "81",
    pages = "134509",
    year = "2010"
}

@article{Ayyar2015,
    author = "Ayyar, Venkitesh and Chandrasekharan, Shailesh",
    title = "{Massive fermions without fermion bilinear condensates}",
    eprint = "1410.6474",
    archivePrefix = "arXiv",
    primaryClass = "hep-lat",
    doi = "10.1103/PhysRevD.91.065035",
    journal = "Phys. Rev. D",
    volume = "91",
    number = "6",
    pages = "065035",
    year = "2015"
}

@article{He2016,
    author = "He, Yuan-Yao and Wu, Han-Qing and You, Yi-Zhuang and Xu, Cenke and Meng, Zi Yang and Lu, Zhong-Yi",
    title = "{Quantum critical point of Dirac fermion mass generation without spontaneous symmetry breaking}",
    eprint = "1603.08376",
    archivePrefix = "arXiv",
    primaryClass = "cond-mat.str-el",
    doi = "10.1103/PhysRevB.94.241111",
    journal = "Phys. Rev. B",
    volume = "94",
    pages = "241111",
    year = "2016"
}

@article{Wen:2013ppa-1305.1045,
    author = "Wen, Xiao-Gang",
    title = "{A lattice non-perturbative definition of an SO(10) chiral gauge theory and its induced standard model}",
    eprint = "1305.1045",
    archivePrefix = "arXiv",
    primaryClass = "hep-lat",
    doi = "10.1088/0256-307X/30/11/111101",
    journal = "Chin. Phys. Lett.",
    volume = "30",
    pages = "111101",
    year = "2013"
}

@article{Wang:2018jkc-1809.11171,
    author = "Wang, Juven and Wen, Xiao-Gang",
    title = "{Nonperturbative definition of the standard models}",
    eprint = "1809.11171",
    archivePrefix = "arXiv",
    primaryClass = "hep-th",
    doi = "10.1103/PhysRevResearch.2.023356",
    journal = "Phys. Rev. Res.",
    volume = "2",
    number = "2",
    pages = "023356",
    year = "2020"
}

@article{Chen2013,
    author = "Chen, Chen and Giedt, Joel and Poppitz, Erich",
    title = "{On the decoupling of mirror fermions}",
    eprint = "1211.6947",
    archivePrefix = "arXiv",
    primaryClass = "hep-lat",
    doi = "10.1007/JHEP04(2013)131",
    journal = "JHEP",
    volume = "2013",
    number = "04",
    pages = "131",
    year = "2013"
}

@article{WilczekZeePLB1979,
    author = "Wilczek, Frank and Zee, A.",
    title = "{Conservation or Violation of B-L in Proton Decay}",
    reportNumber = "UPR-0135T",
    doi = "10.1016/0370-2693(79)90475-1",
    journal = "Phys. Lett. B",
    volume = "88",
    pages = "311--314",
    year = "1979"
}

@article{Wilczek1979hcZee,
	author = {Wilczek, Frank and Zee, A.},
	doi = {10.1103/PhysRevLett.43.1571},
	journal = {Phys. Rev. Lett.},
	pages = {1571--1573},
	reportnumber = {Print-79-0709 (PRINCETON)},
	title = {{Operator Analysis of Nucleon Decay}},
	volume = {43},
	year = {1979}}

@book{Weinberg:1996kr-book,
    author = "Weinberg, Steven",
    title = "{The quantum theory of fields. Vol. 2: Modern applications}",
    doi = "10.1017/CBO9781139644174",
    isbn = "978-1-139-63247-8, 978-0-521-67054-8, 978-0-521-55002-4",
    publisher = "Cambridge University Press",
    month = "8",
    year = "2013"
}

@article{Halperin1982,
    author = "Halperin, B. I.",
    title = "{Quantized Hall conductance, current-carrying edge states, and the existence of extended states in a two-dimensional disordered potential}",
    doi = "10.1103/PhysRevB.25.2185",
    journal = "Phys. Rev. B",
    volume = "25",
    pages = "2185--2190",
    year = "1982"
}

@article{Niu1985,
    author = "Niu, Qian and Thouless, D. J. and Wu, Yong-Shi",
    title = "{Quantized Hall conductance as a topological invariant}",
    doi = "10.1103/PhysRevB.31.3372",
    journal = "Phys. Rev. B",
    volume = "31",
    pages = "3372--3377",
    year = "1985"
}

@article{Wen1990,
    author = "Wen, X. G.",
    title = "{Chiral Luttinger liquid and the edge excitations in the fractional quantum Hall states}",
    doi = "10.1103/PhysRevB.41.12838",
    journal = "Phys. Rev. B",
    volume = "41",
    pages = "12838--12844",
    year = "1990"
}

@article{TKNN1982,
    author = "Thouless, D. J. and Kohmoto, M. and Nightingale, M. P. and den Nijs, M.",
    title = "{Quantized Hall Conductance in a Two-Dimensional Periodic Potential}",
    doi = "10.1103/PhysRevLett.49.405",
    journal = "Phys. Rev. Lett.",
    volume = "49",
    pages = "405--408",
    year = "1982"
}

@misc{seifnashri2026,
  title={Exactly Solvable 1+1d Chiral Lattice Gauge Theories},
  author={Seifnashri, Sahand},
  eprint={2601.14359},
  archivePrefix={arXiv},
  primaryClass={hep-th},
  year={2026},
  url={https://arxiv.org/abs/2601.14359}
}

@misc{LuShao2026,
author = "Lu, Zhiyao and Shao, Shu-Heng",
    title = "{Fermionic Villain model with exact lattice chiral symmetries}",
    eprint = "2608.02728",
    archivePrefix = "arXiv",
    primaryClass = "hep-th",
    reportNumber = "MIT-CTP/6081",
    month = "8",
    year = "2026"
}

@article{Zakharov2026,
  title={Symmetric mass generation of interacting chiral fermions on a one-dimensional lattice without fermion doubling},
  author={Zakharov, V. A. and Ueda, Atsushi and Verstraete, Frank and Beenakker, C. W. J.},
  journal={Phys. Rev. Res.},
  volume={8},
  pages={043009},
  year={2026},
  doi={10.1103/2v9h-25qs},
  eprint={2606.24713},
  archivePrefix={arXiv},
  primaryClass={hep-th},
  url={https://arxiv.org/abs/2606.24713}
}

@misc{Xu2025staggered,
  title={Staggered Fermions with Chiral Anomaly Cancellation},
  author={Xu, Ling-Xiao},
  year={2025},
  eprint={2501.10837},
  archivePrefix={arXiv},
  primaryClass={hep-lat},
  url={https://arxiv.org/abs/2501.10837}
}

\clearpage 
\appendix 
\renewcommand{\theHfigure}{\thesection.\arabic{figure}}
\renewcommand{\theHtable}{\thesection.\arabic{table}}
\onecolumngrid 
\setcounter{figure}{0}
\renewcommand{\thefigure}{A\arabic{figure}}
\section{Ground-state and low-energy excitations in the strong-interaction limit}
\label{app:A}

\subsection{Edge-B ground state analysis in the \texorpdfstring{$g\to\infty$}{g to infinity} limit}
\label{app:A-ground}

 We first consider edge~B with periodic boundary conditions. In the strong-interaction limit it decouples dynamically and its Hamiltonian reduces to
  $H^{1}_{\mathrm{int}}+H^{2}_{\mathrm{int}}$.
The symmetry generator $U^{R}_{(1,4)}$ exchanges the two interaction channels, while $U^{R}_{(1,3)}$ and particle-hole symmetry $U_{\mathcal{C}}$ preserve each channel separately. $[H^{1}_{\mathrm{int}},H^{2}_{\mathrm{int}}]=0$, so the two channels share an eigenbasis. This allows us to focus on $H^{1}_{\mathrm{int}}$ first, the remaining ground states can be derived by symmetry transformation.

\begin{figure}[!htbp]
    \centering
    \includegraphics[width=0.42\textwidth]{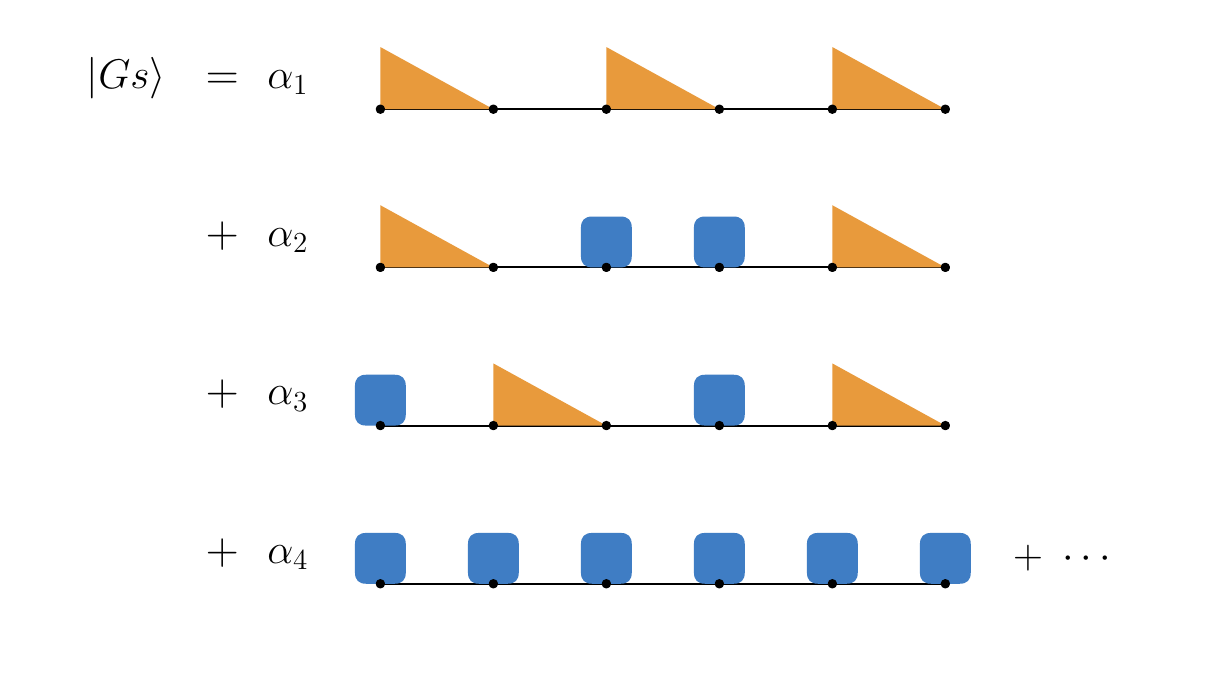}
    \caption{The ground state $|\mathrm{GS}\rangle$ is written as a superposition of low-energy monomer--dimer basis states for the strong-interaction limit of edge~$B$ ($L_x=6$). Blue squares denote monomers and orange triangles denote dimers. The coefficients $\alpha_i$ label the amplitudes of the corresponding basis states.\label{figA1}}
\end{figure}

The six-body interaction admits a monomer--dimer rewriting,
\begin{equation}
H^{1}_{\mathrm{int}} = g_1 \sum_{r=1}^{L_x}
    \left(\Psi_{1,r} \Psi^{\dagger}_{2,r} \Psi^{\dagger}_{2,r+1}
    \Psi_{3,r} \Psi_{4,r} \Psi_{4,r+1} + \text{h.c.}\right) = g_1 \sum_{r=1}^{L_x} \left(M_r^{\dagger} M_{r+1}^{\dagger} D_r + \text{h.c.}\right),
\label{Hint_A}
\end{equation}
with
\begin{equation}
M_r = \Psi_{2,r}^{B}, \qquad D_r = \Psi_{1,r}^{B} \Psi_{3,r}^{B} \Psi_{4,r}^{B} \Psi_{4,r+1}^{B}.
\label{Hint_appendix}
\end{equation}
Thus $M_r^{\dagger}$ creates a single flavor-$2$ fermion at site $r$, and $D_r^{\dagger}$ creates a
four-fermion composite on the bond $(r,r{+}1)$, with flavors $1$, $3$ and $4$ at site $r$ and flavor $4$ at site $r{+}1$. Every term converts one
dimer into two adjacent monomers and back. This conversion preserves the non-overlap of
monomers and dimers, so $H^{1}_{\mathrm{int}}$ acts invariantly on the span of non-overlapping
monomer--dimer coverings.
By Table~\ref{tab1} every covering carries the same charge $(Q^{(1)}_B,Q^{(2)}_B)=(4L_x,5L_x)$,
so the whole manifold lies in one sector. We propose a superposition of
coverings as ground-state ansatz,
\begin{equation}
    |\mathrm{GS}\rangle = \sum_{C}\alpha_C |C\rangle,\qquad |C\rangle = \prod_{r\in\mathcal{M}(C)} M_r^{\dagger}\prod_{\langle r,r+1\rangle\in\mathcal{D}(C)} D_r^{\dagger}|\Omega\rangle,
\end{equation}
with $\mathcal{M}(C)$ and $\mathcal{D}(C)$ the monomer sites and dimer bonds. A
representative covering is drawn in Fig.~\ref{figA1}.

In order to solve within this subspace, we introduce a dual description from monomer-dimer coverings to a spin-1/2 chain, where we assign to each bond a
two-level system, $|{\uparrow}\rangle_r$ for a dimer and $|{\downarrow}\rangle_r$ for none.
The non-overlap condition forbids dimers on adjacent bonds, which is exactly the PXP
constraint~\cite{Fendley2004,Lesanovsky2012}, and the projected Hamiltonian is
\begin{equation}
H_{\mathrm{eff}} = g_1 \sum_{r=1}^{L_x} P_{r-1}\, \sigma^x_r\, P_{r+1},
\qquad P_r=(1-\sigma^z_r)/2.
\end{equation}
A related reinterpretation as an effective lattice gauge theory has been discussed
recently~\cite{Mark2025O2510.11679}. An $L_x$-site chain maps to a PXP chain of $L_x$
(for periodic boundary conditions) or $L_x-1$ (for open boundary conditions) sites, with constrained dimensions given by Lucas and Fibonacci
numbers~\cite{Turner2018}.

We label the ground state of the finite PXP chain within the covering manifold by
$|G_1\rangle$. It is the state $|\mathrm{GS}\rangle$ introduced above. It has energy $E_0$ and charge
$(Q^{(1)}_B,Q^{(2)}_B)=(4L_x,5L_x)$. The
second channel annihilates this state, since
$H^{2}_{\mathrm{int}}|C\rangle=0$ for every covering and hence
$H^{2}_{\mathrm{int}}|G_1\rangle=0$. Therefore, $|G_1\rangle$ is an exact eigenstate of the
full edge-B Hamiltonian $H^{1}_{\mathrm{int}}+H^{2}_{\mathrm{int}}$ with energy $E_0$, and we
verify numerically below that it is in fact the ground state. We first analyze how the $D_4$
symmetry acts on it.

Writing $W\equiv U^{R}_{(1,4)}U^{R}_{(1,3)}$, the eight elements of $D_4$ are the
four unitary rotations and the four antiunitary reflections,
\begin{equation}
D_4=\{\,1,\ W,\ U_{\mathcal{C}}=W^{2},\ W^{3}\,\}\cup
\{\,U^{R}_{(1,3)},\ U_{\mathcal{C}}U^{R}_{(1,3)},\ U^{R}_{(1,4)},\ U_{\mathcal{C}}U^{R}_{(1,4)}\,\}.
\label{eq:D4elements}
\end{equation}
Under $U^{R}_{(1,3)}$, the occupations transform as
$(n_1,n_2,n_3,n_4)\mapsto(n_3,1-n_4,n_1,1-n_2)$, leaving the monomer $(0,1,0,0)$ and the two
dimer sites $(1,0,1,1)$ and $(0,0,0,1)$ unchanged. Thus $U^{R}_{(1,3)}$ preserves the
covering manifold and its charge sector. Since it also preserves $H^{1}_{\mathrm{int}}$, the
nondegeneracy of the finite PXP ground state implies
$U^{R}_{(1,3)}|G_1\rangle=e^{\ii\phi}|G_1\rangle$. The other six elements change the charge
sector by particle--hole conjugation or channel exchange and therefore do not stabilize
$|G_1\rangle$. Thus only the identity and $U^{R}_{(1,3)}$ stabilize $|G_1\rangle$.
Since $D_4$ has eight elements, the orbit--stabilizer theorem gives the four states listed below.
\begin{equation}
\begin{array}{llll}
\toprule
 & \text{monomer} & \text{dimer } (\text{left},\ \text{right}) & (Q^{(1)}_B,Q^{(2)}_B)\\
\midrule
G_1 & \psi_2 & (\psi_1\psi_3\psi_4,\ \psi_4) & (4L_x,5L_x)\\
G_2 & \psi_3 & (\psi_1\psi_2\psi_4,\ \psi_1) & (5L_x,4L_x)\\
G_3 & \psi_1\psi_3\psi_4 & (\psi_2,\ \psi_1\psi_2\psi_3) & (8L_x,7L_x)\\
G_4 & \psi_1\psi_2\psi_4 & (\psi_3,\ \psi_2\psi_3\psi_4) & (7L_x,8L_x)\\
\bottomrule
\end{array}
\label{eq:corners}
\end{equation}

\begin{figure}[!htbp]
    \centering
    \includegraphics[width=\textwidth]{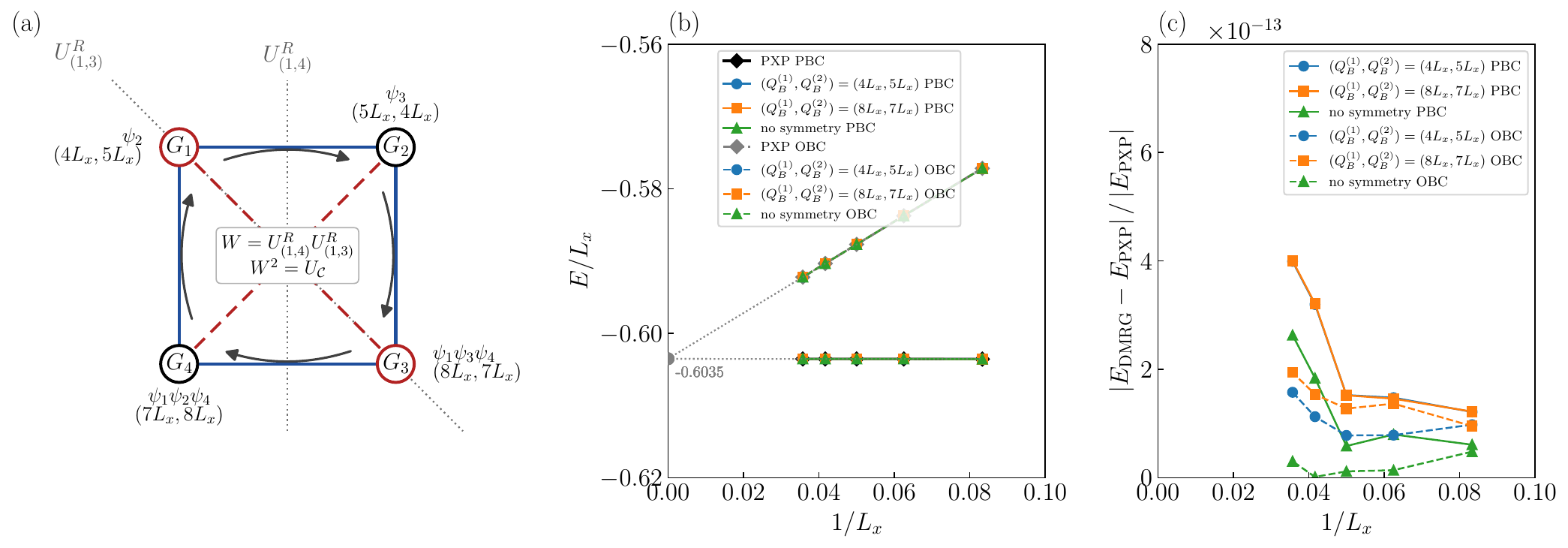}
    \caption{(a) The four states of Eq.~\eqref{eq:corners} and their $D_4$ transformations.
    The solid arrows denote $W=U^{R}_{(1,4)}U^{R}_{(1,3)}$, the dashed diagonals denote
    $U_{\mathcal{C}}=W^{2}$, and the dotted lines denote the two antiunitary reflections.
    (b) Ground-state energy per site of $H^{1}_{\mathrm{int}}$ obtained from the effective PXP
    model and from DMRG in the $(4L_x,5L_x)$ and $(8L_x,7L_x)$ sectors and without symmetry
    restriction, with energies in units of $g_1$, for periodic boundary conditions (lower line) and open boundary conditions (upper line). (c) Relative energy difference
    between the PXP and DMRG results.}
    \label{figA2}
\end{figure}

Because $D_4$ commutes with $H^{1}_{\mathrm{int}}+H^{2}_{\mathrm{int}}$, the four states
are exactly degenerate and, being in distinct charge sectors, mutually orthogonal. Their
transformations under $D_4$ are shown in Fig.~\ref{figA2}(a). In particular,
$U^{R}_{(1,3)}$ fixes $G_1$ and $G_3$, while $U_{\mathcal{C}}U^{R}_{(1,3)}$ fixes $G_2$ and
$G_4$. These residual symmetries constrain the flavor densities: for $G_1$ and $G_3$,
$\langle n_1\rangle=\langle n_3\rangle$ and $\langle n_2\rangle+\langle n_4\rangle=1$;
for $G_2$ and $G_4$, $\langle n_2\rangle=\langle n_4\rangle$ and
$\langle n_1\rangle+\langle n_3\rangle=1$.

We now verify the ground-state claim numerically. First, exact diagonalization of
$H^{1}_{\mathrm{int}}+H^{2}_{\mathrm{int}}$ on periodic rings at $L_x=4$ and $6$ finds a fourfold-degenerate
lowest level, with one state in each of the four charge sectors of Eq.~\eqref{eq:corners},
in agreement with the symmetry-generated multiplet. Fig.~\ref{figA2} benchmarks the
PXP reduction against charge-resolved DMRG for
$H^{1}_{\mathrm{int}}$ in the sectors $(4L_x,5L_x)$ and $(8L_x,7L_x)$ and without symmetry
restriction, under both open boundary conditions and periodic boundary conditions. The constrained manifold is small, so $E_0$ is obtained by
direct diagonalization, and extrapolates to $E_0/L_x\to-0.60356\,g$. All DMRG energies agree
with the PXP values to machine precision for the sizes shown, supporting the PXP reduction and
the absence of a lower state outside the covering subspace. Together, these results provide
numerical evidence that the fourfold multiplet of Eq.~\eqref{eq:corners} is the periodic edge-B
ground-state manifold in the $g\to\infty$ limit.

\subsection{Edge-B excitation analysis}
\label{app:A-excite}

We next determine the low-energy excitations, taking the sector $(8L_x,7L_x)$ of $G_3$
as the reference state. Because $D_4$ is a symmetry of the Hamiltonian, its action maps
the four ground states and their corresponding excitation ladders onto one another without
changing the energy. Diagonalizing the full Hilbert space at $L_x=6$ shows
that the low-energy excitations fall into two families and their bound states, and both are
built on the uniform monomer covering of $G_3$. A particle excitation removes or adds
fermions on a single site. A domain-wall excitation replaces the covering on $d$
consecutive sites by that of another ground state, producing a pair of walls.

\begin{figure}[!htbp]
    \centering
    \includegraphics[width=\textwidth]{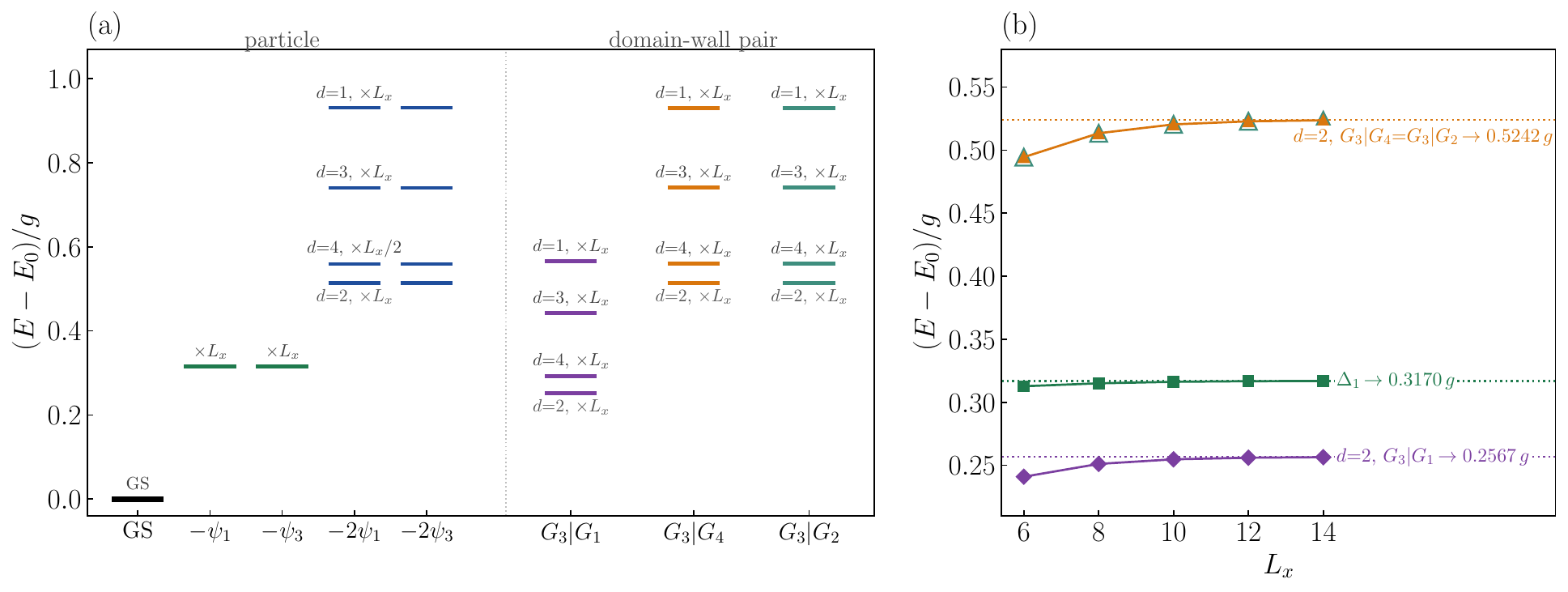}
    \caption{Selected low-lying excitations of the isolated edge-B ring above $G_3$, sector
    $(8L_x,7L_x)$, at $L_x=8$, from exact diagonalization of the Hilbert-space fragments. Each
    level is labeled by the degeneracy of the specified excitation. (a) Particle excitations (left), including pairs of
    them at separation $d$, and domain-wall pairs (right) on rings with
     $d\le L_x/2$ sites of $G_3$ and $L_x-d$ sites of $G_1$, $G_4$ or $G_2$. (b) Size dependence of
    $\Delta_1$ and of the two $d=2$ domains, with dotted lines marking their limits and
    open triangles marking $G_3|G_2$, which coincides with $G_3|G_4$. }
     \label{figA3}
\end{figure}

At larger sizes we start from a single configuration, the monomer covering of $G_3$
with one defect inserted, and collect every configuration reached from it by repeated
action of $H^{1}_{\mathrm{int}}+H^{2}_{\mathrm{int}}$. These configurations span an invariant
subspace that is small enough to diagonalize exactly. Translating the defect gives
disconnected copies of this subspace that are related by symmetry, and the number of copies
gives the degeneracies in Fig.~\ref{figA3}(a). The indicated degeneracies count the distinct positions of each specified excitation. All energies are proportional to $g$.

A particle excitation interrupts the dimer resonance. Removing $\psi_1$ or $\psi_3$ from a
site blocks the bond to its right and leaves an open PXP chain of $L_x-1$ sites, so
$\Delta_1=E^{\mathrm{open}}_{\mathrm{PXP}}(L_x-1)-E^{\mathrm{ring}}_{\mathrm{PXP}}(L_x)\to0.317045\,g$,
where $E^{\mathrm{open}}_{\mathrm{PXP}}(m)$ and $E^{\mathrm{ring}}_{\mathrm{PXP}}(L_x)$ are the
ground energies of the open $m$-site and periodic $L_x$-site chains. Two
defects do not move, so each separation $d$ gives its own flat level, with even $d$ bound
below $2\Delta_1$ and odd $d$ above it.

The domain-wall excitations form three families, labeled by the ground state $G_j$,
$j=1,2,4$, that fills the replaced segment. A member of the $G_3|G_j$ family is a ring with
$d$ sites in $G_3$ and $L_x-d$ sites in $G_j$, bounded by two $G_3|G_j$ walls. We obtain
each family by diagonalizing the interaction Hamiltonian in the subspace generated from
$G_3^dG_j^{L_x-d}$ for every $d\leq L_x/2$. The lowest level of the $G_3|G_1$ family is the
$d=2$ domain with $\Delta_{\mathrm{pair}}\to0.2567\,g$. Since $U^{R}_{(1,3)}$ fixes $G_3$
and exchanges $G_2$ and $G_4$, the $G_3|G_2$ and $G_3|G_4$ families are exactly degenerate at
every $d$, and their lowest level, also at $d=2$, is $0.5242\,g$. In every family odd $d$
costs more than even $d$. Figure~\ref{figA3}(b) shows how the
lowest level of each family converges with system size. For comparison, the bulk gap of the free layers is
$\Delta_{\mathrm{bulk}}=\min(8t_2,4t_1)=4t_1$~\cite{Wang2023}.
The corresponding particle-addition or removal gap at half-filling is
$\Delta_{\mathrm{bulk}}/2=2t_1=2$.

\subsection{Full system at finite large \texorpdfstring{$g$}{g}}
\label{app:A-finite}

We now embed the edge-B analysis into the full two-dimensional system. In the strict $g\to\infty$ limit, bulk--edge~B hopping is parametrically small compared with the interaction scale, so edge~B dynamically decouples and a ground state in each edge-B sector can be chosen to factorize,
\begin{equation}
    |\mathrm{GS}\rangle = |\mathrm{GS}_B\rangle \otimes |\mathrm{GS}_{\mathrm{bulk}}\rangle,
\end{equation}
where $|\mathrm{GS}_B\rangle$ lies in the fourfold multiplet of Eq.~\eqref{eq:corners} and $|\mathrm{GS}_{\mathrm{bulk}}\rangle$ is the ground state of four decoupled Chern insulators on the remaining $L_x\times(L_y-1)$ geometry. Total $\mathrm{U}(1)^{(1)}\times\mathrm{U}(1)^{(2)}$ conservation fixes the bulk charge once the edge-B sector is specified, $(Q^{(1)}_{\mathrm{bulk}},Q^{(2)}_{\mathrm{bulk}}) = (Q^{(1)}_{\mathrm{tot}} - Q^{(1)}_B, Q^{(2)}_{\mathrm{tot}} - Q^{(2)}_B)$. At half filling, the corresponding bulk ground states are related by symmetry, so the fourfold edge-B degeneracy persists in the full system.

Each of the four ground states retains only its antiunitary stabilizer,
$\{1,\,U^{R}_{(1,3)}\}$ for $G_1$ and $G_3$ and $\{1,\,U_{\mathcal{C}}U^{R}_{(1,3)}\}$ for
$G_2$ and $G_4$, so the strong-coupling regime breaks $D_4$ down to a single antiunitary
$\mathbb{Z}_2$. The four states are labeled by the two broken symmetries
$\mathbb{Z}_2^{R}$ and $\mathbb{Z}_2^{\mathcal{C}}$, and the question is in which order
these two symmetries break as $g$ increases.

  For a nested symmetry-breaking sequence, group theory allows two scenarios. Any residual symmetry group along such a
 breaking path must contain the final stabilizer $\{1,U^{R}_{(1,3)}\}$, and the subgroups of
$D_4$ that contain it are the stabilizer itself,
$\{1,\ U^{R}_{(1,3)},\ U_{\mathcal{C}},\ U_{\mathcal{C}}U^{R}_{(1,3)}\}$, and $D_4$. Either both
symmetries break in a single step, or $\mathbb{Z}_2^{R}$ breaks first and the residual
group is the four-element one. The particle-hole symmetry $\mathbb{Z}_2^{\mathcal{C}}$ can
never break first.

At finite $g$, hopping between edge~B and the bulk hybridizes the four decoupled ground
states. Turning one ground state into another moves $\mathcal{O}(L_x)$ fermions between
edge~B and the bulk, one hop at a time. Every intermediate state contains an edge-B defect
whose energy is proportional to $g$, so a path of $n$ hops contributes $t^{n}/g^{n-1}$, and
the splitting between two ground states is
\begin{equation}
\Delta \;\sim\; g\,(t/g)^{n}, \qquad n=\mathcal{O}(L_x).
\label{eq:splitting}
\end{equation}
It is a power of $g$ at fixed size and becomes exponentially small in $L_x$ at fixed $g$.

\subsection{Exact diagonalization of the full system at \texorpdfstring{$L_x=4$, $L_y=2$}{Lx=4, Ly=2}}

\begin{figure}[!htbp]
    \centering
    \includegraphics[width=\textwidth]{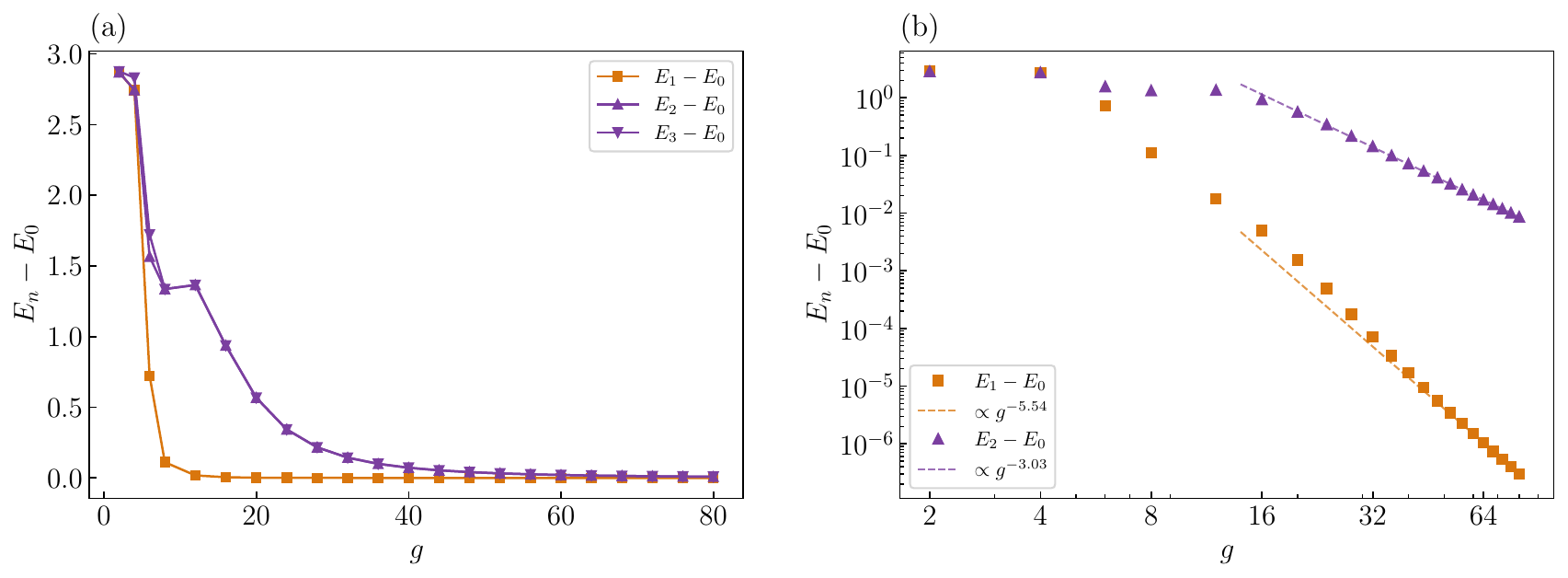}
    \caption{Exact spectrum at $L_x=4$, $L_y=2$, sector $(48,48)$, with
    antiperiodic boundary conditions along $x$.
    (a) Excitation energies $E_n-E_0$ of the three excited states among the four
    lowest levels.
    (b) The two lowest gaps $E_1-E_0$ and $E_2-E_0$ on a log-log scale,
    each fitted to a power of $g$ on its last four points.}
    \label{figA4}
\end{figure}

To test this picture we exactly diagonalize the full model at $L_x=4$,
$L_y=2$ with antiperiodic boundary conditions along $x$, which
lift the free-fermion degeneracies at half filling, and compute the four
lowest levels of the half-filled sector
$(Q^{(1)},Q^{(2)})=(6L_xL_y,6L_xL_y)=(48,48)$ for $2\le g\le 80$.
Figure~\ref{figA4}(a) shows the excitation energies $E_n-E_0$ for
$n=1,2,3$. At strong coupling the four levels group into two
quasi-degenerate pairs, $(E_0,E_1)$ and $(E_2,E_3)$, which approach one another as $g$ increases and form the four-state
low-energy multiplet. Figure~\ref{figA4}(b) shows the two relevant gaps: for
$E_2-E_0$, the separation between these pairs scales as $g^{-3}$, whereas
$E_1-E_0$, the splitting within the lowest pair, decays faster, with an
effective exponent between $5$ and $6$. Both behaviors agree with
Eq.~\eqref{eq:splitting}.

The order-parameter matrix elements in Tables~\ref{tabA1} and~\ref{tabA2} identify the
low-energy pairs associated with the two symmetries. The $U^{R}_{(1,4)}$-odd channel imbalance
  $\hat D=(H^{1}_{\mathrm{int}}-H^{2}_{\mathrm{int}})/g$ couples only $(E_0,E_1)$
and $(E_2,E_3)$, identifying the pairs coupled by the channel imbalance. The splitting of the lowest pair is $E_1-E_0$. The
$U_{\mathcal C}$-odd edge-B charge
$\hat S=\tfrac12(\hat Q^{(1)}_B+\hat Q^{(2)}_B)-6L_x$ instead couples the two pairs.
It therefore identifies $E_0, E_1, E_2, E_3$ as
the relevant four-state manifold, with separation $E_2-E_0$ between its two pairs. The origin of the two
strong-coupling power laws is explained below from the shortest hopping paths.

Having identified the gaps, we can understand their scaling from the shortest
hopping path between the two ground-state manifolds, $n$ hops giving
$t^{n}/g^{n-1}$. The coverings of $G_3$ and $G_1$ share the site contents
$\psi_2$ and $\psi_1\psi_3\psi_4$, so the dimerized covering of $G_3$ reaches
the monomer covering of $G_1$ by emptying its two dimer right ends into the
bulk,
\begin{equation}
(\psi_2,\ \psi_1\psi_2\psi_3,\ \psi_2,\ \psi_1\psi_2\psi_3)
\;\xrightarrow{\;-\psi_1,\,-\psi_3\ \text{on sites 2 and 4}\;}\;
(\psi_2,\ \psi_2,\ \psi_2,\ \psi_2),
\end{equation}
four hops and hence $g^{-3}$. The coverings of $G_3$ and $G_4$ share no site
content, so all four sites must change. The charge transfer $(-L_x,L_x)$
needs four hops to the bulk, and two hops along edge~B complete the
conversion,
\begin{align}
&(\psi_2,\ \psi_1\psi_2\psi_3,\ \psi_1\psi_3\psi_4,\ \psi_1\psi_3\psi_4)\notag\\
&\xrightarrow{\substack{\psi_1\ \text{from site 2 to site 1}\\[1pt] \psi_2\ \text{from site 2 to site 3}}}
(\psi_1\psi_2,\ \psi_3,\ \psi_1\psi_2\psi_3\psi_4,\ \psi_1\psi_3\psi_4)\notag\\
&\xrightarrow{\substack{+\psi_4\ \text{on site 1},\ -\psi_1\ \text{on site 3}\\[1pt] -\psi_3,\,+\psi_2\ \text{on site 4}}}
(\psi_1\psi_2\psi_4,\ \psi_3,\ \psi_2\psi_3\psi_4,\ \psi_1\psi_2\psi_4),
\end{align}
six hops and hence $g^{-5}$. Paths with seven and eight hops contribute
$g^{-6}$ and $g^{-7}$, and the observed exponent between $5$ and $6$ is the
crossover between these channels.

\begin{table}[!htbp]
  \begin{minipage}[t]{0.48\textwidth}
  \centering
  \caption{$|\langle m|\hat D|n\rangle|$ for
    $\hat D = (H^{1}_{\mathrm{int}}-H^{2}_{\mathrm{int}})/g$ in the basis of the
  four lowest eigenstates at $g=16$. Entries below $5\times10^{-9}$ are
  shown as $0$.}
  \label{tabA1}
  \begin{tabular*}{\textwidth}{@{\extracolsep{\fill}}l cccc}
    \toprule
     & $|0\rangle$ & $|1\rangle$ & $|2\rangle$ & $|3\rangle$ \\
    \midrule
    $\langle 0|$ & $0$ & $2.2256$ & $0$ & $0$ \\
    $\langle 1|$ & $2.2256$ & $0$ & $0$ & $0$ \\
    $\langle 2|$ & $0$ & $0$ & $0$ & $2.3374$ \\
    $\langle 3|$ & $0$ & $0$ & $2.3374$ & $0$ \\
    \bottomrule
  \end{tabular*}
  \end{minipage}\hfill
  \begin{minipage}[t]{0.48\textwidth}
  \centering
  \caption{$|\langle m|\hat S|n\rangle|$ for
  $\hat S = \tfrac12(\hat Q^{(1)}_B+\hat Q^{(2)}_B)-6L_x$ in the basis of
  the four lowest eigenstates at $g=72$. Entries below $10^{-7}$ are shown
  as $0$.}
  \label{tabA2}
  \begin{tabular*}{\textwidth}{@{\extracolsep{\fill}}l cccc}
    \toprule
     & $|0\rangle$ & $|1\rangle$ & $|2\rangle$ & $|3\rangle$ \\
    \midrule
    $\langle 0|$ & $0$ & $0$ & $4.2033$ & $4.2049$ \\
    $\langle 1|$ & $0$ & $0$ & $4.2049$ & $4.2033$ \\
    $\langle 2|$ & $4.2033$ & $4.2049$ & $0$ & $0$ \\
    $\langle 3|$ & $4.2049$ & $4.2033$ & $0$ & $0$ \\
    \bottomrule
  \end{tabular*}
  \end{minipage}
\end{table}

\FloatBarrier
\renewcommand{\thefigure}{B\arabic{figure}}
\begingroup
\raggedbottom
\section{Charge and current distributions}
\setcounter{figure}{0}
\label{app:C}

\subsection{Charge and current distributions for different interaction strengths}
\label{app:C-profiles}

Figure~\ref{fig4} in the main text shows the deviations in the fermion densities and currents
at three representative interaction strengths ($g=2.0$, $5.5$, $8.0$), after the addition
of one $I=1$ particle. Here we present the full evolution across six values,
$g = 4.0, 5.0, 6.0, 7.0, 16.0, 80.0$, in Fig.~\ref{figC1}. The results shown here are
obtained by adiabatically continuing the $H^{\text{int}}_1$-dominant strong-coupling sector
toward smaller $g$. Concretely, we run DMRG sequentially: the converged ground state at the
largest $g$ is used as the initial state for the next slightly smaller $g$, and this
procedure is repeated stepwise down to $g=4.0$. This continuation procedure is used to
follow a single low-energy branch across $g$, thereby reducing branch-switching artifacts
in the density and current profiles.

\begin{figure}[!htbp]
    \centering
    \includegraphics[width=0.8\textwidth]{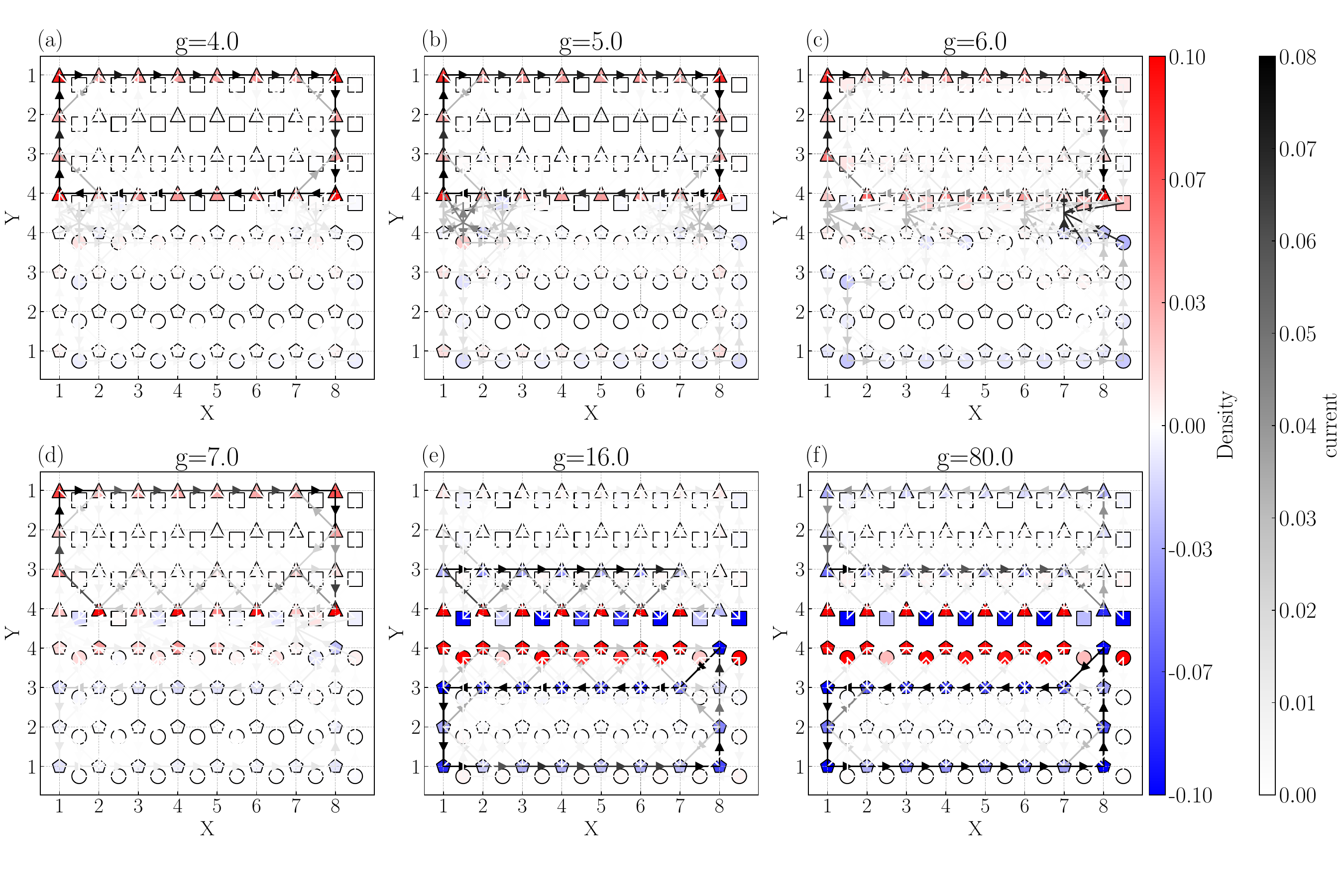}
    \caption{Deviations in the fermion densities and currents for $L_x=8$, $L_y=4$ at interaction strengths $g=4.0$, $5.0$, $6.0$, $7.0$, $16.0$, and $80.0$ under open boundary conditions,
    after the addition of one $I=1$ particle. The blue-red color scale encodes the density deviation from the half-filling value $0.5$. The arrows show the measured currents without background subtraction. For visibility, the contribution from the nonlocal interaction term is multiplied by~$5$. Fermion flavors $I=1,2,3,4$ are represented by triangles, squares, pentagons, and circles, respectively. The system is at half-filling with one additional fermion of flavor $I=1$ and charge $(Q^{(1)},Q^{(2)})=(3,0)$.
    \label{figC1}}
\end{figure}

At weak coupling ($g=4.0$), the injected fermion remains localized on the $I=1$ layer and forms a chiral current loop around the perimeter, as expected for a gapless boundary mode. The edge response is therefore still dominated by the original chiral fermion channel.

At $g=5.0$ and $5.5$, the pattern changes qualitatively. The excess charge is no longer confined to the $I=1$ layer: positive density develops in the $I=3$ layer and negative density in the $I=4$ layer, while pronounced interaction-induced currents appear at the two endpoints of edge~B. Thus the incoming $(3,0)$ charge is redistributed through the interacting edge rather than simply propagating as a free chiral mode. In this regime the two interaction channels $H^{1}_{\text{int}}$ and $H^{2}_{\text{int}}$ contribute comparably, which is consistent with the interpretation of edge~B as an SMG edge.

At $g=6.0$, the pattern is still distinct from the weak-coupling edge, but it becomes visibly more asymmetric: the currents associated with $H^{1}_{\text{int}}$ are stronger and the density profile starts to resemble that of the strong-coupling $H^{1}_{\text{int}}$-dominant sector. This is consistent with a partially gapped regime.

By $g=7.0$ the profile is already close to the strong-coupling pattern, and the
 flavor-mixing current on edge~B is strongly suppressed. At $g=16.0$ and $g=80.0$, edge~B is fully interaction-dominated, and the density-deviation profile is controlled by the strong-coupling branch labeled $(8L_x,7L_x)$ identified in Appendix~\ref{app:A}. The remaining mobile part of the injected charge is then carried by the emergent near-boundary bulk mode rather than by the original edge~B mode.

To sharpen this interpretation, we compare the symmetric line $g_1=g_2=g$ with the single-channel case $g_1=g$, $g_2=0$, shown in Fig.~\ref{figC2}. The purpose of this comparison is to distinguish two effects: genuine SMG physics, which requires the interplay of both interaction channels, and the simpler tendency of the system to fall into an $H^{1}_{\text{int}}$-dominated strong-coupling sector.

At weak coupling ($g=4.0$), the symmetric and single-channel calculations are qualitatively similar, as expected, because the interaction remains too weak. The physically important comparison is in the intermediate regime. At $g=5.5$, the $g_2=0$ state differs clearly from the symmetric $g_1=g_2$ state (see main text): The single-channel calculation does not reproduce the flavor-resolved charge redistribution and endpoint currents observed on the symmetric line. Within the finite-size data studied here, this indicates that the characteristic pattern at $g=5.5$ belongs to the symmetric intermediate regime itself rather than arising from simple $H^{1}_{\text{int}}$ dominance.

By contrast, at $g=6.0$ the symmetric state becomes much closer to the $g_2=0$ state. In other words, once the interaction is increased beyond the SMG regime, the edge response begins to look like that of a system in which one interaction channel dominates. This is consistent with a partially gapped regime on the finite ladder, in which the bosonic mode associated with $H^{1}_{\text{int}}$ is preferentially pinned while another bosonic degree of freedom remains gapless. From this perspective, $g \simeq 6.0$ may already lie on the $\mathbb{Z}_2^{R}$-broken side of the transition, but has not yet reached the fully decoupled strong-coupling limit.

\begin{figure}[!htbp]
    \centering
    \includegraphics[width=1.0\textwidth]{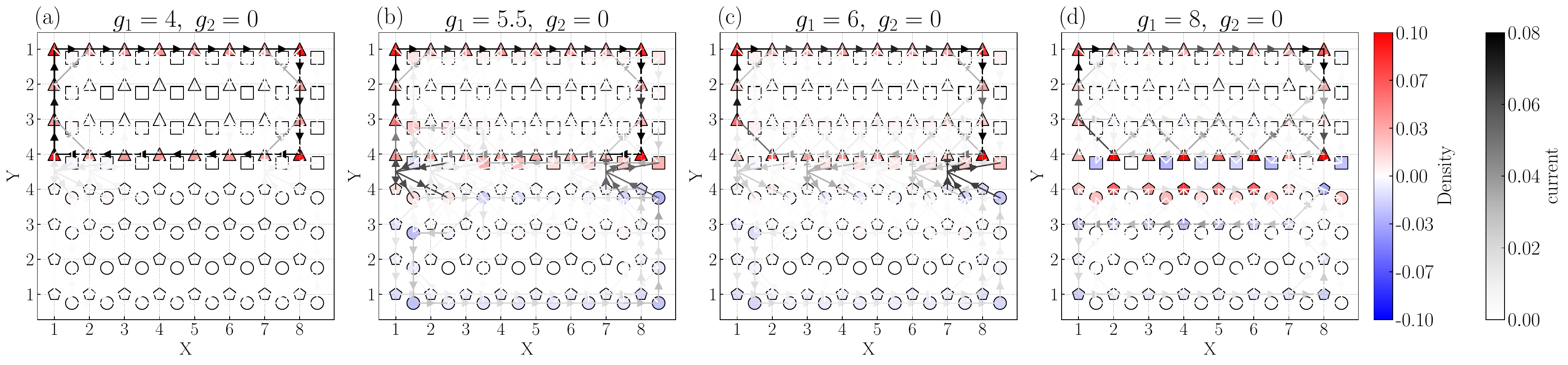}
    \caption{Deviations in the fermion densities and currents for $g_1=g$ and $g_2=0$ at interaction strengths $g=4.0,5.5,6.0,8.0$, for $L_x=8$ and $L_y=4$ under open boundary conditions. The system is at half-filling with one additional fermion of flavor $I=1$ and charge $(Q^{(1)},Q^{(2)})=(3,0)$. All conventions are the same as in Fig.~\ref{figC1}.
    \label{figC2}}
\end{figure}

\subsection{Charge and current distributions from different strong-coupling sectors}
\label{app:C-sectors}
The four symmetry-related branches of Appendix~\ref{app:A} have equal energies in the decoupled limit at half filling, where the total charge sector is invariant under the full $D_4$. With one fermion added, no nontrivial element of $D_4$ preserves the total charge sector ($U^{R}_{(1,3)}$, $U^{R}_{(1,4)}$ and $U_{\mathcal{C}}$ map the added charge $(3,0)$ to $(5,4)$, $(0,3)$ and $(-3,0)$), so the four strong-coupling configurations become inequivalent branches already in the strict $g\to\infty$ limit: their edge parts remain degenerate, but the bulk absorbs a different charge in each. We therefore first construct the analytical profiles in the $g\to\infty$ limit and then compare them with DMRG states targeted in each branch at $g=80.0$.

We fix the total charge to $(Q^{(1)}_{\mathrm{tot}}, Q^{(2)}_{\mathrm{tot}}) = (6L_xL_y + 3,\; 6L_xL_y)$, corresponding to half-filling plus one additional fermion of flavor $I=1$. For each strong-coupling branch, charge conservation gives $(Q^{(1)}_{\mathrm{bulk}}, Q^{(2)}_{\mathrm{bulk}}) = (Q^{(1)}_{\mathrm{tot}} - Q^{(1)}_B,\; Q^{(2)}_{\mathrm{tot}} - Q^{(2)}_B)$, and the corresponding bulk profile is the free-fermion ground state of the decoupled Chern insulators at that filling. The four analytical profiles are shown in the upper row of Fig.~\ref{figC3}, each labeled by its edge-B charge pair.

Under open boundary conditions there is a subtlety: with $H^{1}_{\mathrm{int}}$ acting on bonds $(r,r{+}1)$ with $r=1,\ldots,L_x{-}1$, the six-body terms do not reach the rightmost edge-B site $r=L_x$ for flavors $I=1,3$. These corner sites therefore remain coupled to the adjacent bulk row by residual hopping even in the strong-interaction limit, and are most naturally grouped with the bulk rather than with the interaction-locked edge-B interior. For the purpose of computing the analytical charge profile, we accordingly treat the ``bulk'' as the $(L_y-1)\times L_x$ bulk rows augmented by the rightmost edge-B site for $I=1,3$, with the analogous $I=2,4$ corners included in the bulk for the $H^{2}_{\mathrm{int}}$-dominant branches. The edge-B charge pairs label the four branches. Since each flavor of the decoupled Chern-insulator bulk is noninteracting, its charge profile on this augmented bulk geometry is the free-fermion ground state at the corresponding filling.

As a concrete example, consider $L_x=8$, $L_y=4$ in the edge-B sector $(Q^{(1)}_B,Q^{(2)}_B)=(8L_x,7L_x)$. Subtracting the edge-B contribution gives reference bulk fillings of $5/12$ for $I=1,3$ and $1/2$ for $I=2,4$. For the finite system, we choose the nearest charge-compatible integer occupations, accounting for the added fermion and corner sites. Applied at other sizes, the same construction reproduces the $g=80$ DMRG density profiles at $L_x=4,6$ (Appendix~\ref{app:H}).
The same construction applies to the remaining three edge-B sectors in the $D_4$ multiplet of Eq.~\eqref{eq:corners}.

The resulting analytical density-deviation profile for $(8L_x, 7L_x)$ agrees with the $g=80$ DMRG data in Fig.~\ref{figC1}(f), consistent with this branch assignment for the adiabatically tracked large-$g$ state in Fig.~\ref{figC1}. For completeness, we also run DMRG at $g=80.0$ from initial states in the three remaining branches, and show the full set of profiles in Fig.~\ref{figC3}. The close agreement between the infinite-$g$ construction (upper row) and the finite-$g$ DMRG data (lower row) validates both the four-branch construction and the analytical solution of Appendix~\ref{app:A}.

The four branches are not degenerate. Measuring $\langle H\rangle$ on the four $g=80$
states orders them with $(8L_x,7L_x)$ lowest and $(4L_x,5L_x)$ highest, the remaining two
lying close together in between, the ordering following from where the added fermion is
absorbed in each branch. The added fermion occupies the gapless boundary mode that emerges after edge~B decouples from the bulk. At the sizes studied here, the four branches have well-resolved energy splittings, with the adiabatically tracked branch of Fig.~\ref{figC1} having the lowest energy.

\begingroup
\setlength{\intextsep}{6pt plus 2pt minus 2pt}
\begin{figure}[!htbp]
    \centering
    \includegraphics[width=0.90\textwidth]{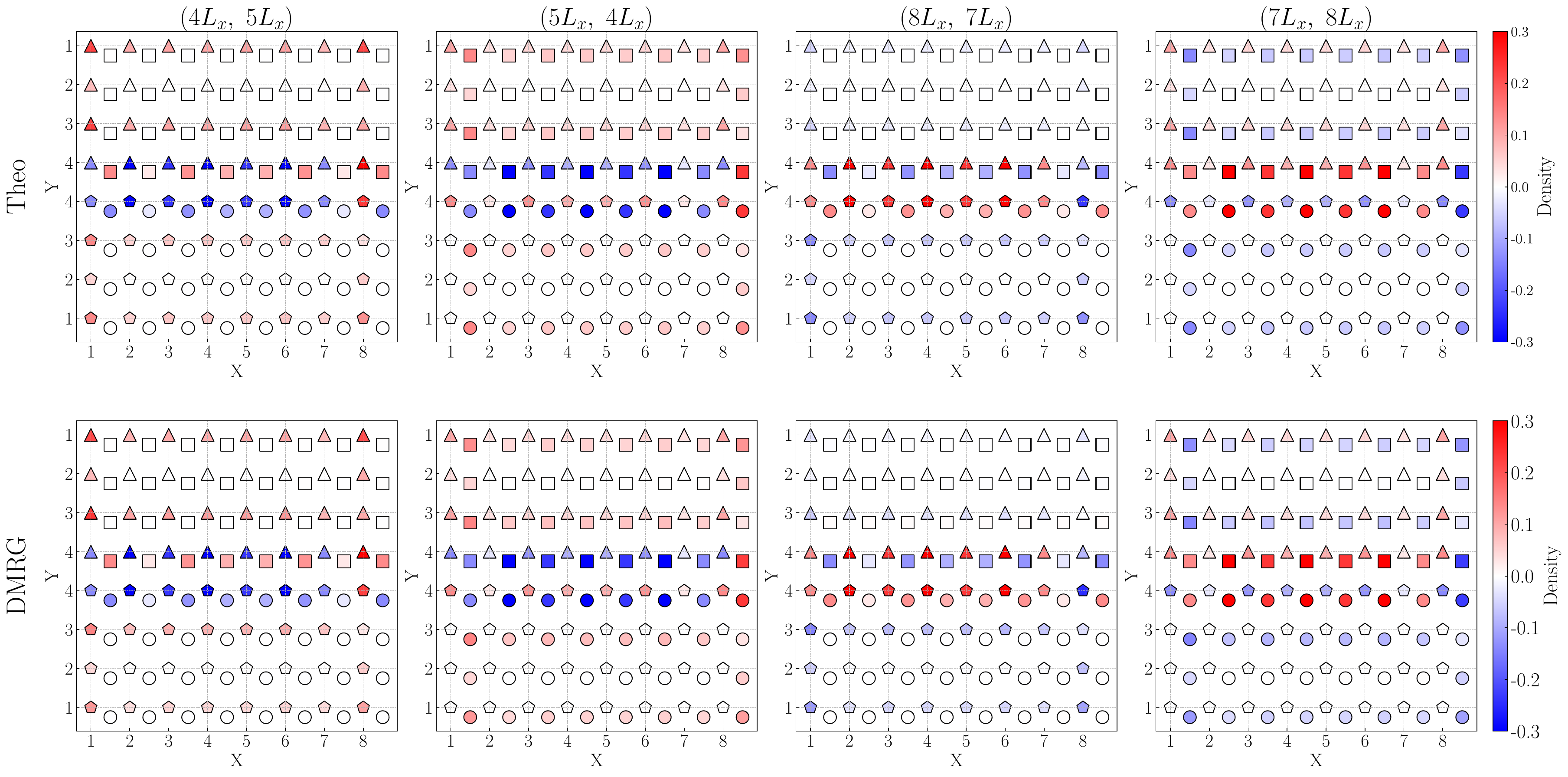}
 \caption{Deviations in the fermion densities for the four strong-coupling branches, labeled by their edge-B charge pairs: $(Q_B^{(1)}, Q_B^{(2)}) = (4L_x, 5L_x)$, $(5L_x, 4L_x)$, $(8L_x, 7L_x)$, and $(7L_x, 8L_x)$. Upper row: analytical predictions from the infinite-interaction limit. Lower row: DMRG results at $g=80.0$ from different initial states. The color scale encodes the density deviation from the half-filling value $0.5$. System: $L_x=8$, $L_y=4$ at half-filling with one additional fermion of charge $(Q^{(1)},Q^{(2)})=(3,0)$ under open boundary conditions. All conventions are the same as in Fig.~\ref{figC1}.
    \label{figC3}}
\end{figure}

\begin{figure}[!htbp]
    \centering
    \includegraphics[width=0.87\textwidth]{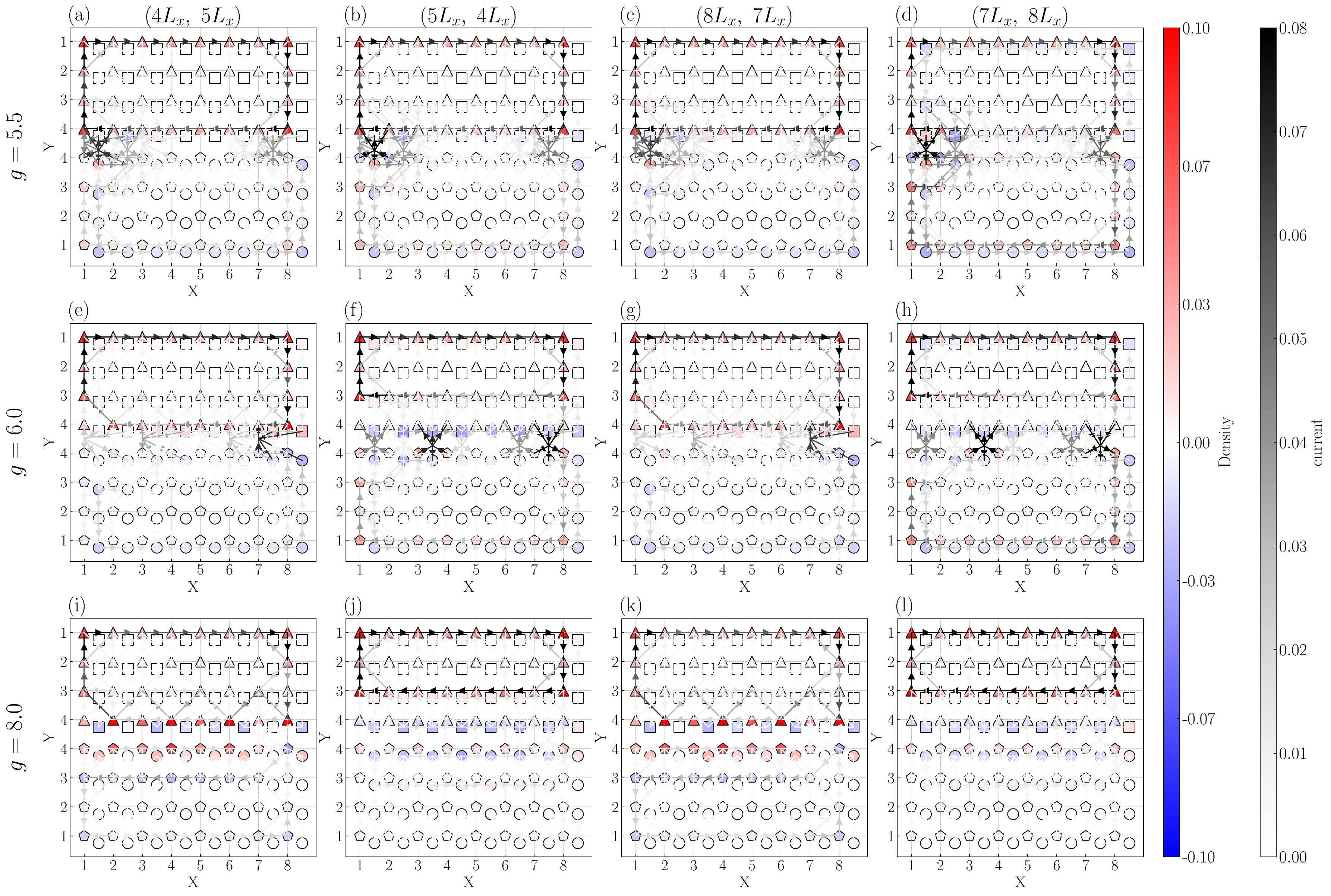}
    \caption{Deviations in the fermion densities and currents at $g=5.5,6.0,8.0$, obtained by adiabatically reducing $g$ from $g=80.0$ starting from the four strong-coupling branches labeled by their edge-B charge pairs: $(Q_B^{(1)}, Q_B^{(2)}) = (4L_x, 5L_x)$, $(5L_x, 4L_x)$, $(8L_x, 7L_x)$, and $(7L_x, 8L_x)$. System: $L_x=8$, $L_y=4$ at half-filling with one additional fermion of charge $(Q^{(1)},Q^{(2)})=(3,0)$ under open boundary conditions. All conventions are the same as in Fig.~\ref{figC1}.
    \label{figC4}}
\end{figure}

At finite $g$, bulk--edge hybridization can connect branches with
$U_{\mathcal{C}}$-related edge configurations through charge transfer between edge~B and the
bulk. The four strong-coupling branches can therefore merge into two profile classes, with one profile class corresponding to each interaction channel.

To test this expectation, we initialize DMRG at $g=80.0$ in each of the four strong-coupling sectors and then reduce $g$ stepwise to smaller values. The resulting density profiles are shown in Fig.~\ref{figC4}. At $g=8.0$, the four DMRG trajectories converge into two distinct profile classes, each associated with one interaction channel. This two-class structure persists at $g=6.0$, although the separation is weaker, consistent with the proposed $\mathbb{Z}_2^{R}$-broken, partially gapped regime before the fully decoupled strong-coupling pattern develops. At $g=5.5$, by contrast, the profiles become substantially more similar. The common qualitative pattern of charge redistribution is consistent with the symmetric SMG regime.

\endgroup

\FloatBarrier
\endgroup
\renewcommand{\thefigure}{C\arabic{figure}}
\begingroup
\raggedbottom
\setlength{\textfloatsep}{6pt plus 1pt minus 1pt}
\setlength{\intextsep}{6pt plus 1pt minus 1pt}
\setlength{\floatsep}{6pt plus 1pt minus 1pt}
\setlength{\abovecaptionskip}{4pt}
\setlength{\belowcaptionskip}{0pt}
\section{Quantized charge pumping: bulk origin and numerical details}
\setcounter{figure}{0}
\label{app:D}
\subsection{Pumping setup and bulk prediction}
\label{app:D-bulk}
As in the main text, we insert flux by continuously interpolating the $\mathbb{Z}_{10}^{(2)}$ twist along $\ell_2=(2,1,-2,1)$. Appendix~\ref{app:D-embeddings} compares this with a second interpolation path. Threading $p$ flux quanta corresponds to the boundary twist
\begin{equation}
\Psi_{I,L_x+1}=e^{\ii\theta\ell_{2,I}}\Psi_{I,1},
\qquad \theta=2\pi p/10.
\label{eq:twistBC}
\end{equation}
In the DMRG calculations at $L_x=8$ and $L_y=4$, this twist is applied adiabatically, relative to antiperiodic boundary conditions that lift the free-fermion degeneracies at half-filling. For Fig.~\ref{fig5}, we decrease $p$ from $0$ to $-1$, removing one flux quantum. For the interpolation along $\ell_2$, the bulk Hall response gives
\begin{equation}
\Delta Q_A^{(a)}=\frac{\theta}{2\pi}\,q^{(a)\intercal}K^{-1}\ell_2,
\qquad a=1,2.
\end{equation}
Here $K=\mathrm{diag}(1,1,-1,-1)$ encodes the edge-mode chiralities~\cite{Wen1990}.
The mixed-anomaly coefficients $q^{(a)\intercal}K^{-1}\ell_2$ are $(20,10)$.
Removing one $\mathbb{Z}_{10}^{(2)}$ flux quantum, $\theta=-2\pi/10$, therefore gives
$(\Delta Q_A^{(1)},\Delta Q_A^{(2)})=(-2,-1)$.

To test this prediction we measure the charge transported to each row. The $\mathrm{U}(1)^{(1)}$ charge of row $Y$ is
\begin{equation}
Q^{(1)}_{Y} = \sum_{r \in Y} \left( 3n_{1,r} + 4n_{2,r} + 5n_{3,r} + 0\, n_{4,r} \right),
\label{eq:Q1row}
\end{equation}
with $n_{I,r} = \Psi_{I,r}^{\dagger}\Psi_{I,r}$ the particle-number operator for flavor $I$ at site $r$, and we report its deviation from half-filling $\Delta Q^{(1)}_{Y} = Q^{(1)}_{Y} - Q_{\text{bg}}$, with $Q_{\text{bg}} = 6L_x$ the background charge of a half-filled row. At $t_2=0.5$ the edge modes are tightly localized on the outermost row, so $\Delta Q^{(1)}_{\text{edge A}}$ already captures the full pumped charge. For smaller $t_2$, where the edge modes extend further into the bulk, one would instead need to sum $\Delta Q^{(1)}_Y$ over the first few rows.

\subsection{Uniformity at integer multiples of the \texorpdfstring{$\mathbb{Z}_{10}^{(2)}$}{Z10(2)} flux quantum}
\label{app:D-uniformity}
Figure~\ref{figD1} shows the charge density under different threaded fluxes and interaction strengths. At integer $p$, the charge density is translationally uniform provided that the state preserves the dressed translation symmetry. The uniformity follows from a dressed one-site magnetic translation. The magnetic translation $T$ acts for $x=2,\ldots,L_x$ as
\begin{equation*}
T^{\dagger}\Psi_{I,x,y}T=
\begin{cases}
\ii\lambda(x,y)\Psi_{I,x-1,y}, & I=1,2,\\
\ii\lambda^{*}(x,y)\Psi_{I,x-1,y}, & I=3,4.
\end{cases}
\end{equation*}
Here $x=1,\ldots,L_x$ and $y=1,\ldots,L_y$, with $y=1$ on edge~A, and $\lambda(x,y)=(-1)^{x-1}(-\ii)^{(x+y)\bmod2}$ is the same staggered phase as in Eq.~\eqref{eq:lambda}. Define a local phase rotation $U_p$ acting on all fermions in the last column,
\begin{equation}
U_p^{\dagger} \Psi_{I,L_x} U_p = e^{\ii\theta\ell_{2,I}}\Psi_{I,L_x}.
\end{equation}
With $T$ defined, the dressed translation $\tilde{T}=T U_p^{\dagger}$ has the same magnetic phases and an additional factor $e^{-\ii\theta\ell_{2,I}}$ in its action at $x=1$. The hopping Hamiltonian is invariant for any $\theta$. The two interaction vertices carry total $\ell_2$ charges $0$ and $10$, so the phases introduced by the boundary twist cancel when $10\theta\in2\pi\mathbb{Z}$. Thus $[\tilde{T},H]=0$ at integer $p$. Absent spontaneous symmetry breaking of $\tilde{T}$, the ground-state density matrix obeys $[\tilde{T},\rho]=0$. Combined with $\tilde{T}^{\dagger} n_{I,x} \tilde{T} = n_{I,x-1}$, where $n_{I,0}\equiv n_{I,L_x}$, this yields
\begin{equation}
\langle n_{I,x} \rangle = \mathrm{Tr}(\rho n_{I,x}) = \mathrm{Tr}(\tilde{T} \rho \tilde{T}^{\dagger} n_{I,x}) = \mathrm{Tr}(\rho\, \tilde{T}^{\dagger} n_{I,x} \tilde{T}) = \langle n_{I,x-1} \rangle,
\end{equation}
a uniform charge density at integer $p$. For nonzero $g_2$ and noninteger $p$, this dressed translation is broken along $\ell_2$, allowing the nonuniform profiles in Fig.~\ref{figD1}.

\begin{figure}[!htbp]
    \centering
    \includegraphics[width=1.0\textwidth]{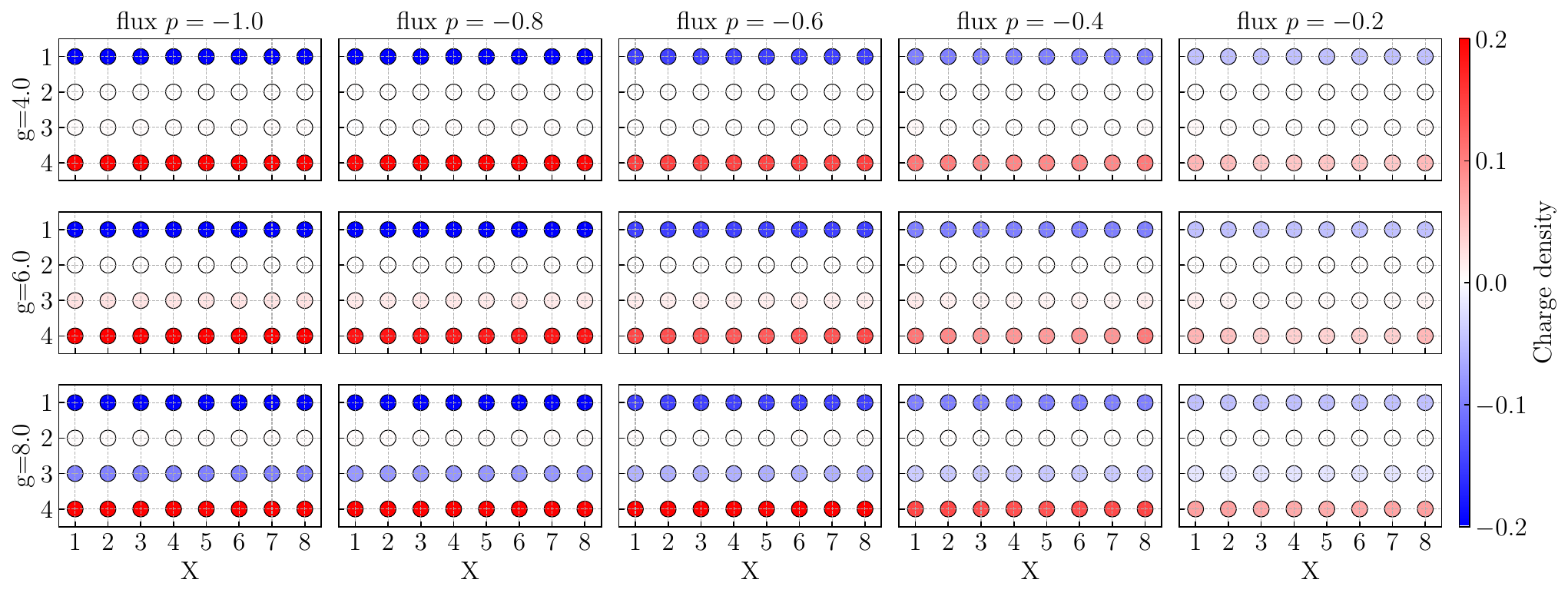}
    \caption{Charge density distributions for different interaction strengths $g$ and insertions of $\mathbb{Z}_{10}^{(2)}$ flux. Rows correspond to different values of $g$, while columns correspond to different flux values. The column labels give the inserted flux $p$ in units of one $\mathbb{Z}_{10}^{(2)}$ flux quantum with $\theta = 2\pi p/10$.
    \label{figD1}}
\end{figure}

\subsection{Edge-B charge accumulation in the strong-interaction limit}
\label{app:D-accumulation}
As shown in Fig.~\ref{fig5}(a) of the main text, the charge pumped onto edge~A is proportional to the inserted flux at every interaction strength considered, while the response on edge~B depends sensitively on $g$. Figure~\ref{figD2}(a) reports the total charge accumulated on edge~B as a function of the inserted flux $\theta$ for $g=4.0,\,5.0,\,6.0,\,8.0$ and $g=16.0$. In all cases the response is linear for small $\theta$, but its slope varies with $g$, revealing three distinct regimes.

At $g=4.0$, the pumped charge remains on gapless edge~B and matches the response of edge~A. At $g=5.0$--$6.0$, part of the charge spreads into the adjacent bulk row, reducing the edge-B response. At $g=8.0$ and $g=16.0$, the low-energy channel shifts to that row. After one flux quantum, edge~B can accumulate more than $2$ units of charge, while the combined charge on the two rows remains approximately $2$. We explain this excess using a two-level effective theory of the $\mathbb{Z}_2^{R}$-broken strong-coupling state.

As discussed in Appendix~\ref{app:A}, in the $g\to\infty$ limit the $H^{1}_{\mathrm{int}}$-dominant $\mathbb{Z}_2^{R}$-SSB sector hosts two degenerate edge-B ground states related by the PH symmetry $U_{\mathcal{C}}$, with charges $(Q^{(1)}_B, Q^{(2)}_B) = (8L_x,7L_x)$ and $(4L_x,5L_x)$, with $U_{\mathcal{C}}|(8L_x,7L_x)\rangle = |(4L_x,5L_x)\rangle$. At finite $g$, coupling to the bulk lifts this degeneracy through virtual hopping across the edge-B/bulk interface. Projecting onto the subspace spanned by $|(8L_x,7L_x)\rangle$ and $|(4L_x,5L_x)\rangle$ gives the effective Hamiltonian (omitting the constant shift proportional to the identity)
\begin{equation}
H_{\mathrm{eff}}(\theta) = \varepsilon(\theta)\,\sigma_z + \Delta(\theta)\,\sigma_x,
\end{equation}
where $\theta$ is the inserted flux, $\sigma_z = |(8L_x,7L_x)\rangle\langle(8L_x,7L_x)| - |(4L_x,5L_x)\rangle\langle(4L_x,5L_x)|$ distinguishes the two PH-related charge sectors, and $\sigma_x$ encodes the tunneling between them. It requires that $g$ be large enough for the two sectors to stay separated from the rest of the spectrum, and the flux dependence is carried by $\varepsilon(\theta)$ and $\Delta(\theta)$.

Virtual edge--bulk hopping with characteristic amplitude $t$ generates local energy shifts of order $t^2/g$. Assuming a size-independent local coupling to the pumped charge, we estimate $\varepsilon(\theta)\sim(t^2/g)\theta$. In contrast, tunneling between the two sectors requires $O(L_x)$ virtual hops, making $\Delta$ exponentially small in $L_x$ at strong coupling (Appendix~\ref{app:A}). PH exchanges the sectors and reverses $\theta$, so $\varepsilon$ is odd and $\Delta$ is even in $\theta$. Thus $\Delta(\theta)=\Delta(0)+O(\theta^2)$, with $\Delta(0)\sim g(t/g)^n$ and $n=O(L_x)$ [Eq.~\eqref{eq:splitting}].

To extract $\varepsilon$ from ED, we rotate the low-energy eigenstates into approximate branch states by diagonalizing a linear combination of the projected edge-B charge and channel-difference operators. For the PH-related pair in the $H^1_{\mathrm{int}}$ channel, $\varepsilon$ is half the difference between the diagonal matrix elements of the Hamiltonian in this basis, and $\Delta$ is the magnitude of the off-diagonal matrix element. Figure~\ref{figD3} supports the $1/g$ scaling of the bias. These sizes are too small to test whether $\varepsilon$ is independent of $L_x$.

The antisymmetric edge-B charge response in Fig.~\ref{figD2}(b) illustrates the odd dependence of $\varepsilon(\theta)$ on flux. For the parameters shown, $\theta<0$ favors $|(8L_x,7L_x)\rangle$ and $\theta>0$ favors $|(4L_x,5L_x)\rangle$, shifting the edge-B charge above and below $6L_x$, respectively.
 The non-linear dependence of the edge B charge on $\theta$ seen in Fig.~\ref{figD2}(a) is consistent with a model where $\epsilon$ varies linearly with $\theta$ but $\Delta$ is constant.

We note, in addition, that for $g_1 = g_2$, the true ground state for a finite system should be an equal linear combination of the ground states corresponding to the two sectors of
$\langle \Delta H_{\rm{int}}\rangle $, separated in energy from the orthogonal combination by an amount that vanishes exponentially with the system size.

\begin{figure}[!htbp]
    \centering
    \includegraphics[width=0.45\textwidth]{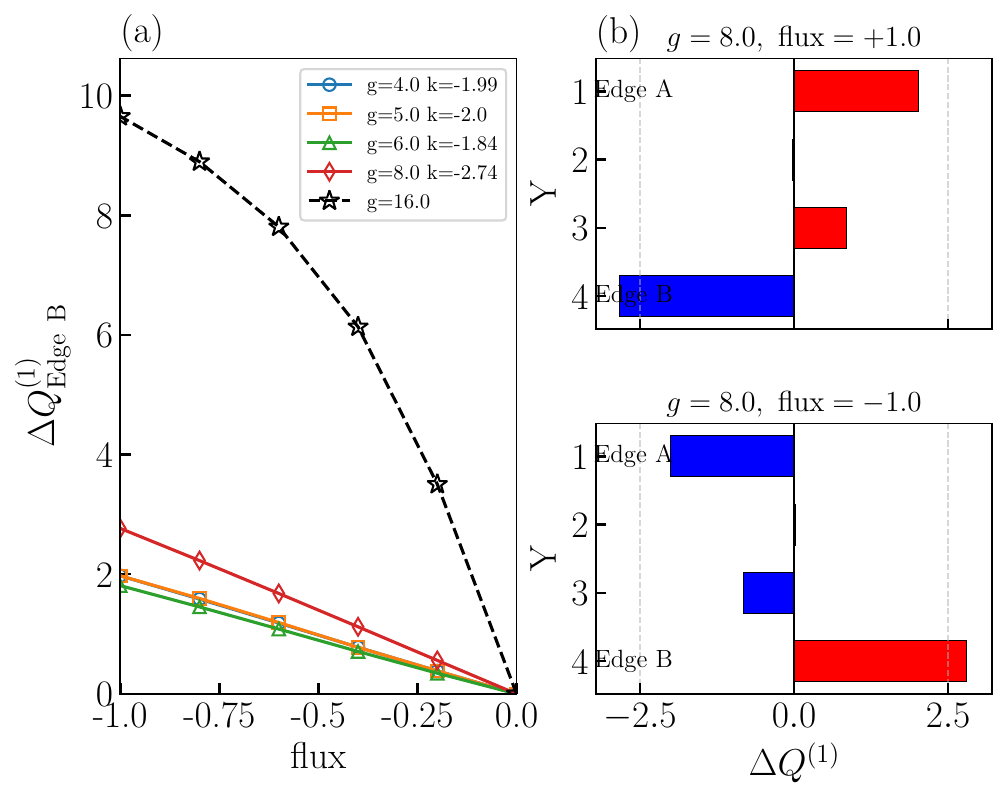}
    \caption{(a) Charge deviation on edge~B as a function of the inserted flux for different interaction strengths $g$. (b) Integrated charge deviation $\Delta Q^{(1)}_Y$ for each row $Y$ after insertion of one positive or one negative $\mathbb{Z}_{10}^{(2)}$ flux quantum.
    \label{figD2}}
\end{figure}

\begin{figure}[!htbp]
    \centering
    \includegraphics[width=0.30\textwidth]{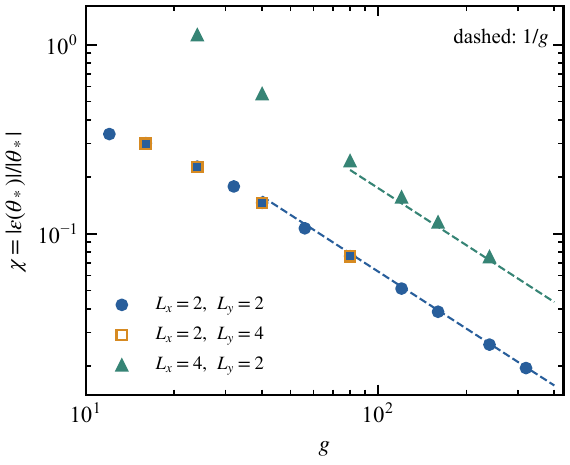}
    \caption{Flux-induced bias from exact diagonalization. The finite-flux ratio $\chi=|\varepsilon(\theta_*)|/|\theta_*|$ is shown versus $g$ for the indicated sizes, with $t_1=1$, $t_2=0.5$, and $|\theta_*|=\pi/10$ measured from the antiperiodic starting point. Dashed lines are guides proportional to $1/g$.
    \label{figD3}}
\end{figure}

\subsection{Species-resolved pumped charge distribution}
\label{app:D-species}
The row-resolved deviation $\Delta Q^{(1)}_Y$ in Fig.~\ref{fig5}(b) of the main text combines the four flavors with their $\mathrm{U}(1)^{(1)}$ charge weights $(3,4,5,0)$ into a single bar per row
for each value of $g$.
To expose how the individual flavors contribute, Fig.~\ref{figD4} shows the unweighted density deviation $\sum_{x}(\langle n_I\rangle-\tfrac12)$ of each flavor $I\in\{1,2,3,4\}$ (carrying $\mathrm{U}(1)^{(1)}$ charges $3,4,5,0$), resolved per folded row, after removal of one $\mathbb{Z}_{10}^{(2)}$ flux quantum. The rows of the grid correspond to increasing interaction strength $g\in\{4,5,6,8\}$ and the columns to the four flavors. Within each panel the bars run from edge~A (top) to edge~B (bottom).

In the weak-coupling regime, the pumped density is carried by the gapless chiral modes and is localized on the two edges. In the intermediate SMG regime, the species pattern is close to the weak-coupling one, except that the edge-B charge now penetrates into the adjacent bulk row. In the strong-coupling regime, the edge-B charge instead redistributes among flavors according to the selected $\mathbb{Z}_2^{R}$-broken charge sector, consistent with the edge-B analysis above.

\begin{figure}[!htbp]
    \centering
    \includegraphics[width=0.8\textwidth]{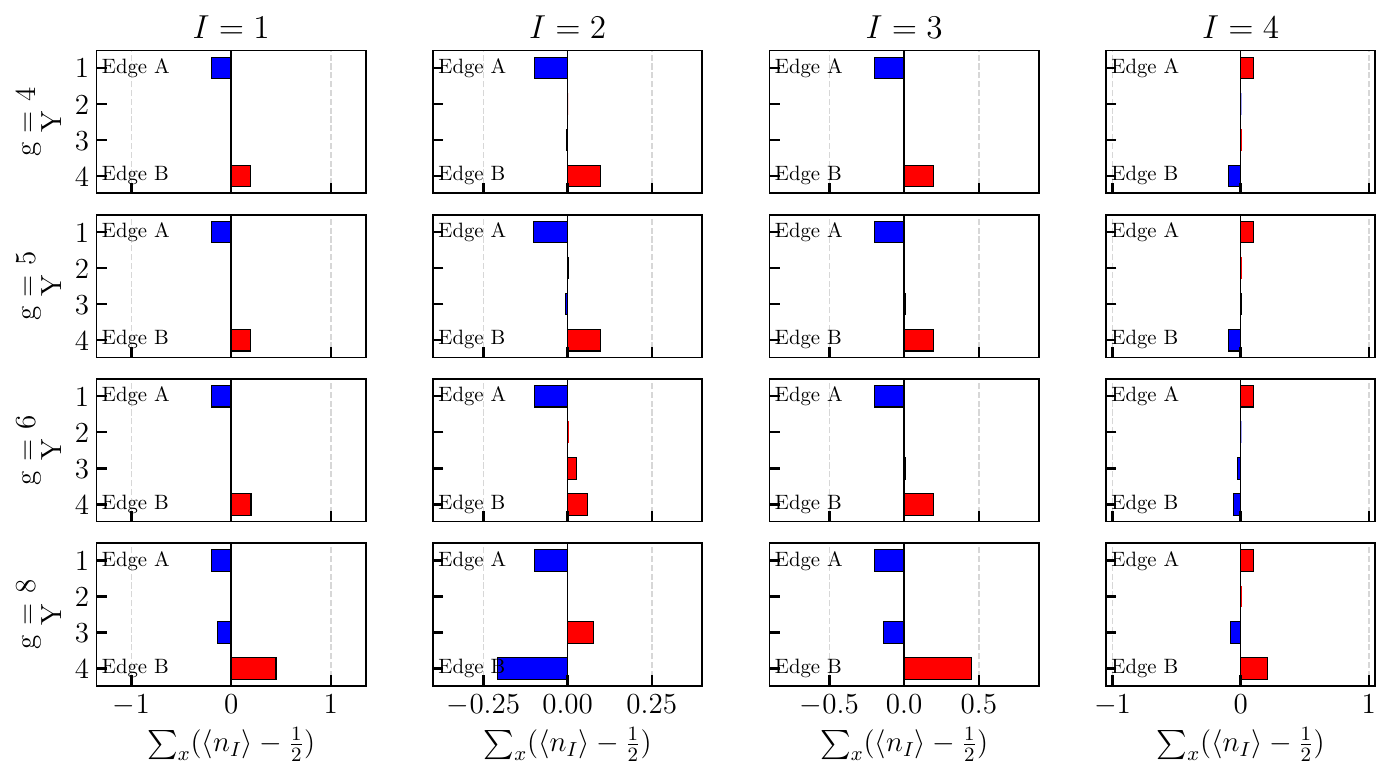}
    \caption{Species-resolved density deviation $\sum_{x}(\langle n_I\rangle-\tfrac12)$ per folded row after removal of one $\mathbb{Z}_{10}^{(2)}$ flux quantum. Rows correspond to interaction strengths $g\in\{4,5,6,8\}$ and columns to the four flavors $I\in\{1,2,3,4\}$ (with $\mathrm{U}(1)^{(1)}$ charges $3,4,5,0$). Within each panel edge~A is at the top and edge~B at the bottom, with red (blue) bars denoting positive (negative) deviation. Panels in each column use the same horizontal-axis scale, also used in Fig.~\ref{figD6}.
    \label{figD4}}
\end{figure}

\begingroup
\subsection{Two interpolation paths for the same symmetry twist}
\label{app:D-embeddings}

We compare two continuous interpolations of the same $\mathbb{Z}_{10}^{(2)}$ symmetry twist, with boundary phases $e^{\ii\theta\ell_{2,I}}$ and $e^{\ii\theta(4q^{(1)}_I-3q^{(2)}_I)}$, respectively. The Hamiltonians coincide at integer $p$ but differ at intermediate flux. The second path lies within the preserved $\mathrm{U}(1)^{(1)}\times\mathrm{U}(1)^{(2)}$ symmetry.

The continuous $\mathrm{U}(1)$ symmetries used for these interpolations have mixed-anomaly coefficients $(20,10)$ and $(0,0)$ with the two conserved symmetries, respectively. The bulk Hall response therefore predicts
\begin{equation}
\left(\Delta Q_A^{(1)},\Delta Q_A^{(2)}\right)
=\begin{cases}
p(2,1),&\ell_2,\\
(0,0),&4q^{(1)}-3q^{(2)}.
\end{cases}
\end{equation}
The zero mixed Hall contribution along the second path allows changes in individual flavor densities, while their charge-weighted contributions cancel.

Figure~\ref{figD5} compares the charge and energy along both paths, starting from the corresponding zero-flux states and decreasing $p$ from $0$ to $-1$. Along both paths, we use two two-site DMRG sweeps per step of $0.2$ in $p$, as in Fig.~\ref{fig5}. The $g=0$ curves are an independent free-fermion calculation that follows the occupied single-particle levels continuously through their crossings along each path. The curves describe states followed along each path, whose endpoint Hamiltonians are identical. The interacting results are consistent with this free-fermion calculation along both paths. At $p=-1$, the $\ell_2$ path gives $\Delta Q_A^{(1)}\simeq-2$ for $g=0$ and every interacting $g$, whereas the $(12,1,8,-9)$ path gives a charge change nearly zero in all cases. The latter path reaches a higher-energy branch of the common endpoint Hamiltonian.

\begin{figure}[!htbp]
\centering
\includegraphics[width=0.9\textwidth]{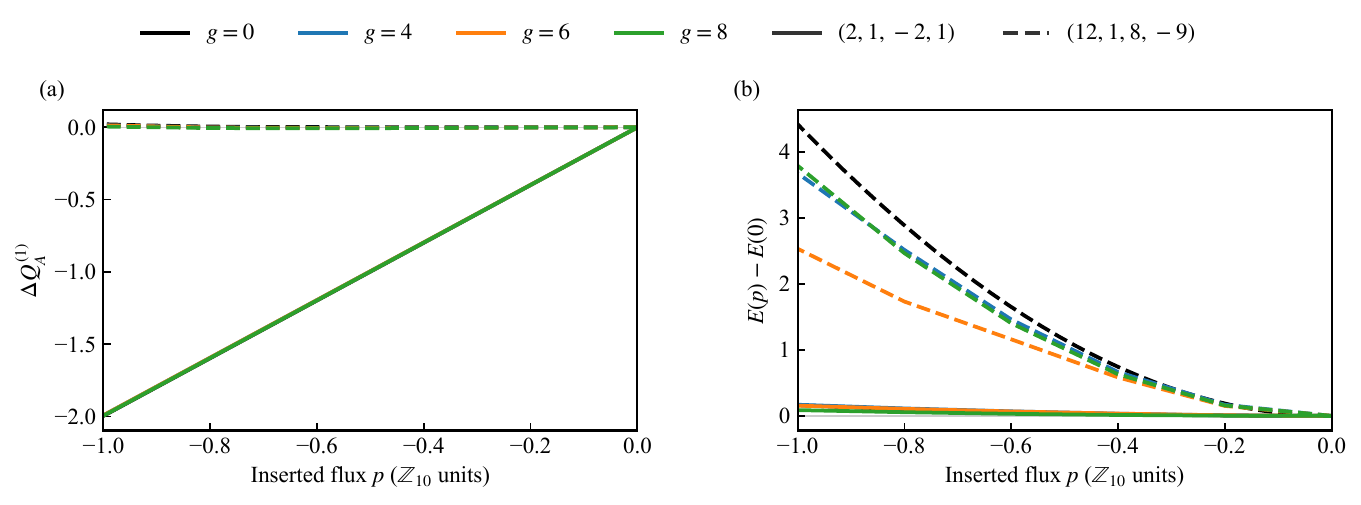}
\caption{(a) Edge-A charge change $\Delta Q_A^{(1)}=Q_A^{(1)}(p)-Q_A^{(1)}(0)$ and (b) energy change $E(p)-E(0)$ as one $\mathbb{Z}_{10}^{(2)}$ flux quantum is removed along two interpolation paths of the same twist, at $L_x=8$ and $L_y=4$. Solid lines use the boundary phase $e^{\ii\theta\ell_{2,I}}$ with $\ell_2=(2,1,-2,1)$, and dashed lines use $e^{\ii\theta(4q^{(1)}_I-3q^{(2)}_I)}$ with $4q^{(1)}-3q^{(2)}=(12,1,8,-9)$. Colors indicate the interaction strength $g=0$, $4$, $6$, and $8$.
\label{figD5}}
\end{figure}

Figure~\ref{figD6} shows the flavor-resolved densities at $p=-1$ along the second path, measured relative to half-filling. On edge~A the profiles nearly coincide across $g$, consistent with the bulk response. On edge~B they differ because the interactions transfer density between flavors while conserving $Q^{(1)}$ and $Q^{(2)}$. Such transfers are generated by $\ell_1=(1,-2,1,2)$ and $\ell_2=(2,1,-2,1)$, with their relative contributions depending on the selected channel. Without interactions, edge~B would mirror edge~A.

\begin{figure}[!htbp]
\centering
\includegraphics[width=0.85\textwidth]{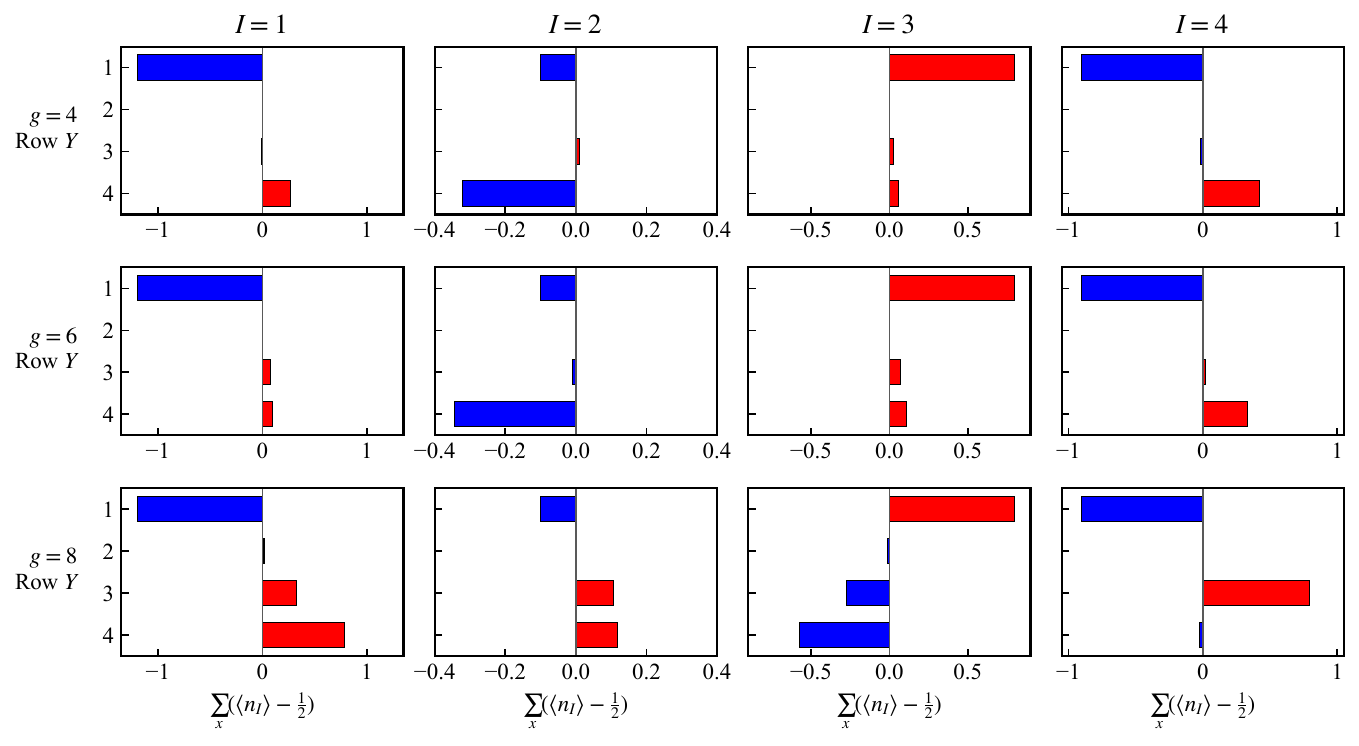}
\caption{Flavor-resolved density deviation $\sum_x(\langle n_{I,x,Y}\rangle-\tfrac12)$ of each folded row $Y$ after one $\mathbb{Z}_{10}^{(2)}$ flux quantum is removed with the boundary phase $e^{\ii\theta(4q^{(1)}_I-3q^{(2)}_I)}$, where $4q^{(1)}-3q^{(2)}=(12,1,8,-9)$, at $L_x=8$ and $L_y=4$. Columns correspond to the flavors $I=1,2,3,4$ and rows to $g=4$, $6$ and $8$. Within each panel edge~A is at the top and edge~B at the bottom, with red (blue) bars denoting positive (negative) deviation.
\label{figD6}}
\end{figure}
\endgroup

\FloatBarrier
\endgroup
\renewcommand{\thefigure}{D\arabic{figure}}
\section{ Single-particle correlations for different \texorpdfstring{$g$}{g}}
\setcounter{figure}{0}
\label{app:E}
Figure~\ref{figE1} shows the single-particle correlations for interaction strengths $g=4.0$, $5.5$, $6.0$, $8.0$, $16$, and $80$, computed for a system of size $L_x=8$, $L_y=4$ with antiperiodic boundary conditions. We plot $|\langle\Psi^\dagger_{I,x_0,Y}\Psi_{I,x_0+r,Y}\rangle|$ from the central reference site $x_0$ at separations $r=1,\ldots,4$ along each row. The correlations on edge~A remain consistent with power-law decay across the interaction strengths shown.

\begin{figure}[!htbp]
    \centering
    \includegraphics[width=0.95\textwidth]{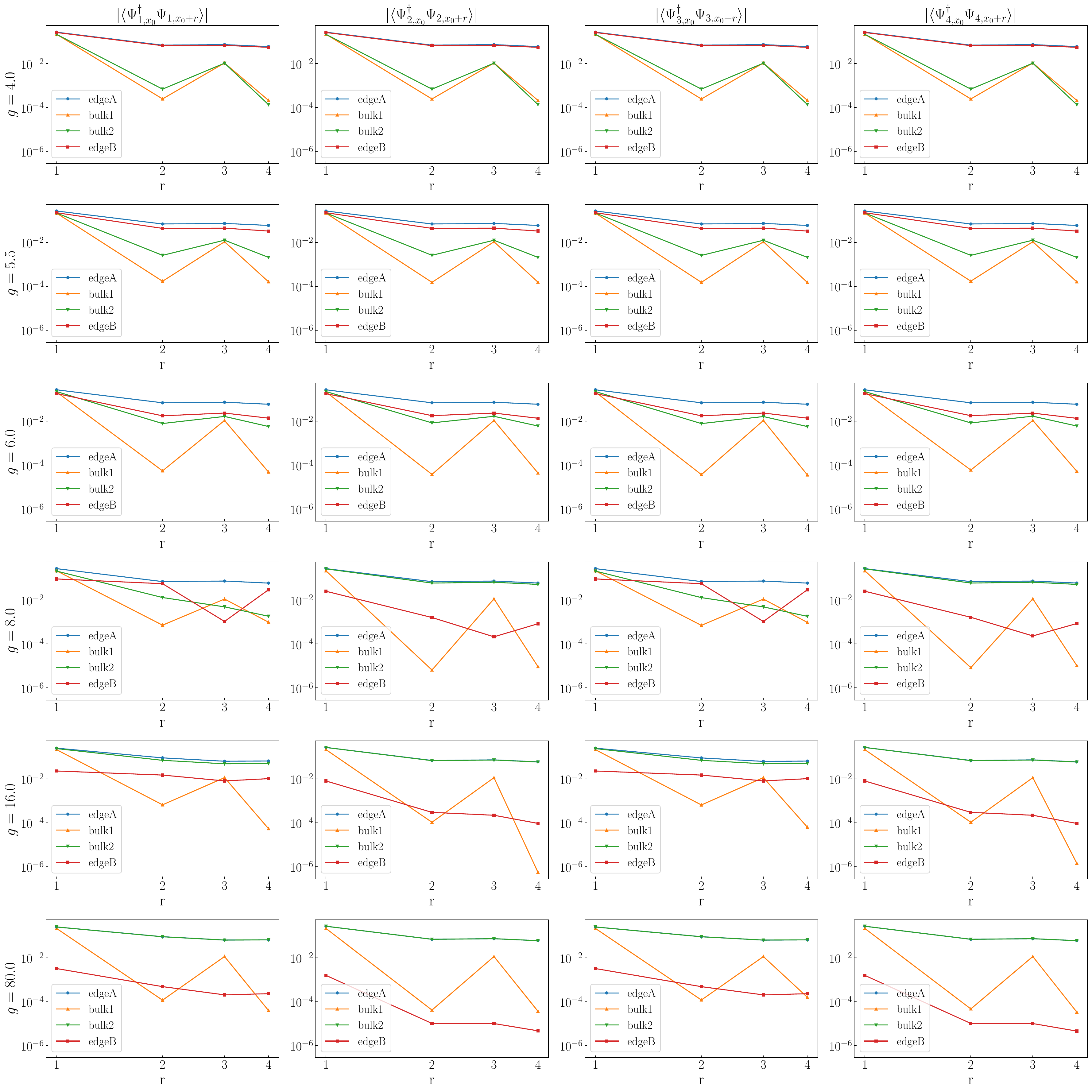}
    \caption{ Absolute values of the single-particle correlations for various flavors and interaction strengths $g$. System size: $L_x=8$, $L_y=4$ with antiperiodic boundary conditions. Correlations are measured from $x_0=4$ at separations $r=1,\ldots,4$ along each row. The coordinates $x$ and $Y$ run from $1$ to $L_x$ and from $1$ to $L_y$, respectively. The legend labels correspond, from top to bottom in the geometry, to edge~A ($Y=1$), bulk$_1$ ($Y=2$), bulk$_2$ ($Y=3$), and edge~B ($Y=4$).
    \label{figE1}}
\end{figure}

For $g=4.0$, the bulk correlations decay much faster than those on the edges, consistent with a large bulk gap. On edge~B, the correlation length is comparable to that on edge~A, indicating a gapless boundary mode. At the intermediate coupling $g=5.5$, the edge~B correlation begins to fall below that of edge~A while still decaying more slowly than the bulk, consistent with the onset of the SMG gap in the symmetric intermediate regime. For $g=6.0$, the correlation on edge~B is clearly reduced, while the neighboring bulk row remains more weakly decaying. The edge-B correlations are suppressed, but the spatial pattern remains distinct from that in the fully decoupled strong-coupling limit. For $g=8.0$, the adjacent bulk row nearly matches edge~A for flavors $I=2,4$, while edge~B is strongly suppressed. For flavors $I=1,3$, the bulk-row correlations still decay appreciably, indicating that the two flavor pairs approach the decoupled limit at different rates. At $g=16$ and $g=80.0$, the edge-B correlations are strongly suppressed relative to edge~A, while the adjacent bulk row (bulk$_2$) is much less suppressed than edge~B, consistent with the decoupled picture in the strong-coupling regime of Appendix~\ref{app:A}. The correlations show similar overall behavior within each flavor pair, $(\psi_1,\psi_3)$ and $(\psi_2,\psi_4)$, consistent with the residual symmetry $U^{R}_{(1,3)}$ of the strong-coupling phase.

\FloatBarrier
\renewcommand{\thefigure}{E\arabic{figure}}
\raggedbottom
\section{Convergence with system size \texorpdfstring{$L_x$}{Lx}}
\setcounter{figure}{0}
\label{app:H}

The main-text results and the preceding appendices use $L_x=8$, $L_y=4$. In this appendix we repeat the key measurements at $L_x=4$ and $L_x=6$ (keeping $L_y=4$) and compare them with the $L_x=8$ data and with the $g\to\infty$ analytical construction. The comparison shows that the strong-interaction charge sector of Appendix~\ref{app:A}, the quantized charge pumping, and the phase-dependent real-space charge-current distribution, are not artifacts of one particular $L_x$.

\subsection{Strong-coupling density versus the infinite-interaction prediction}

\begin{figure}[!htbp]
    \centering
    \includegraphics[width=0.85\textwidth]{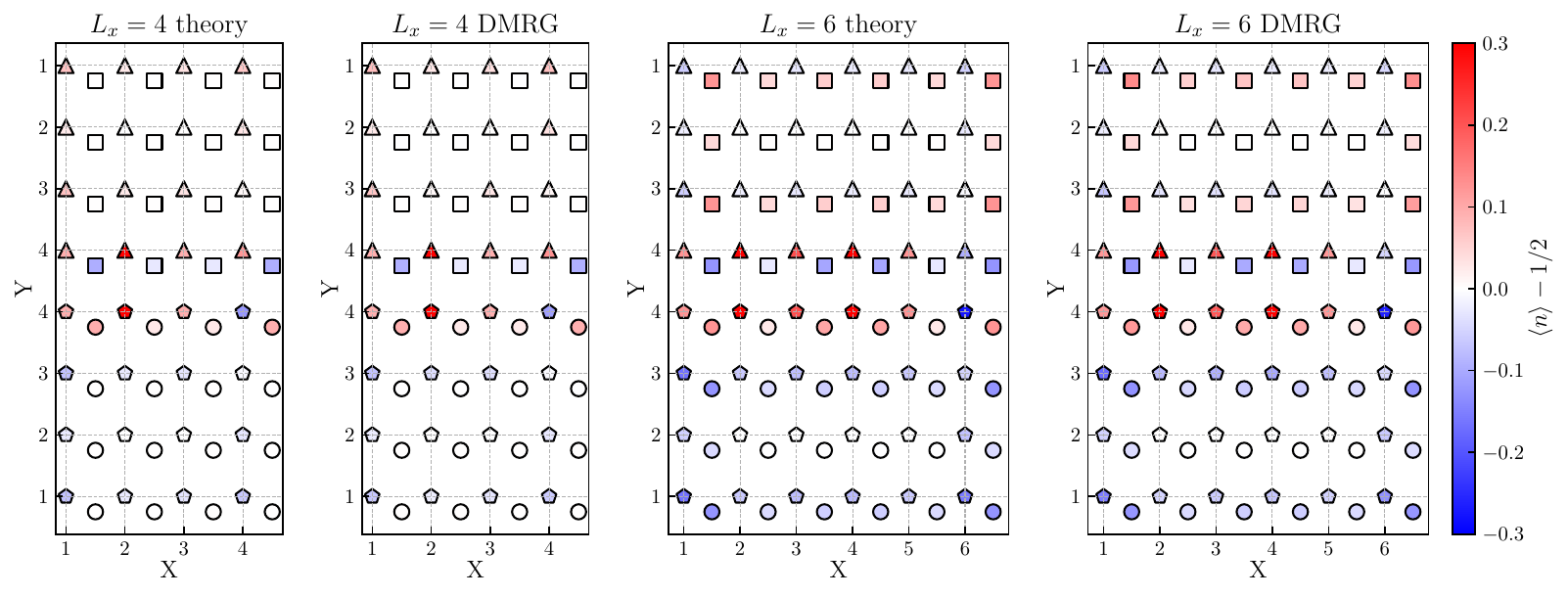}
    \caption{Deviations in the fermion densities for $L_x=4$ (two left panels) and $L_x=6$ (two right panels), $L_y=4$, at half-filling with one additional fermion of charge $(Q^{(1)},Q^{(2)})=(3,0)$ under open boundary conditions. Comparison of the $g\to\infty$ analytical profiles (Appendix~\ref{app:C}) with DMRG at $g=80.0$ on the branch labeled by the edge-B charge pair $(8L_x,7L_x)$. The corner-site correction follows Appendix~\ref{app:C}. Symbols follow Fig.~\ref{figC1}: triangles, squares, pentagons, and circles denote flavors $I=1,2,3,4$.
    \label{figF1}}
\end{figure}

We first test the strong-coupling picture of Appendices~\ref{app:A} and~\ref{app:C} at small $L_x$. For each size, Fig.~\ref{figF1} places the analytical infinite-interaction density of the adiabatically tracked branch labeled by the edge-B charge pair $(8L_x,7L_x)$, obtained from the construction of Appendix~\ref{app:C} evaluated at the corresponding width, next to the DMRG density at $g=80.0$: the four panels show, from left to right, $L_x=4$ theory, $L_x=4$ DMRG, $L_x=6$ theory, and $L_x=6$ DMRG.
As in Appendix~\ref{app:C}, we have added one $I=1$ fermion to the half-filled state and use open boundary conditions.
At both sizes the DMRG density profile is close to the analytical one, supporting the monomer--dimer structure and augmented-bulk charge assignment of Appendix~\ref{app:C}.

\begingroup
\setlength{\textfloatsep}{10pt plus 2pt minus 2pt}
\setlength{\intextsep}{8pt plus 2pt minus 2pt}
\subsection{Charge pumping}
\label{app:H-pumping}

We next repeat the flux-pumping protocol of the main text on the small cylinders. For $L_x=4$ and $L_x=6$ at $L_y=4$, $\mathrm{QN}=0$, we thread $\mathbb{Z}_{10}^{(2)}$ flux, warm-starting each flux step from the converged state at the previous one, at couplings $g=4.0, 6.0, 8.0$ (and additionally $g=10.0$ for $L_x=6$). Figure~\ref{figF2} reports, for each size, the edge-A charge deviation $\Delta Q^{(1)}_{\mathrm{edge\,A}}$ as a function of the inserted flux, together with the row-resolved charge deviation $\Delta Q^{(1)}_Y$ after threading one flux quantum.

The edge-A curves nearly coincide across couplings in panels (a) and (c). Approximately $1.88$ and $1.98$ charges are transferred per flux quantum at $L_x=4$ and $6$, respectively, approaching the quantized value $2$ seen at $L_x=8$ in Fig.~\ref{fig5}(a).

\begin{figure}[!htbp]
    \centering
    \includegraphics[width=0.78\textwidth]{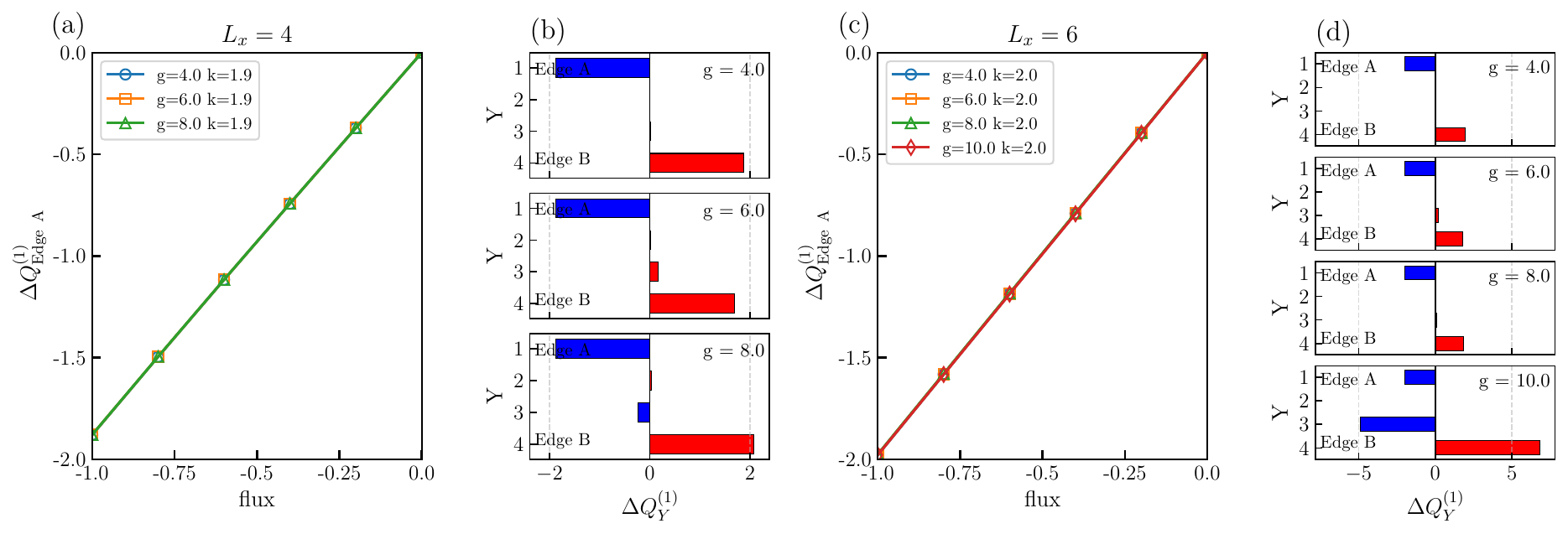}
    \caption{Charge pumping under adiabatic insertion of $\mathbb{Z}_{10}^{(2)}$ flux for $L_x=4$ (panels (a),(b)) and $L_x=6$ (panels (c),(d)), $L_y=4$, $\mathrm{QN}=0$, at $g=4.0, 6.0, 8.0$, and additionally $g=10.0$ for $L_x=6$. Panels (a),(c): edge-A charge deviation $\Delta Q^{(1)}_{\mathrm{edge\,A}}$ versus inserted flux. Panels (b),(d): row-resolved charge deviation $\Delta Q^{(1)}_Y$ per folded row after threading one $\mathbb{Z}_{10}^{(2)}$ flux quantum, one bar group per $g$, with edge~A at the top and edge~B at the bottom. Red (blue) bars denote positive (negative) deviation. All conventions follow Fig.~\ref{fig5}.
    \label{figF2}}
\end{figure}

\subsection{Charge and current distributions}
\label{app:H-profiles}

Figure~\ref{figF3} shows the injected $(3,0)$ fermion under open boundaries at $L_x=4$ (top) and $6$ (bottom). As at $L_x=8$ in Fig.~\ref{figC1}, the excess charge at $g=4$ resides mainly on flavor $I=1$ with a perimeter current. At $g=5.5$, charge redistributes among flavors $I=1,3,4$ and interaction-induced endpoint currents develop. At $g=8$, the edge-B current is suppressed.

\begin{figure}[!htbp]
    \centering
    \includegraphics[width=0.82\textwidth]{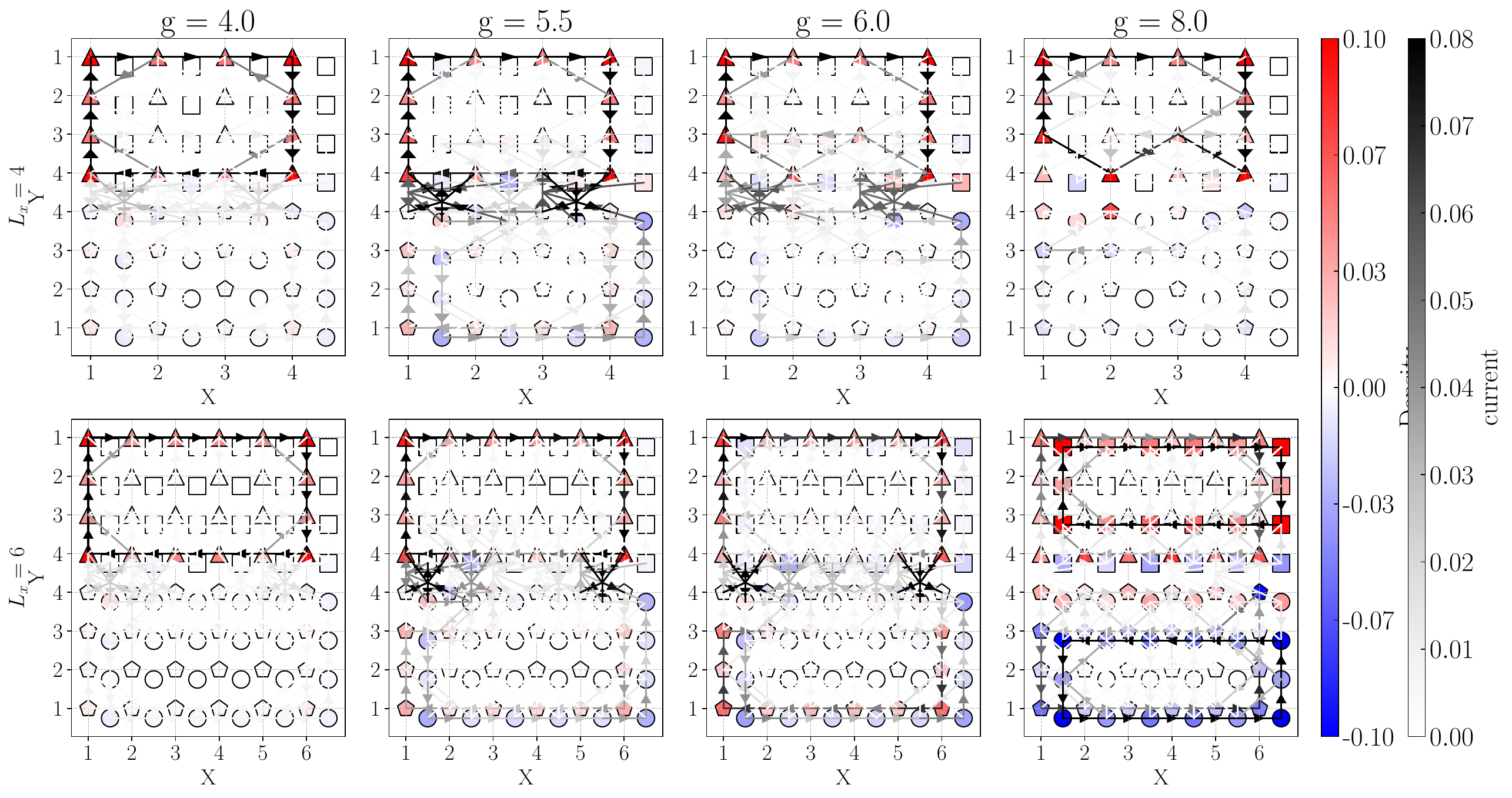}
    \caption{Deviations in the fermion densities and currents for $L_x=4$ (top row) and $L_x=6$ (bottom row), $L_y=4$, at interaction strengths $g=4.0, 5.5, 6.0, 8.0$ under open boundary conditions, with one additional fermion of charge $(Q^{(1)},Q^{(2)})=(3,0)$. The blue-red color scale encodes the density deviation from $0.5$, and the arrows show the measured currents (the interaction-current magnitudes are multiplied by~$5$ in the arrow color scale).
    \label{figF3}}
\end{figure}

\endgroup
\end{document}